\documentclass[11pt, preprint]{aastex63}
\usepackage{bm}
\usepackage{amsmath}

\newcommand{\ii}{\mathrm{in}}
\newcommand{\oo}{\mathrm{out}}

\newcommand{\kep}{\mathrm{Kep}}
\newcommand{\bin}{\mathrm{bin}}

\newcommand{\pzo}{t_\mathrm{i}}
\newcommand{\ooz}{{\oo,0}}
\newcommand{\iiz}{{\ii,0}}

\received{2020 February 25}
\revised{2020 May 5}
\accepted{2020 May 26}
\submitjournal{ApJ}

\shorttitle{Radial-velocity variation of a star orbiting a binary
  black hole}
\shortauthors{Hayashi \& Suto}
\graphicspath{{./}{figures/}}
\begin{document}

\title{Radial-velocity variation of a tertiary star orbiting a
    binary black hole in coplanar and noncoplanar triples: short- and
    long-term anomalous behavior}

\correspondingauthor{Toshinori Hayashi}
\email{toshinori.hayashi@phys.s.u-tokyo.ac.jp}

\author[0000-0003-0288-6901]{Toshinori Hayashi}
\affiliation{Department of Physics, The University of Tokyo,  
Tokyo 113-0033, Japan}

\author[0000-0002-4858-7598]{Yasushi Suto}
\affiliation{Department of Physics, The University of Tokyo,  
Tokyo 113-0033, Japan}
\affiliation{Research Center for the Early Universe, School of Science,
The University of Tokyo, Tokyo 113-0033, Japan}

\begin{abstract}
 A number of ongoing surveys are likely to discover star-black hole binaries in our Galaxy in the near future. A fraction of
  them may be triple systems comprising an inner binary, instead of a
  single black hole, which might be progenitors of binary black holes
  (BBHs) routinely discovered now from the gravitational wave. We
  extend our previous proposal to locate inner BBHs from the
  short-term radial-velocity (RV) variation of a tertiary star in
  coplanar triples, and we consider noncoplanar triples and their
  long-term RV variations as well. Specifically, we assume coplanar
  and noncoplanar triples with an inner BBH of the total mass
  $20~M_\odot$, whose outer and inner orbital periods are 80 days and 10 days, respectively. We perform a series of N-body
simulations and compare the results with analytic approximate
  solutions based on quadrupole perturbation theory. For coplanar
  triples, the pericenter shift of the outer star can be used to
  detect the hidden inner BBH. For noncoplanar triples, the total RV
  semi-amplitude of the outer star is modulated periodically on the
  order of $100$km/s due to its precession over roughly the
  Kozai-Lidov oscillation timescale. Such long-term modulations would
  be detectable within a decade, independent of the short-term RV
  variations on the order of of $100$ m/s at roughly twice the orbital
  frequency of the inner binary.  Thus the RV monitoring of future
  star-black hole binary candidates offers a promising method for searching
  for their inner hidden BBHs in optical bands.
\end{abstract}


\section{Introduction}
\label{sec:intro}

Astronomy is a science triggered and advanced by a series of
surprising discoveries. Well-known examples include black holes (BHs)
and neutron stars, which had been {\it predicted} by physics but
regarded for a long time as merely theoretical concepts.  No law of
physics prohibits the presence of Hot Jupiters and massive binary BHs
(BBHs), but they had not been seriously considered to be detectable in
reality, nor even to exist at all. Their discoveries
\citep{Mayor1995,Abbott2016}, however, have brought revolutions in
astronomy.

A recent discovery of a star--BH binary system, LB-1 \citep{Liu2019},
might be the case as well. The mass of the central BH was originally
claimed to be $68^{+11}_{-13} M_\odot$, which is too large according
to conventional theories of BH formation
\citep[e.g.,][]{Leung2019_2}. An exciting possibility is that LB-1 is
indeed a triple system comprising an inner BBH and an outer orbiting
star.

Several subsequent studies pointed out that the original claim should
be revised; the mass of the inner BH is more likely to be smaller and
between $5 M_\odot$ and $20 M_\odot$
\citep{Abdul-Masih2020,El-Badry2020}, and the presence of a possible
inner BBH in LB-1 is severely constrained \citep{Shen2019}.

 While we revised this paper according to the referee report,
  however, \citet{Shenar2020} reported that LB-1 is unlikely to
  contain a BH, but rather consists of a stripped primary star of
  $\sim1.5~M_\odot$ and a fast-rotating B3 Ve star of $\sim7~M_\odot$
  from their latest spectra observed with HERMES and FEROS.
  Throughout this paper, we still adopt the set of parameters for
  triples inspired from the parameters originally estimated by
  \citet{Liu2019}, \citet{Abdul-Masih2020}, and \citet{El-Badry2020}.
  Nevertheless, our results presented below are applicable to the
  star-BH binaries with similar architecture in general, and provide
  useful strategies for searching for BBHs.

Since it is quite possible that our Galaxy hosts abundant
  star-BH binaries, there are many proposals to search for star-BH
  binaries with {\it Gaia}
  \citep[e.g.][]{Breivik2017,Kawanaka2016,Mashian2017,Yamaguchi2018,
    Shikauchi2020_2} and {\it TESS} \citep[e.g.][]{Masuda2019} among
  others.  A number of such star--BH binaries are likely to be
  detected in the near future, and a fraction of them may turn out to
  be a star -- BBH triple in reality.

Observationally, more than 70 percent of OBA stars and 50
  percent of KGF stars are in binaries or higher multiples
  \citep[][]{Raghavan2010,Sana2012}.  \citet{Rose2019}, for instance,
  performed secular simulations of triples, assuming many initial
  distribution models for orbital parameters. They found that the
  final inner-period distribution after $10$ Myr is statistically
  consistent with the observed distribution of massive binaries in
  \citet{Sana2012} and \citet{Kobulnicky2014}. Thus, it is indeed
  possible that there are abundant triple systems consisting of a star
  and an inner compact binary.

Our previous paper \citep[][hereafter Paper I]{Hayashi2020},
  showed that the short-term radial velocity (RV) variations provide a useful probe of a
  hidden inner BBH in a coplanar triple system.  The present paper
extends the work, and considers the noncoplanar and unequal mass
cases as well.  We perform a series of
  N-body simulations, and model the resulting RV variations by
  generalizing analytic formulae based on the quadrupole perturbation
  theory \citep{Morais2008}.  For coplanar triples, we find that the
  precession of the argument of pericenter is a useful probe of an inner
  hidden binary. For noncoplanar triples, the long-term variations of
  the RV semi-amplitude induced by the nodal precession and the
  Kozai-Lidov oscillation can be used to search for an inner binary,
  as can the short-term RV variations.

The rest of the paper is organized as follows. Section
\ref{sec:constraints} presents stability constraints on an inner BBH
for a hypothetical star-BBH triple inspired by the set of
  parameters originally proposed for the LB-1 system \citep{Liu2019},
  using the approximate RV formula in coplanar orbits by
  \citet{Morais2008}.

Then we predict the RV variations of the outer star around the inner
BBH in section \ref{sec:radial-velocity}. We first consider coplanar
orbits, and find that the numerical results are reasonably well
reproduced by the analytic approximation for the residual RV velocity
component by \citet{Morais2008} and \citet{Morais2011} even including
the eccentricity effect as long as the quasi-Keplerian motion is
extracted properly.  Next we examine noncoplanar cases from numerical
simulations. Due to the precession of the inner and outer orbits in
noncoplanar systems, the amplitude of the stellar RV changes
significantly over roughly the Kozai-Lidov timescale. In
  section \ref{sec:discussion}, we discuss possible effects of the
  general relativistic correction on the orbital evolution, and also
  possible formation channels of star-BBH triples.  Section
\ref{sec:summary} is devoted to the conclusions of this paper.
Appendix A discusses the long-term behavior of noncoplanar star-BBH
triples on the basis of the secular perturbation theory.

\section{Constraints on a possible inner binary
  in the hypothetical triple inspired by the previous
    estimate for the LB-1 system \label{sec:constraints}}

\begin{figure*}
\begin{center}
\includegraphics[clip,width=12.0cm]{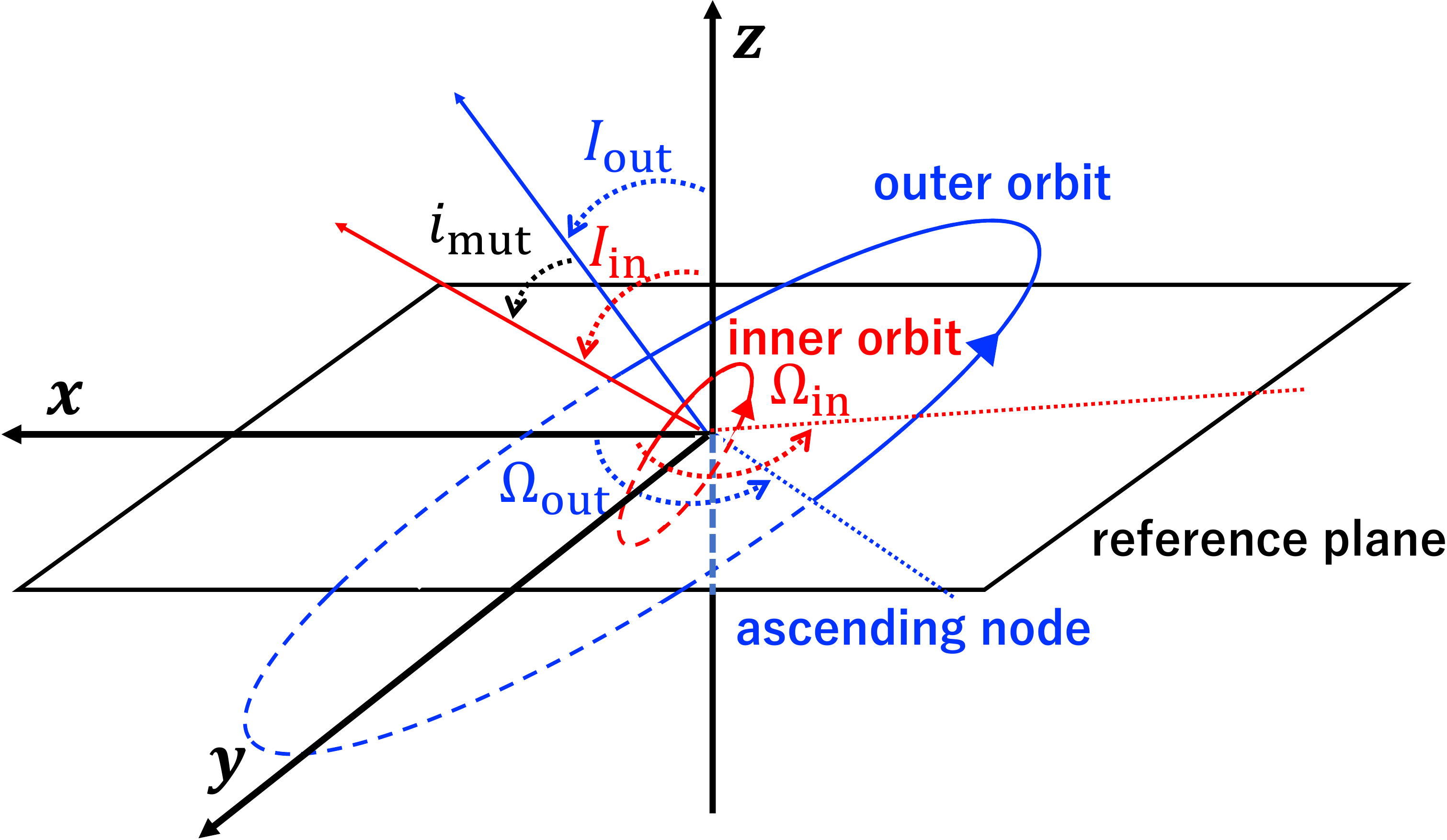}
\end{center}
\caption{Schematic illustration of a triple system that we consider in
  the present paper.  The orbital angles are defined with respect to
  the reference Cartesian frame whose origin is set to be the
  barycenter of the inner orbit.
  \label{fig:orbits}}
\end{figure*}   

Figure \ref{fig:orbits} shows a schematic configuration of a triple
system consisting of an inner BBH and an outer orbiting star. In what
follows, we use the subscript $j (=\ii$ and $\oo)$ to distinguish
between those variables of the inner and outer orbits, respectively.
With respect to the reference coordinate system, the inner and outer
orbits are specified by the instantaneous longitudes of the ascending
nodes $\Omega_j$, semi-major axes $a_j$, eccentricities $e_j$,
arguments of pericenter $\omega_j$, orbital inclinations $I_j$, and
their mutual inclination $i_{\rm mut}$. Note that our reference plane
in Figure \ref{fig:orbits} is arbitrary while it is often chosen as
the invariant plane of the triple system.

In the case of a coplanar and near-circular hierarchical triple system,
\citet{Morais2008} found that the RV of an outer star ($m_*$)
orbiting an inner binary ($m_1$ and $m_2$) is approximately decomposed
to the three terms for a distant observer along the $z$-axis
in Figure \ref{fig:orbits}:
\begin{eqnarray}
\label{eq:RV}
V_\mathrm{RV}(t) = V^{(0)}_\kep(t) + \delta V_\kep(t) + V_\bin(t).
\end{eqnarray}

The first term in the right-hand-side of equation (\ref{eq:RV})
corresponds to the unperturbed Keplerian motion of the star around the
barycenter of the system:
\begin{eqnarray}
\label{eq:V0Kep}
V^{(0)}_\kep(t) &=& K_0\sin{I_\oo}\cos[\nu_\oo t+f_\ooz+\omega_\oo], \\
\label{eq:K0}
K_0 &\equiv& \frac{m_1+m_2}{m_{1}+m_2+m_*} a_\oo \nu_\oo ,
\end{eqnarray}
where $K_0$ is the semi-amplitude of the unperturbed Keplerian RV for
an edge-on observer, $\nu_\oo$ and $\omega_\oo$ denote the mean motion
and argument of pericenter of the outer star, and $f_\ooz$ is the initial true
anomaly of the star at $t=0$.  Since orbits in a triple system should
have a non-vanishing eccentricity, $\omega_\oo$ in
the above expressions is well defined in general.

The second term is the lowest-order perturbation correction to the
stellar Keplerian motion due to the inner binary:
\begin{eqnarray}
\label{eq:delta-VKep}
\delta V_\kep(t)  &=& K_1 \sin{I_\oo}\cos[\nu_\oo t+f_\ooz+\omega_\oo], \\
\label{eq:K1}
K_1 &\equiv& \frac{3}{4}K_0\left(\frac{a_\ii}{a_\oo}\right)^{2}
\frac{m_1m_2}{(m_1+m_2)^2} .
\end{eqnarray}

Finally the third term is the RV variation of the star with roughly 
twice the orbital frequency of the inner binary:
\begin{eqnarray}
\label{eq:RV_bin}
V_\bin(t) &=&
-\frac{15}{16}K_\bin \sin{I_\oo}
\cos[(2\nu_\ii-3\nu_\oo)t+2(f_\iiz+\omega_\ii)-3(f_\ooz+\omega_\oo)] \cr
&&  + \frac{3}{16}K_\bin \sin{I_\oo}
\cos[(2\nu_\ii-\nu_\oo)t+2(f_\iiz+\omega_\ii)-(f_\ooz+\omega_\oo)], \\
\label{eq:KBBH}
K_\bin &\equiv& \frac{m_1 m_2}{(m_1+m_2)^2}
\sqrt{\frac{m_{1}+m_2+m_*}{m_{1}+m_2}}
\left(\frac{a_\ii}{a_\oo}\right)^{7/2}K_0,
\end{eqnarray}
where $K_\bin$ is the characteristic semi-amplitude of the RV variation of our primary interest, $\nu_\ii$ and $\omega_\ii$ denote the mean motion
and argument of pericenter of the inner binary, and $f_\iiz$ is the initial true
anomaly of the inner binary at $t=0$.
Equation (\ref{eq:RV_bin}) indicates that the RV variation indeed
consists of two slightly different frequency modes around $2\nu_\ii$:
\begin{eqnarray}
\label{eq:nu-3}
\nu_{-3} &\equiv& 2\nu_\ii-3\nu_\oo, \\
\label{eq:nu-1}
\nu_{-1} &\equiv& 2\nu_\ii-\nu_\oo .
\end{eqnarray}
Since we are interested in the case of $\nu_\ii \gg \nu_\oo$, the
above two modes may be degenerate unless the observational duration
is sufficiently long, and the cadence is sufficiently high.

In the case of $m_1=m_2 \gg m_*$, the ratio of the above three
semi-amplitudes is simplified as
\begin{eqnarray}
  K_0 : K_1: K_\bin
  = 1 : \frac{3}{16}\left(\frac{a_\ii}{a_\oo}\right)^{2}
  : \frac{1}{4}\left(\frac{a_\ii}{a_\oo}\right)^{7/2} .
\end{eqnarray}
We note also that the above expressions for a prograde triple can be
applied to a retrograde triple of the same orbits if $\nu_\ii$,
$\omega_\ii$, and $f_\iiz$ are replaced by $-\nu_\ii$, $-\omega_\ii$,
and $-f_\iiz$, respectively. In the retrograde triple, therefore, we
define the
mean motions of the two modes:
\begin{eqnarray}
\label{eq:nu+3}
\nu_{+3} &\equiv& 2\nu_\ii+3\nu_\oo, \\
\label{eq:nu+1}
\nu_{+1} &\equiv& 2\nu_\ii+\nu_\oo .
\end{eqnarray}

\begin{figure*}
\begin{center}
 \includegraphics[clip,width=8.0cm]{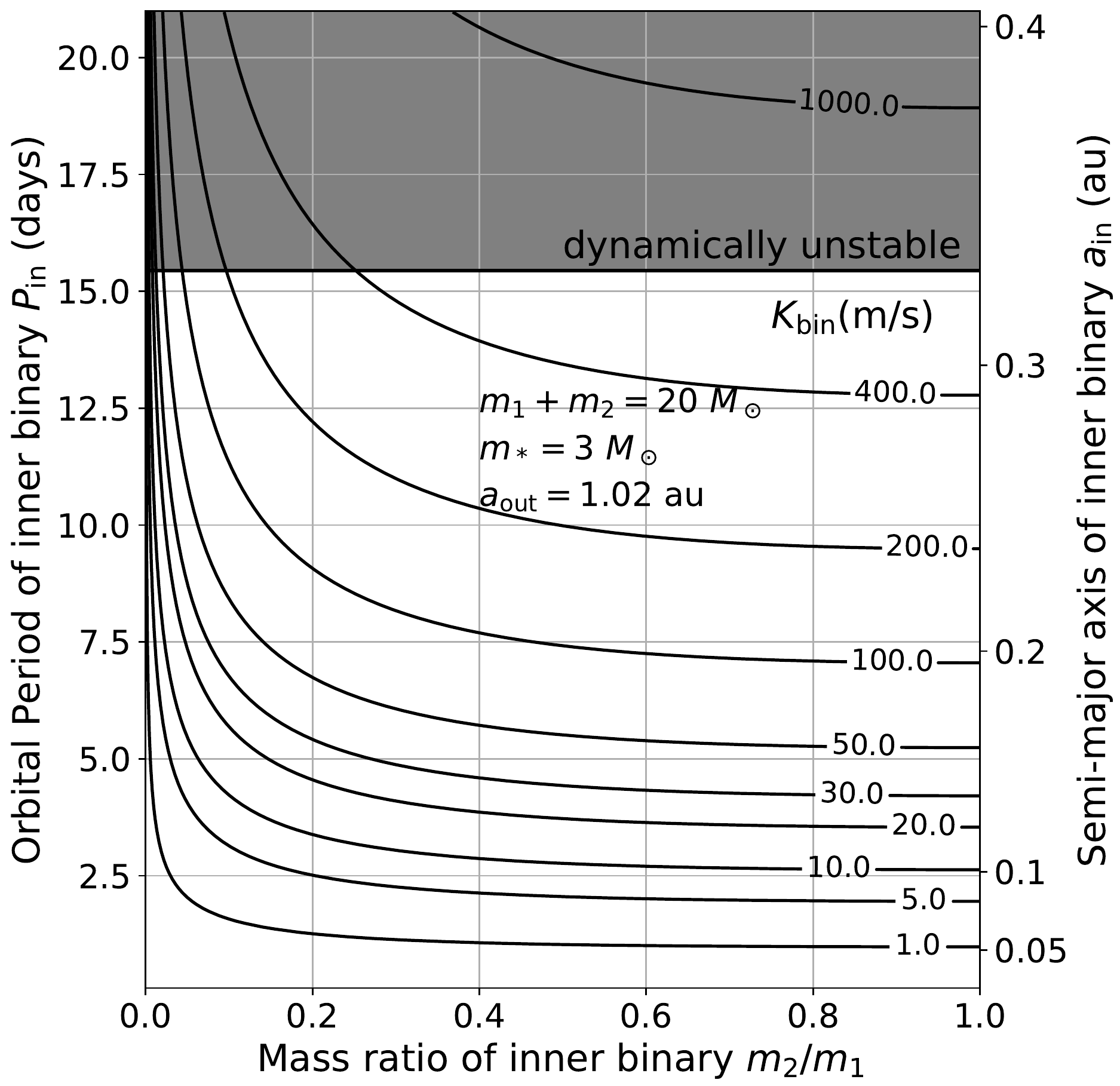} \hspace{15pt}
 \includegraphics[clip,width=8.0cm]{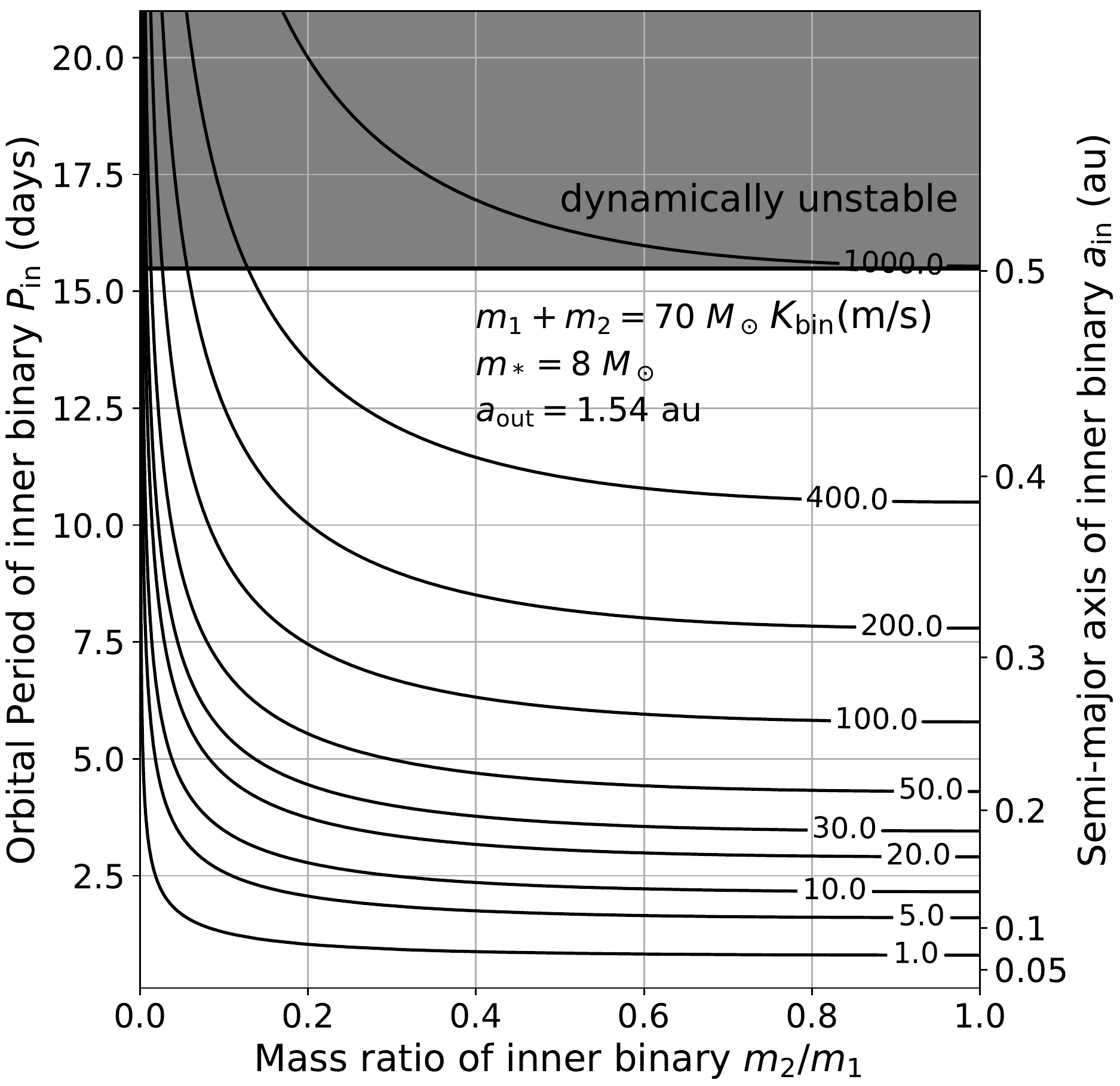}
\end{center}
\caption{Contours of semi-amplitude of RV variations $K_\bin$ expected
  from an inner BBH in the LB-1 system. Each contour curve is labeled
  by the value of $K_\bin$ in units of m/s. The gray areas indicate
  the dynamically unstable region for a coplanar case in
    Newtonian theory; see inequality
  (\ref{eq:instability}). \label{fig:fig1}}
\end{figure*}

As in Paper I, the orbital period and mass ratio of a possible inner
BBH in the LB-1 system are constrained from its dynamical
stability. First note that the mass function of the LB-1 system
\citep{Liu2019} is observationally estimated to be
\begin{equation}
\label{eq:massf}
\frac{m_{12}^3 \sin^3{I_\oo}}{(m_{12}+m_*)^2}
=\frac{P_\oo K_\oo^3}{2\pi \mathcal{G}}{(1-e_\oo^2)^{3/2}}
= 1.02 \pm 0.05 M_\odot,
\end{equation}
where $\mathcal{G}$ is Newton's gravitational constant, $I_\oo$ is the
inclination of the stellar orbit with respect to our line of sight,
and $K_\oo$ is the observed semi-amplitude of the radial velocity. We
denote the mass of the unseen companion of the star by $m_{12}$, which
should be interpreted as $m_1+m_2$ if the LB-1 is a triple system
hosting an inner binary.

We consider two specific examples following the original
  claims for the LB-1 according to
\citet{Abdul-Masih2020,El-Badry2020, Liu2019}: ($m_{12}$, $m_*$,
$I_\oo$) = ($20M_\odot$, $3M_\odot$, $24^\circ$) and ($70M_\odot$,
$8M_\odot$, $15^\circ$), instead of the more recent estimate
  of ($7M_\odot$, $1.5M_\odot$, $39^\circ$) by \citet{Shenar2020}.
The former corresponds to our fiducial model in this paper, but we
also consider the latter just for comparison because it corresponds
roughly to a range of several BBHs detected by LIGO. We fix
$e_\oo=0.03$ and $P_\oo=78.9$ days \citep[][]{Liu2019} for the outer
star. These values are basically the same as those in
  \citet{Shenar2020}; $e_\oo=0.0036$$\pm0.0021$ and
  $P_\oo=78.7999$$\pm0.0097$.

Figure \ref{fig:fig1} plots a contour of $K_\bin$, equation (\ref{eq:KBBH}), on the $m_2/m_1$ --
$P_\ii (\equiv 2\pi/\nu_\ii)$ plane, where we assume coplanar and
near-circular orbits. The upper shaded regions are excluded from the
dynamical instability condition for the three-body system. The
noncoplanarity between the inner and outer orbits generally weakens
the constraint for the coplanar case, and the instability condition is
approximately given as
\citep{Mardling1999,Mardling2001,Aarseth2001,Toonen2016}
\begin{equation}
\label{eq:instability}
\frac{a_\ii}{a_\oo}
> \frac{1-e_\oo}{2.8(1-0.3i_{\rm mut}/\pi)}
\left(\frac{(1+m_*/m_{12})(1+e_\oo)}{\sqrt{1-e_\oo}}\right)^{-\frac{2}{5}}.
\end{equation}
In the above inequality, the factor $(1-0.3i_{\rm mut}/\pi)$ is
empirically added by \citet{Aarseth2001} so as to reproduce the
earlier result by \citet{Harrington1972}.  Figure \ref{fig:fig1}
implies that an inner BBH of an orbital period between a few days and a
couple of weeks is stable and has a potentially detectable RV
variation amplitude. Thus, we adopt $P_\ii=10$ days in the rest
  of the paper. Note that this choice satisfies the empirical limit
  $P_\oo/P_\ii\gtrsim 5$ for stellar triple systems listed
  in \citet{Tokovinin2008}.

\section{Radial-velocity variation induced by an inner binary
  \label{sec:radial-velocity}}

\subsection{Simulation models and method for removing the quasi-Keplerian
  component from the radial velocity  \label{sec:simulation-method}}

\begin{deluxetable}{lcccccc}
\tablecolumns{7}
\tablewidth{1.0\columnwidth} 
\tablecaption{Simulation models} 
\tablehead{ model & $I_\oo$ (deg) & $I_\ii$ (deg) & $i_\mathrm{mut}$ (deg) &
   $m_1~({\rm M_\odot})$ & $m_2~({\rm M_\odot})$ & $e_\ii$}
\startdata
P1010 & $90$ & $90$ & $0$ & $10$ & $10$ & $10^{-5}$ \\
PE1010 & $90$ & $90$ & $0$ & $10$ & $10$ & $0.2$ \\
R1010& $90$ & $270$ & $180$ & $10$ & $10$ & $10^{-5}$ \\
O1010 & $0$ & $90$ & $90$ & $10$ & $10$ & $10^{-5}$ \\
I1010  & $0$ & $45$ & $45$ & $10$ & $10$ & $10^{-5}$ \\
\hline
P0218  & $90$ & $90$ & $0$ & $18$ & $2$ & $10^{-5}$ \\
PE0218  & $90$ & $90$ & $0$ & $18$ & $2$ & $0.2$ \\
R0218  & $90$ & $270$ & $180$ & $18$ & $2$ & $10^{-5}$ \\
O0218 & $0$ & $90$ & $90$ & $18$ & $2$ & $10^{-5}$ \\
I0218  & $0$ & $45$ & $45$ & $18$ & $2$ & $10^{-5}$ \\
\hline
\enddata
\label{tab:tab1}
\tablecomments{P, PE, R, O and I indicate prograde, prograde
  eccentric, retrograde, orthogonal and inclined orbits.}
\end{deluxetable}

\begin{deluxetable}{lc}
\tablecolumns{2}
\tablewidth{1.0\columnwidth} 
\tablecaption{Initial values of the common parameters} 
\tablehead{ parameter & initial value }
\startdata
orbital period $P_\oo$ & $78.9~\mathrm{days}$ \\
orbital period $P_\ii$ & $10.0~\mathrm{days}$ \\
eccentricity $e_\oo$ & $0.03$ \\
argument of pericenter $\omega_\ii$ & $0~\mathrm{deg}$ \\ 
argument of pericenter $\omega_\oo$ & $0~\mathrm{deg}$ \\ 
longitude of ascending node $\Omega_\ii$ & $0~\mathrm{deg}$ \\ 
longitude of ascending node $\Omega_\oo$ & $0~\mathrm{deg}$ \\ 
true anomaly $f_\ii$ & $30~\mathrm{deg}$ \\ 
true anomaly $f_\oo$ & $120~\mathrm{deg}$ \\ 
tertiary mass $m_*$ & $3~M_\odot$ \\
inner binary mass $m_1+m_2$ & $20~M_\odot$ 
\label{tab:tab2}
\enddata
\end{deluxetable}

 As in Paper I, we perform N-body simulations for a set of triple
  configurations (Table \ref{tab:tab1}), using the public N-body package
  {\tt REBOUND} \citep{Rein2012}. While our analysis in this
    paper is based on purely Newtonian gravity, we made sure that the
    correction due to general relativity (GR) does not change the
    conclusions here by repeating a set of runs using {\tt REBOUNDx}
    \citep[][]{Tamayo2020}, the extended package of {\tt REBOUND},
    with GR effects {\tt gr\_full} \citep[][]{Newhall1983}; see
    section \ref{sec:discussion} below.

The initial conditions of the simulations are summarized in Table
\ref{tab:tab2}. For the simulations, we only consider the case ($m_{12}$,
$m_*$) = ($20M_\odot$, $3M_\odot$) because the results are basically
scalable for different mass regimes.  We use {\tt WHFast} integrator
\citep{Rein2015} with a time step of $10^{-6}~\mathrm{yr}/2\pi$.  We
run each model and output the snapshots every 0.1 day over $0< t
<1000P^{(0)}_\oo$, with $P^{(0)}_\oo=78.9$ days being the input
orbital period of the outer star. We confirmed that all of the systems
remain gravitationally bound and stable at least within
$1000P^{(0)}_\oo$.

As we discussed in Paper I, all of the orbital parameters in the present
simulation runs are time dependent, and the information for the inner
BBH imprinted in the RV variations can be reproduced only when if the
quasi-Keplerian RV component is properly extracted.  While equation
(\ref{eq:RV}) provides a reasonably good approximation, it does not
incorporate the back-reaction from the outer star, and cannot be
directly applied to estimate the quasi-Keplerian RV component. Thus,
we first fit the total RV using the public code {\tt RadVel}
\citep[][]{Fulton2018} to extract the quasi-Keplerian RV component.
Then, we compute the residual RV variations due to the inner BBH,
perform the Lomb-Scargle (LS) periodogram analysis, and compare with the
approximate analytic results.

Consider the prograde, coplanar and circular case with
$m_1=m_2=10M_\odot$ (P1010 in Table \ref{tab:tab1}). We use the
initial orbital period of the star, $P_\oo^{(0)} (=78.9~{\rm days})$,
to normalize the time $t$.  Strictly speaking, the initial conditions
of the simulations (Table \ref{tab:tab2}) are not dynamically
consistent for the triple system.  Thus we examine the evolution of
the systems at $t \geq 100 P_\oo^{(0)}$ when the possible initial
transient behavior goes away.

\begin{figure*}
\begin{center}
\includegraphics[clip,width=16.0cm]{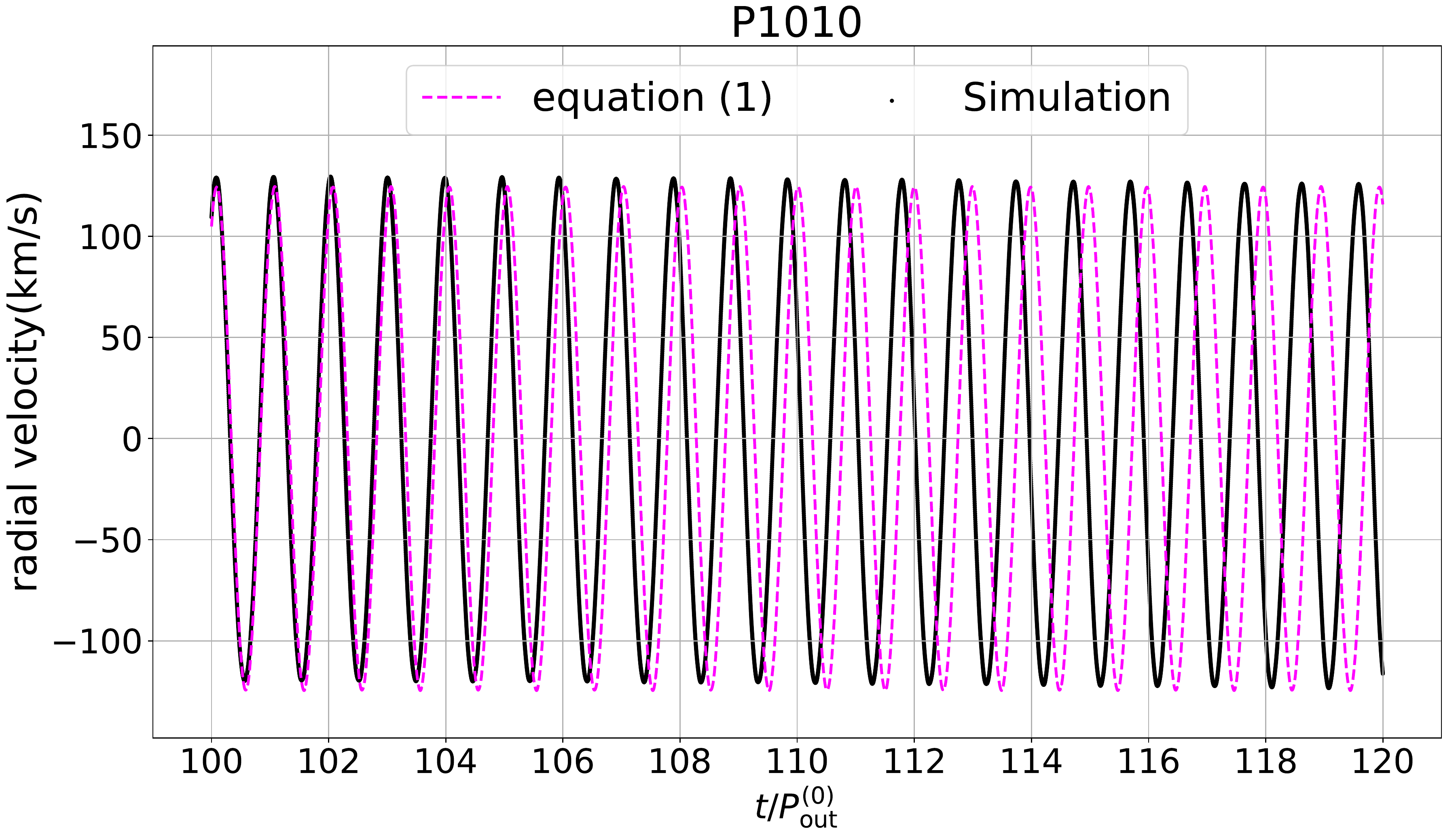}
\end{center}
\caption{Radial velocity(P1010) with 0.1 day cadence. The black points
  and magenta dashed line denote the simulated RV data and RV
  approximate formula (equation (\ref{eq:RV})) evaluated at
  $t=100P^{(0)}_\oo$, respectively.\label{fig:RV-P1010}}
\end{figure*}

Figure \ref{fig:RV-P1010} plots the total RV of P1010 for
$100<t/P^{(0)}_\oo<120$.  The black dots and magenta dashed line
indicate the simulation output and an analytic approximation by
\citet{Morais2008}. For the latter, we evaluate the orbital variables
at $t=100P^{(0)}_\oo$, and substitute those instantaneous values in
equation (\ref{eq:RV}). As expected, Figure \ref{fig:RV-P1010} shows
that the the total RV is dominated by the Keplerian motion, but the
corresponding instantaneous period $P_\oo(t)$ from simulations is
clearly smaller than $P_\oo^{(0)}$.  Note that the orbital period
evaluated with the instantaneous orbital elements at
$t=100P^{(0)}_\oo$ does not differ much from $P^{(0)}_\oo$.

\begin{figure*}
\begin{center}
\includegraphics[clip,width=14.0cm]{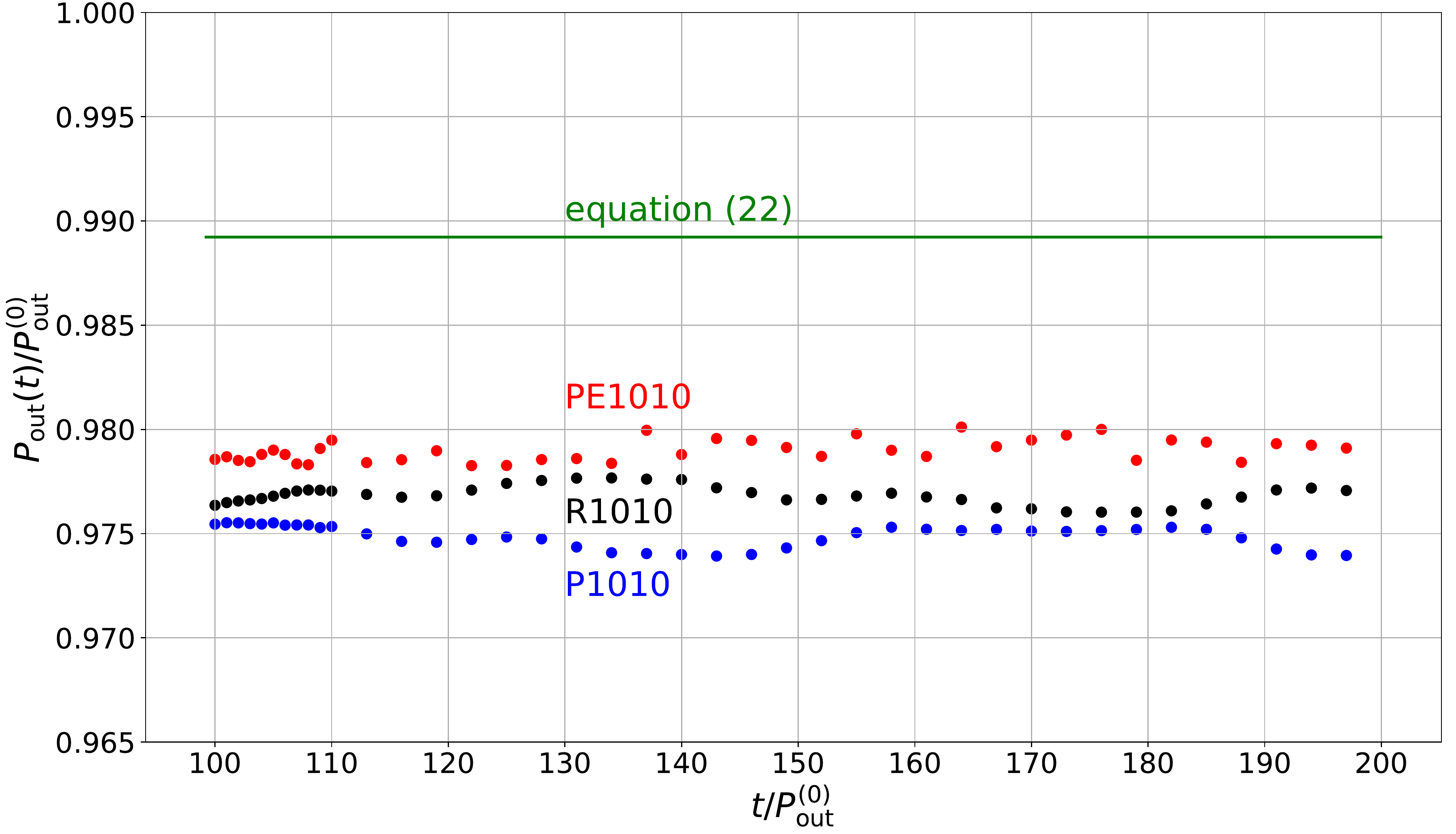}
\end{center}
\caption{Best-fit values of $P_\oo(t_n \equiv nP_\oo^{(0)})$ for
  coplanar systems. They are estimated with {\tt RadVel} using the
  0.1 day cadence simulated RV data over
  $nP_\oo^{(0)}<t<(n+1)P_\oo^{(0)}$ for $100 \leq n < 200$; P1010
  (blue), R1010 (black), and PE1010 (red).  The solid green line
  indicates the analytic prediction that incorporates the average
  time derivative of the argument of pericenter $\omega_\oo(t)$ (see
  equation (\ref{eq:Pout-t-expected})).
 \label{fig:Pout-t}}
\end{figure*}

Therefore, we use a public code {\tt RadVel} \citep{Fulton2018}, and
estimate the value of the quasi-Keplerian period $P_\oo(t_n \equiv
nP_\oo^{(0)})$ by fitting the total RV over $nP_\oo^{(0)}<t
<(n+1)P_\oo^{(0)}$ where $n (\geq 100)$ is an integer.  Figure
\ref{fig:Pout-t} shows the resulting best-fit values of $P_\oo(t_n)$
over $100 \leq n < 200$ for P1010, R1010, and PE1010.

Equations (\ref{eq:RV}) -- (\ref{eq:KBBH}) on the basis of a
perturbation approximation by \citet{Morais2008} assume that both the
outer mean motion $\nu_\oo$ and the argument of pericenter
$\omega_\oo$ are constant. In reality, however, they are dependent on
time due to the perturbation from the inner binary. Let us consider
the following expansions:
\begin{eqnarray}
\label{eq:nuout-t}
  \nu_\oo(t) &=& \nu^{(0)}_\oo + \delta \nu_\oo(t), \\
\label{eq:omegaout-t}
  \omega_\oo(t) &=& \omega^{(0)}_\oo + \delta \omega_\oo(t),
\end{eqnarray}
where
\begin{eqnarray}
\label{eq:nuout-0}
  \nu^{(0)}_\oo \equiv \sqrt{\frac{\mathcal{G}(m_{12}+m_*)}{(a^{(0)}_\oo)^3}}
\end{eqnarray}
is the mean motion expected for the two-body system.

\begin{figure*}
\begin{center}
\includegraphics[clip,width=14.0cm]{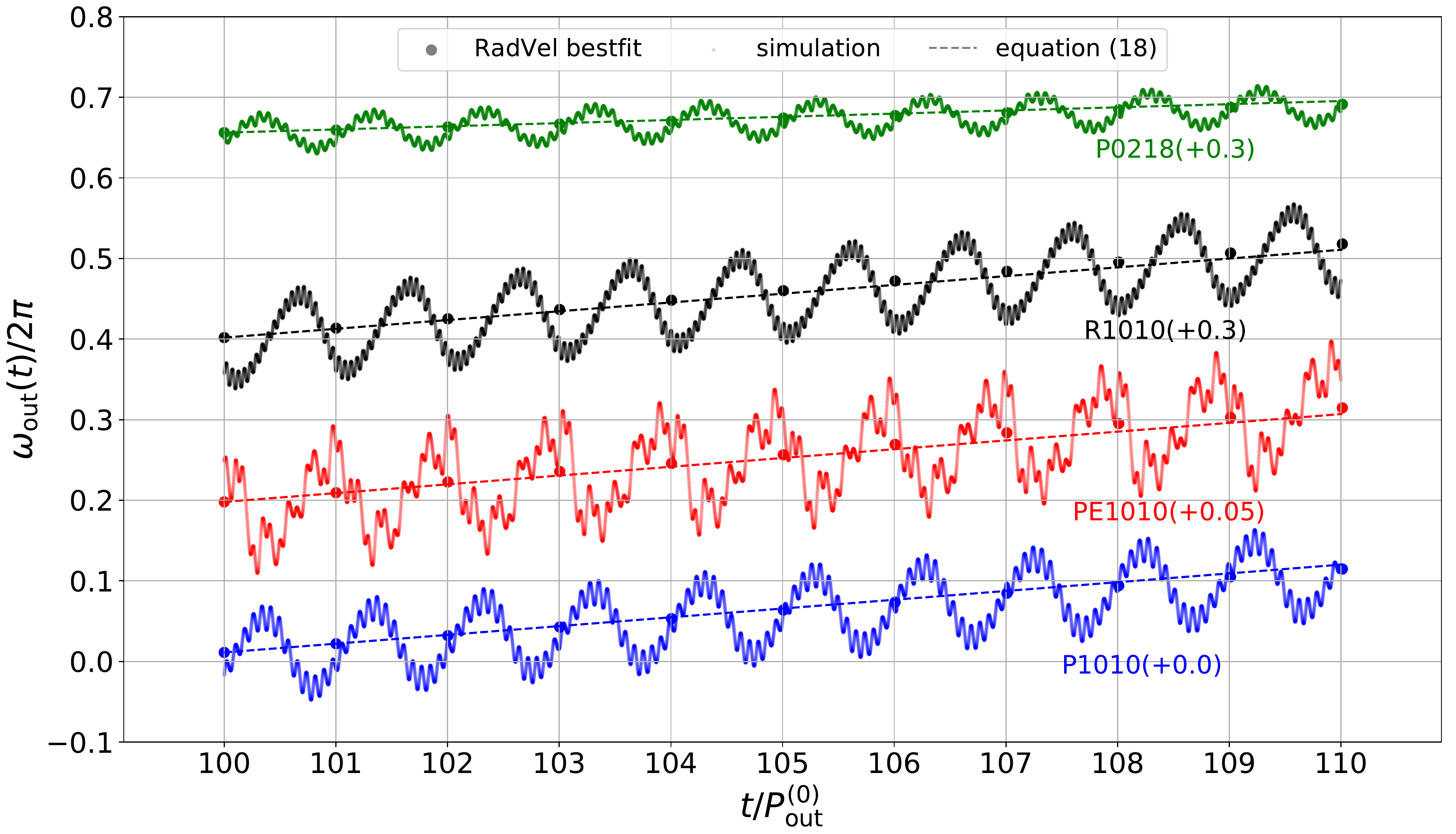}
\end{center}
\caption{ Best-fit values of $\omega_\oo(t)$ for P1010, PE1010, R1010
  and P0218. Each best-fit value is determined with {\tt RadVel} from
  simulation data using their 0.1 day cadence outputs over
  $1P^{(0)}_\oo$ starting at $t$.  The dashed line is calculated using
  the analytic approximate formula of $\dot{\omega}_\oo$ in equation
  (\ref{eq:omega-dot}).  For clarity, the data are translated in the $y$
  direction with the offset value indicated in parentheses.
\label{fig:omega-t}}
\end{figure*}

Figure \ref{fig:omega-t} plots $\omega_\oo(t)$ from the 0.1 day
cadence output of our {\tt REBOUND} run for P1010, PE1010, R1010, and
P0218. It is clearly visible that $\omega_\oo(t)$ exhibits periodic
modulations with frequency roughly corresponding to $\nu_\oo$ and
$\nu_\ii$, in addition to the monotonic increase with $t$.  In order to
remove the oscillation components, we compute the time average of
$\omega_\oo$ over $nP_\oo^{(0)}<t<(n+1)P_\oo^{(0)}$ using {\tt RadVel}
as described in the above, and plot the best-fit values $\langle
\omega_\oo \rangle$ in solid circles at $t=t_n$.

The time derivative of $\omega_\oo$ is given by (see Appendix A for
details)
\begin{eqnarray}
\label{eq:omega-dot}
\frac{\dot{\omega}_\oo}{2\pi}
=  \frac{3}{4}\frac{1}{P_\oo} \left(\frac{a_\ii}{a_\oo}\right)^{2}
  \left(\sqrt{\frac{m_2}{m_1}}+\sqrt{\frac{m_1}{m_2}}\right)^{-2}
  \frac{1}{(1-e^2_\oo)^2},
\end{eqnarray}
for a coplanar triple system with $e^2_\ii \ll 1$.  The slope of
the dashed lines in Figure \ref{fig:omega-t} corresponds to the prediction
of equation (\ref{eq:omega-dot}) evaluating with input values of
orbital parameters (see Tables \ref{tab:tab1} and \ref{tab:tab2}),
which reproduces the behavior of $\langle \omega_\oo \rangle(t)$ very
well. This good agreement indicates that $\delta\omega_\oo(t)$ in equation (\ref{eq:omegaout-t})
averaged over $P_\oo^{(0)}$ is well approximated by $\dot{\omega}_\oo
t$ with equation (\ref{eq:omega-dot}). 

Thus, the pericenter shift itself provides an independent signature of
the presence of the inner binary.  Indeed, this is why a hypothetical
planet Vulcan was proposed by \citet{LeVerrier1859} to explain the
anomalous perihelion shift of Mercury in Newton's theory before
general relativity was discovered by
  \citet[][]{Einstein1915_2}.

This implies that the sinusoidal term in the right-hand side of
equations (\ref{eq:V0Kep}) and (\ref{eq:delta-VKep}) can be written,
to its lowest order, as
\begin{eqnarray}
\label{eq:nuout-t-expand}
\cos[\nu_\oo(t)t + \omega_\oo(t) + f_\ooz]
\approx \cos[(\nu^{(0)}_\oo + \delta\nu_\oo(0)
  + \dot{\omega}_\oo)t + \omega_\oo^{(0)} + f_\ooz].
\end{eqnarray}
Equation (\ref{eq:nuout-t-expand}) suggests that $P_\oo(t)$ averaged
over $P_\oo^{(0)}$ should be
\begin{eqnarray}
\label{eq:Pout-t-expected1}
P_\oo(t) = \frac{2\pi}{\nu^{(0)}_\oo + \delta\nu_\oo(0) + \dot{\omega}_\oo}.
\end{eqnarray}

In the case of a coplanar and circular triple with the equal-mass
inner binary, equation (\ref{eq:omega-dot}) reduces to
\begin{eqnarray}
\label{eq:omega-dot2}
  \frac{\dot{\omega}_\oo}{2\pi} \approx
  \frac{0.011}{P_\oo} \left(\frac{P_\ii}{10~\mathrm{days}}\right)^{4/3}
  \left(\frac{P_\oo}{78.9~\mathrm{days}}\right)^{-4/3}
  \left(\frac{m_1+m_2}{20M_\odot}\right)^{2/3}
  \left(\frac{m_1+m_2+m_*}{23M_\odot}\right)^{-2/3}.
\end{eqnarray}
Therefore, if $\delta\nu_\oo(0)$ can be neglected, equation
(\ref{eq:Pout-t-expected1}) predicts that
\begin{eqnarray}
\label{eq:Pout-t-expected}
\frac{P_\oo(t)}{P_\oo^{(0)}}
\approx 1- \frac{\dot{\omega}_\oo P_\oo^{(0)}}{2\pi} \approx 0.989.
\end{eqnarray}
As plotted in Figure \ref{fig:Pout-t}, however, equation
(\ref{eq:Pout-t-expected}) accounts for approximately one-half of the
systematic decrease of the simulation results, and not entirely. This
may indicate that $\delta\nu_\oo(0)$ cannot be neglected. Indeed, a
different perturbation analysis of the current systems on the basis of
the Lagrange planetary equation seems to be successful in reproducing
the offset of ${P_\oo(t)}/{P_\oo^{(0)}}$ shown in Figure
\ref{fig:Pout-t} \citep[][]{Hayashi2019}; unpublished but posted
in arXiv.1905.07100v1.

In any case, our strategy is to empirically remove the quasi-Keplerian
RV component by local fitting of the data, instead of using the
analytical template. Thus the above offset of
${P_\oo(t)}/{P_\oo^{(0)}}$ does not affect our procedure for extracting
the RV variations due to the inner binary.  To be more specific, we
use the RV data of the simulation runs over
$100P^{(0)}_\oo<$$t<$$101P^{(0)}_\oo$. Then we estimate
$P_\oo(t_{100})$ with {\tt RadVel}, and remove the corresponding
Keplerian component from the data. We analyze the residual RV
variations using the LS periodogram to search for the signal that is due to
the inner binary.  The choice of $100P^{(0)}_\oo<$$t<$$101P^{(0)}_\oo$
is arbitrary, and we made sure that our main conclusion below is not
affected by the choice of the epoch at all.

\subsection{Coplanar orbits  \label{subsec:coplanar}}

\begin{figure*}
\begin{center}
\includegraphics[clip,width=8.0cm]{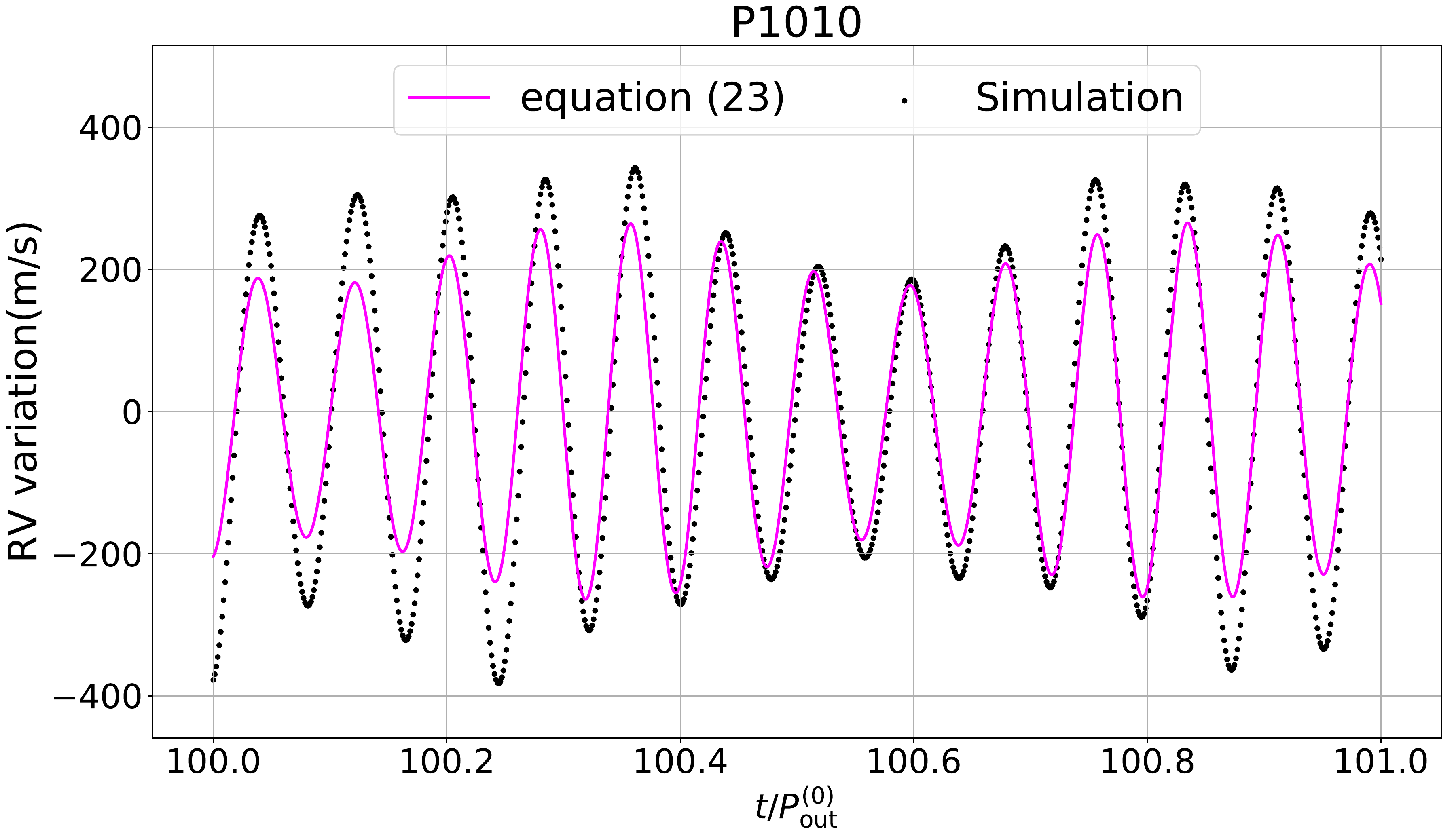}
\includegraphics[clip,width=8.0cm]{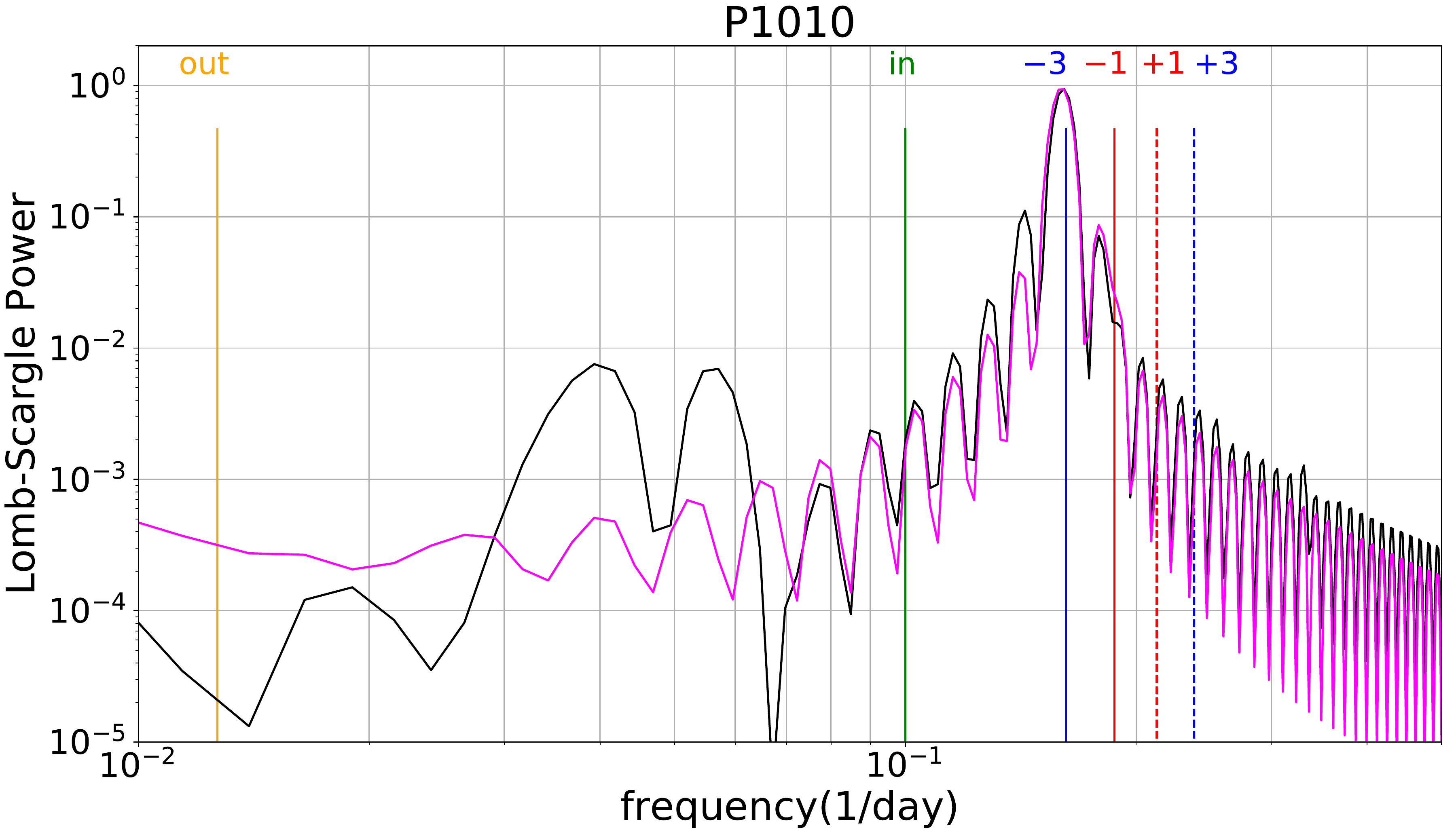}

\includegraphics[clip,width=8.0cm]{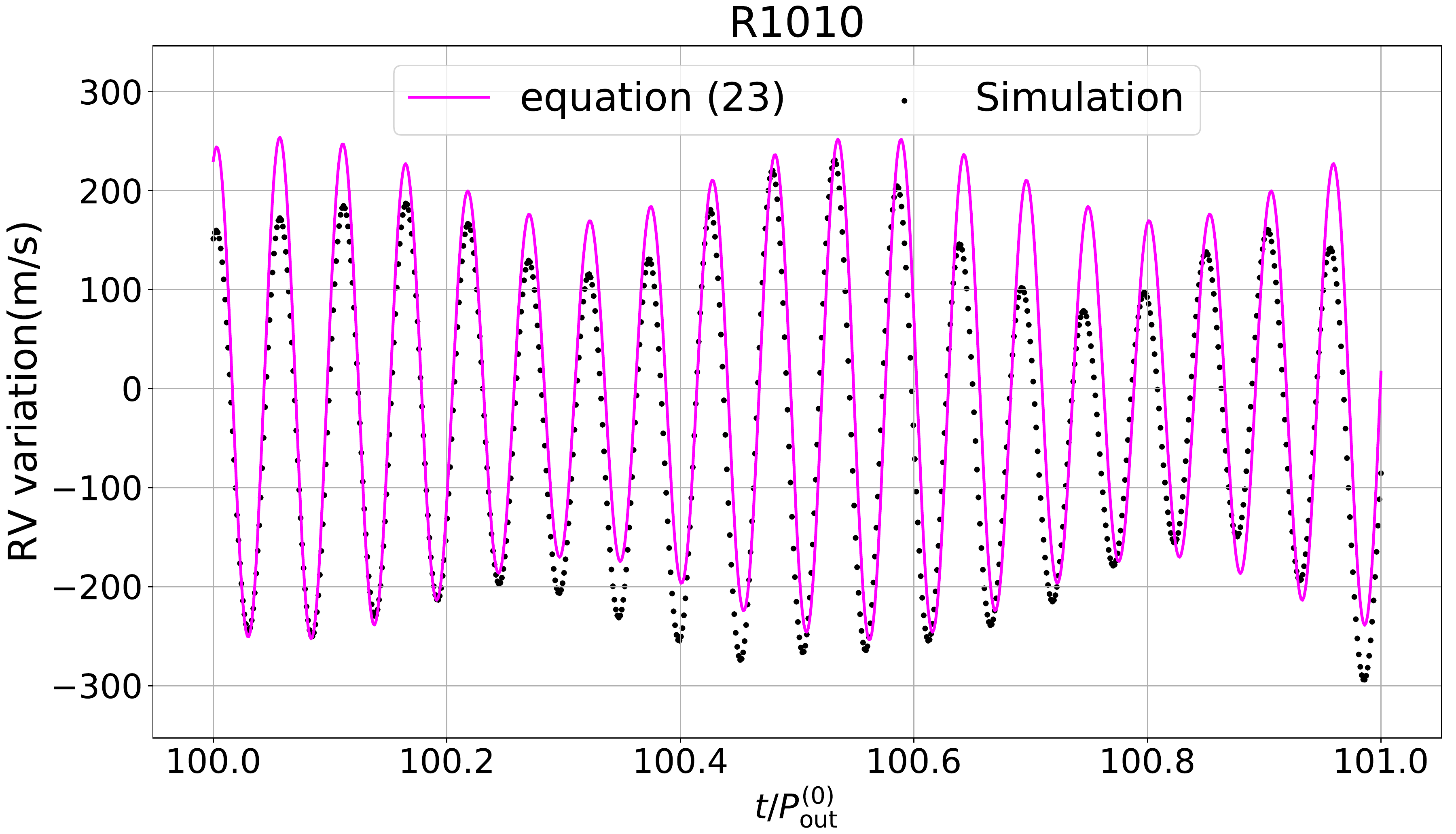} 
\includegraphics[clip,width=8.0cm]{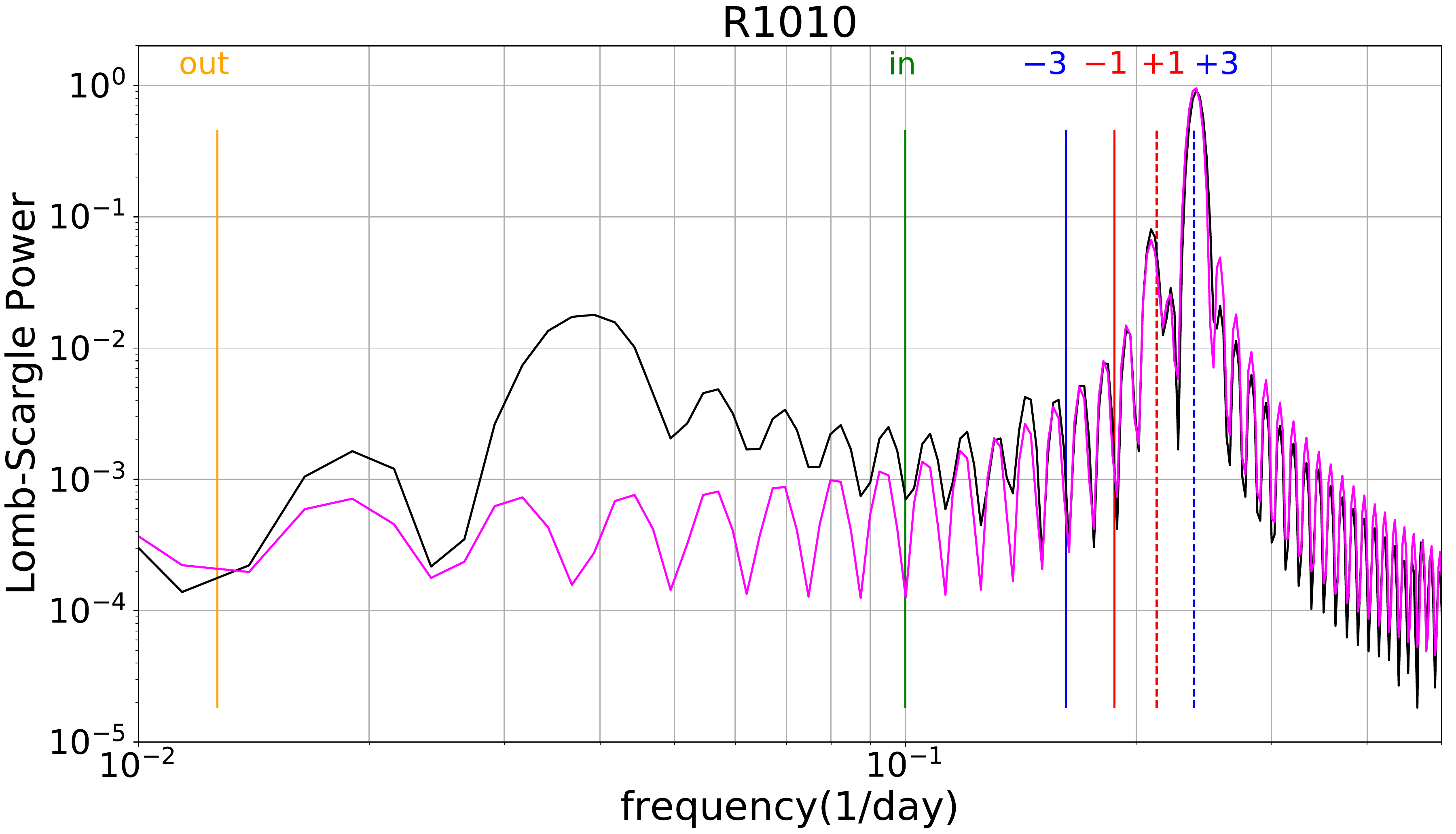}

\includegraphics[clip,width=8.0cm]{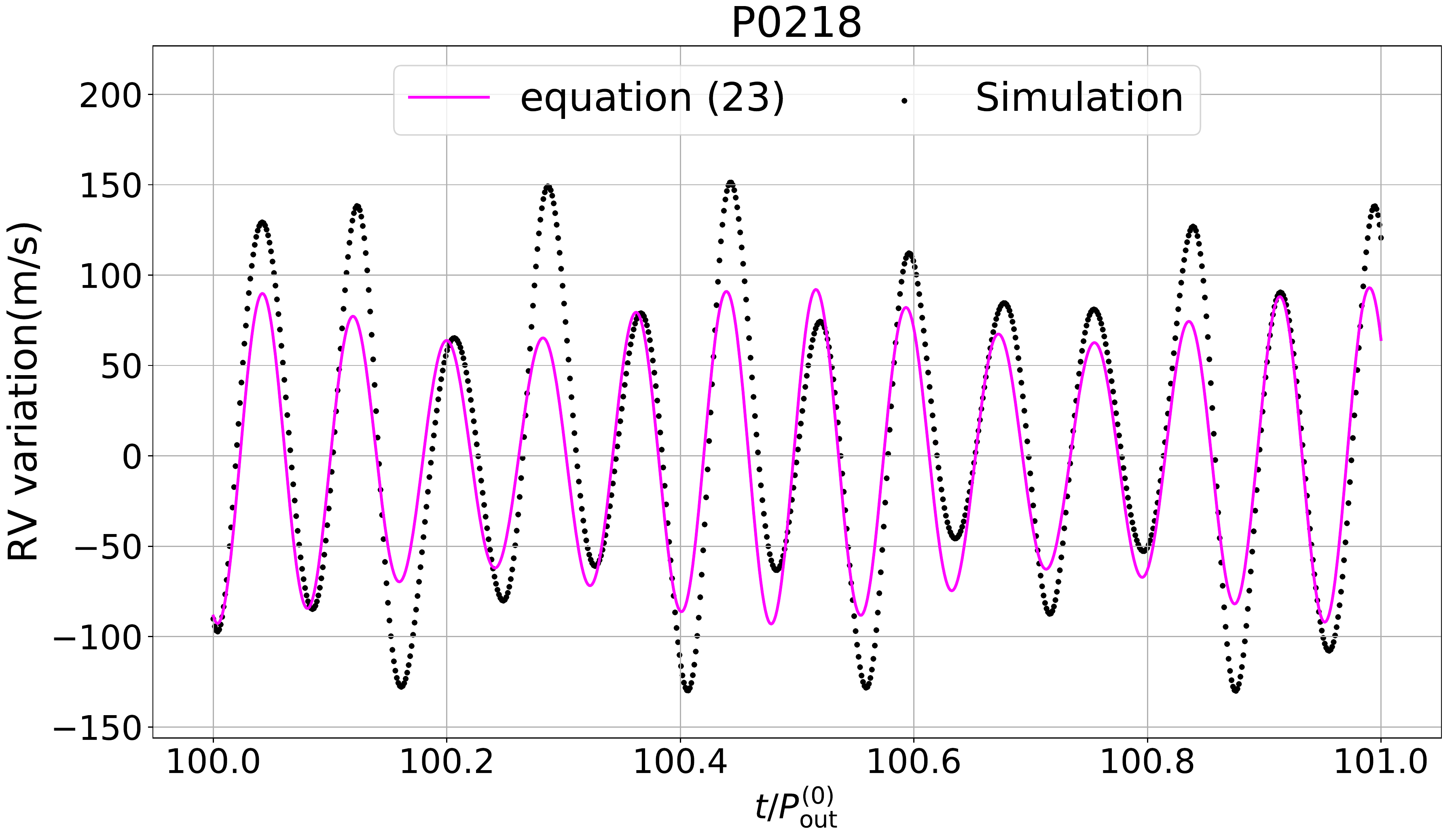} 
\includegraphics[clip,width=8.0cm]{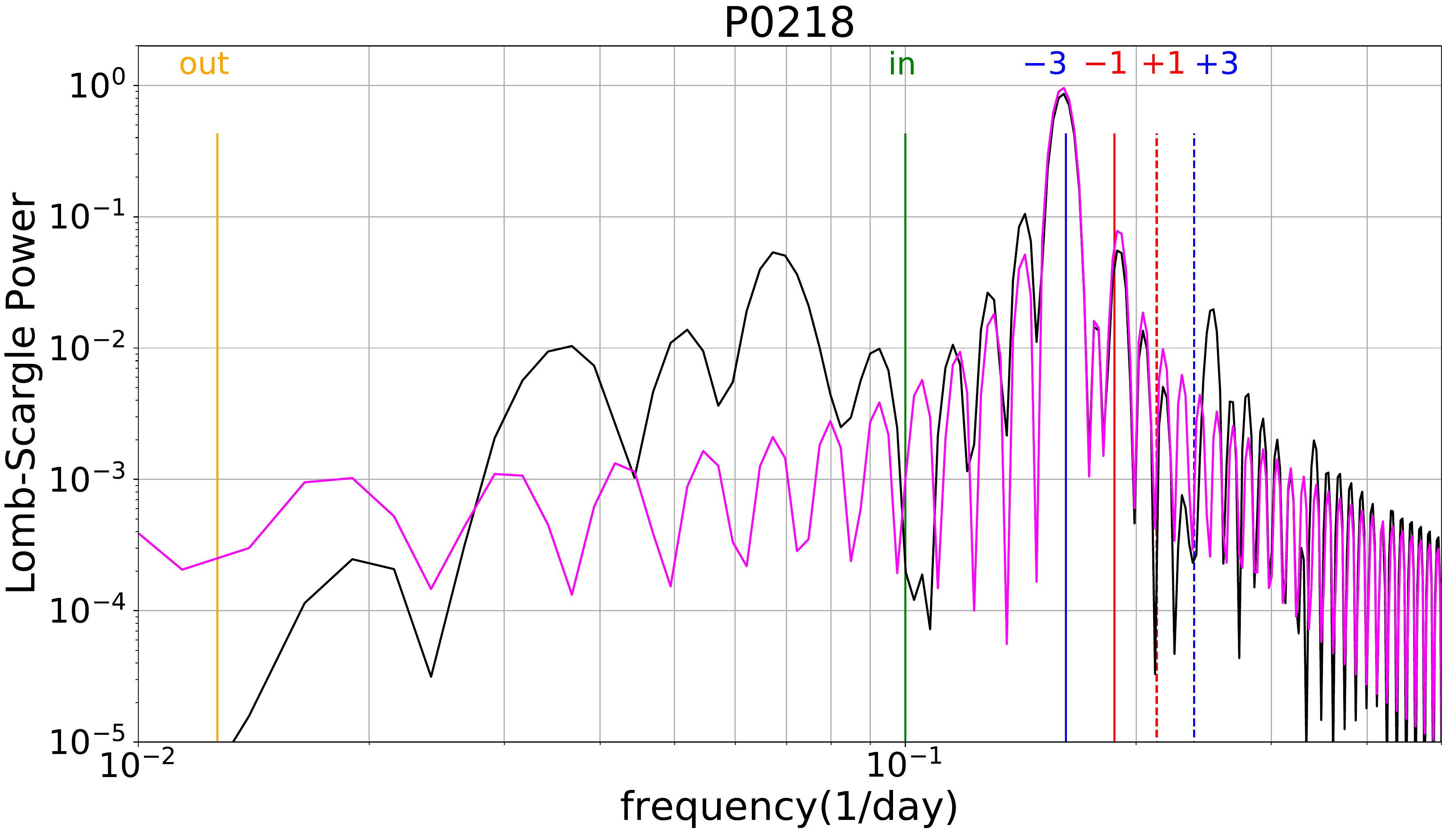}

\end{center}
\caption{RV variations for P1010, R1010, and P0218 with 0.1 day
  cadence: time series (left) and LS periodograms (right). The black
  points indicate the simulated RV variation determined with {\tt
    RadVel}. The magenta lines show the approximate prediction for the
  RV variation, equation (\ref{eq:RV_bin-0}).  In the right panels, 
  the locations of the frequencies at $\nu^{(0)}_\oo$,
  $\nu^{(0)}_\ii$, $\nu^{(0)}_{-3}$, $\nu^{(0)}_{-1}$,
  $\nu^{(0)}_{+1}$, and $\nu^{(0)}_{+3}$ are indicated by
  vertical lines labeled by out, in, -3, -1, +1, +3, respectively.
  \label{fig:RV-variation}}
\end{figure*}

The residual RV variations after removing the empirically fitted
Keplerian component are plotted in Figure \ref{fig:RV-variation} for the 
coplanar and near-circular cases. The top, middle, and bottom panels
correspond to P1010 (prograde and equal-mass binary), R1010
(retrograde and equal-mass binary), and P0218 (prograde and
unequal-mass binary), respectively.

The left panels indicate the RV variations of the simulation runs (dots)
in the time domain every 0.1 days over
$100P^{(0)}_\oo<$$t<$$101P^{(0)}_\oo$. For comparison, magenta curves
show the analytic approximation:
\begin{eqnarray}
\label{eq:RV_bin-0}
V_{\rm bin, \mathrm{i}}(t) &=&
-\frac{15}{16}K^{(\mathrm{i})}_\bin
\cos[\nu^{(\mathrm{i})}_{\mp3}t
  +2(f_\iiz^{(\mathrm{i})}+\omega_\ii^{(\mathrm{i})})
  \mp3(f_\ooz^{(\mathrm{i})}+\omega_\oo^{(\mathrm{i})})] \cr
&&  + \frac{3}{16}K^{(\mathrm{i})}_\bin
\cos[\nu^{(\mathrm{i})}_{\mp1}t+2(f_\iiz^{(\mathrm{i})}+\omega_\ii^{(\mathrm{i})})
  \mp(f_\ooz^{(\mathrm{i})}+\omega_\oo^{(\mathrm{i})})], \\
\nu^{(\mathrm{i})}_{\mp3} &\equiv& 2\nu^{(\mathrm{i})}_\ii\mp3\nu^{(\mathrm{i})}_\oo, \\
\nu^{(\mathrm{i})}_{\mp1}  &\equiv&  2\nu^{(\mathrm{i})}_\ii\mp\nu^{(\mathrm{i})}_\oo,
\end{eqnarray}
where the minus and plus signs are for prograde and retrograde orbits,
respectively. We introduce the superscript $(\mathrm{i})$ so as to
indicate instantaneous orbital elements evaluated at
$\pzo=100P^{(0)}_\oo$. We evaluate equation (\ref{eq:RV}) using the instantaneous orbital elements at $t_{\rm i}$
rather than their input values (Table \ref{tab:tab2}). This is necessary to 
accurately estimate the phases $f_\iiz^{(\mathrm{i})}+\omega_\ii^{(\mathrm{i})}$ and $f_\ooz^{(\mathrm{i})}+\omega_\oo^{(\mathrm{i})}$ in order for the numerical results to reproduce the approximate formula.

Equation (\ref{eq:RV_bin-0}) reproduces the amplitudes of the RV
variations from the simulations (left panels in Figure
\ref{fig:RV-variation}) reasonably well.  Note that the simulated RV
variations are dependent on the empirically removed quasi-Keplerian
component, while equation (\ref{eq:RV_bin-0}) is the lowest-order
perturbation approximation neglecting the back-reaction of the outer
star on the inner orbit. Therefore, the discrepancy between the two
should not be regarded as serious.

Nevertheless, the corresponding LS periodograms (right panels in
Figure \ref{fig:RV-variation}) clearly detect the presence of the
periodic components that are due to the inner binary, especially at the
frequencies of $\nu^{(0)}_{-3}$ and $\nu^{(0)}_{+3}$ for prograde and
retrograde orbits, respectively.  Furthermore, the lower-amplitude
peak at the accompanying frequency ($\nu^{(0)}_{-1}$ or
$\nu^{(0)}_{+1}$) can imply in principle whether the inner and outer
orbits are prograde or retrograde.  The agreement between the
simulations and predictions is degraded for frequencies less than
$\nu_\ii^{(0)}$, which likely results from the uncertainty of the
empirical removal of the underlying quasi-Keplerian RV component, as
mentioned in the above.  The LS periodograms prove, however, that the
frequency modes at $\nu^{(0)}_{\pm 3}$ and $\nu^{(0)}_{\pm 1}$ are
fairly robust against the removal procedure.

Incidentally, the agreement between the simulation and predictions
seems worse for the unequal-mass binary case (P0218).  This is
thought to come from the higher-order perturbation effect; the larger
mass difference of the binary enhances the octupole
\citep[e.g.][]{Mardling2013}, which is neglected in the approximation
by \citet{Morais2008}.

\subsection{Effect of the eccentricity of the inner binary on the stellar
  radial velocity variation
  \label{subsec:eccentricity}}

Both $e_\ii$ and $e_\oo$, the eccentricities of the inner and outer
orbits, sensitively change the RV variations as shown in Paper I.  The
outer stellar orbit could be very eccentric, but we neglect it in the
present paper because $e_\oo$ is estimated to be 0.03 for the LB-1
system \citep[][]{Liu2019}. On the other hand, $e_\ii$ is expected to be not so large for
BBHs that we are primarily interested in, because of the
circularization due to the emission of the gravitational wave,
especially for those with a short orbital period.  Therefore, we focus
on the effect of relatively small $e_\ii$ on the RV variation of the
tertiary star in coplanar triple systems.

\begin{figure*}
\begin{center}
\includegraphics[clip,width=8.0cm]{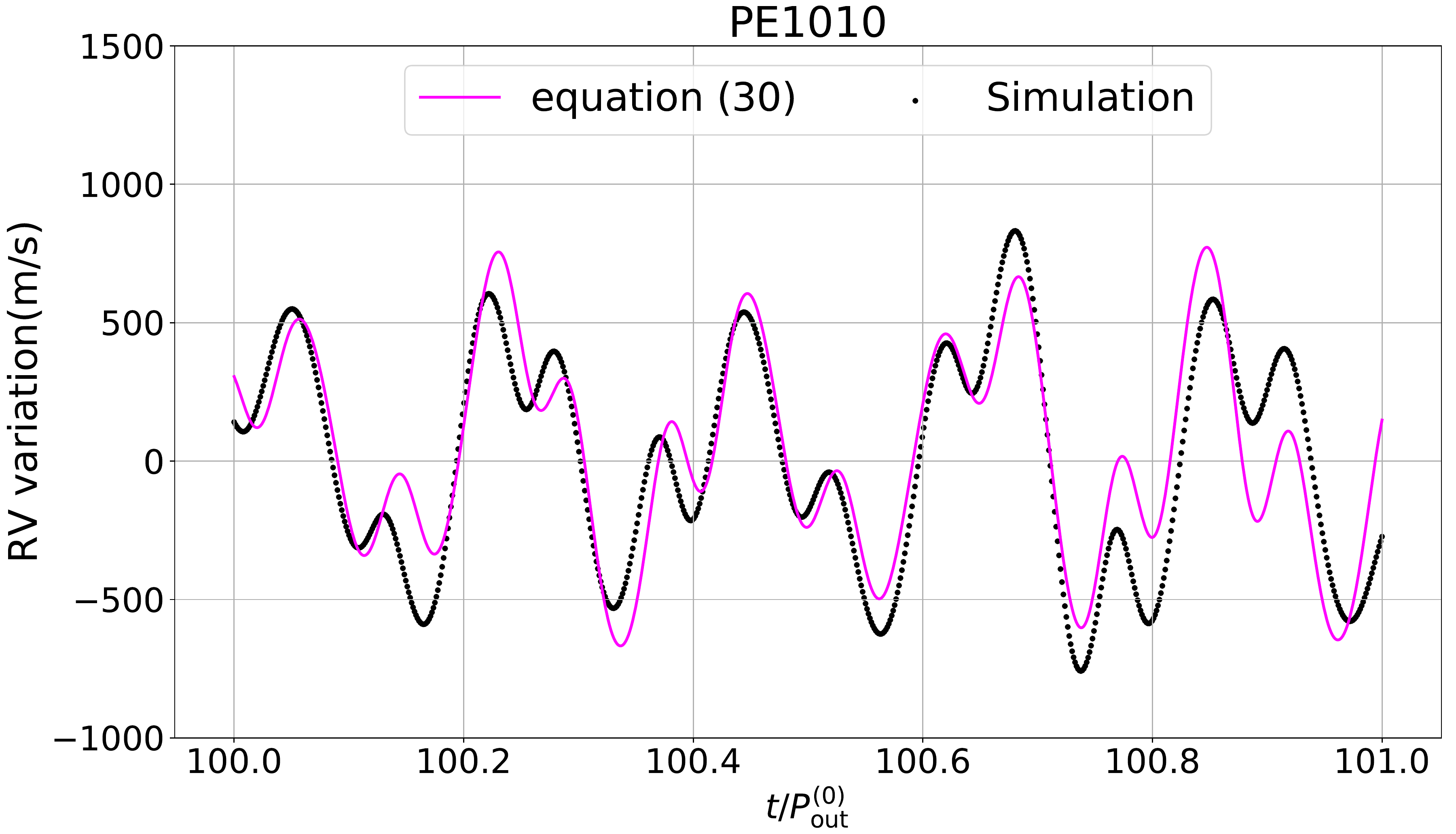} 
\includegraphics[clip,width=8.0cm]{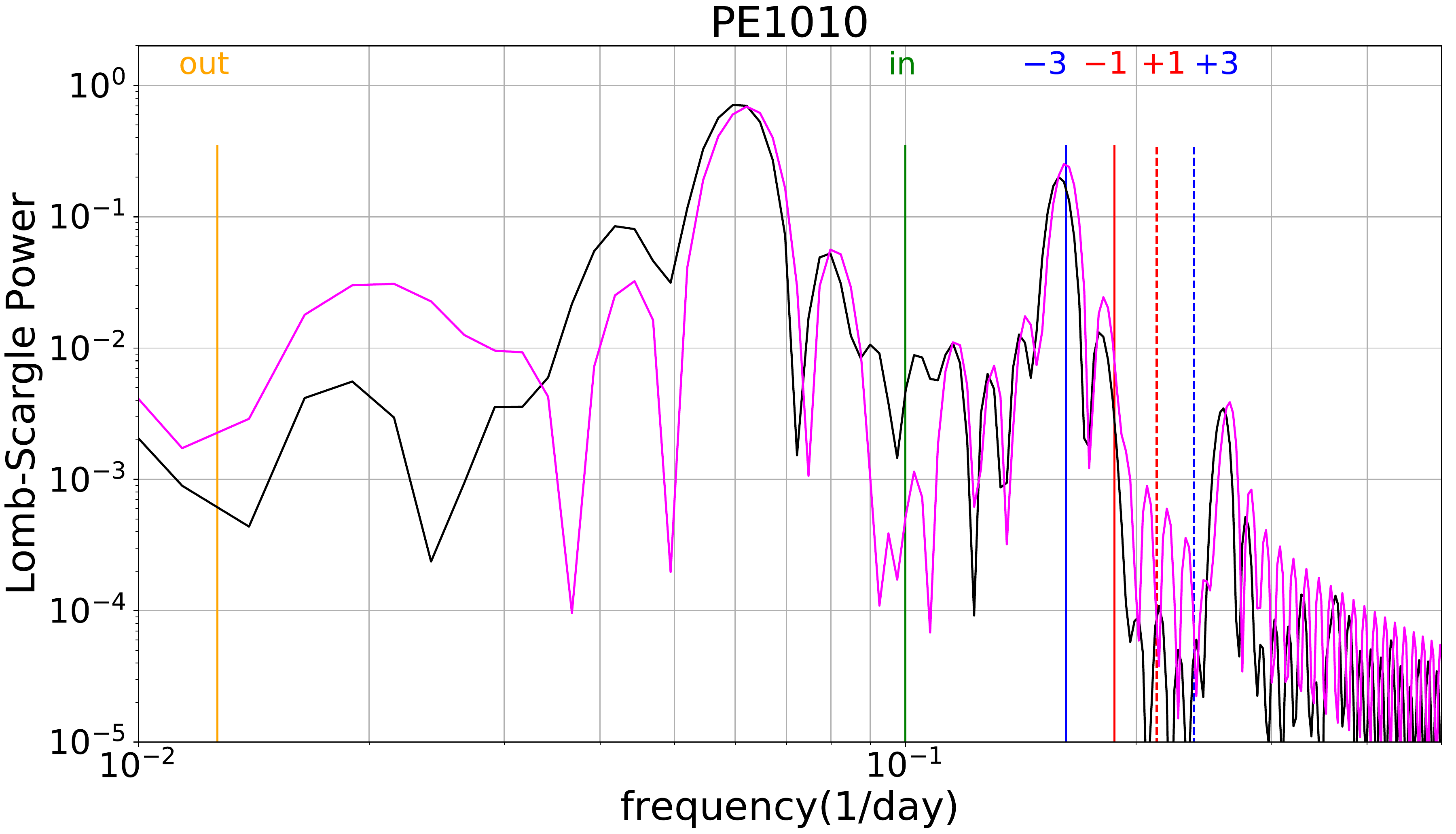}

\includegraphics[clip,width=8.0cm]{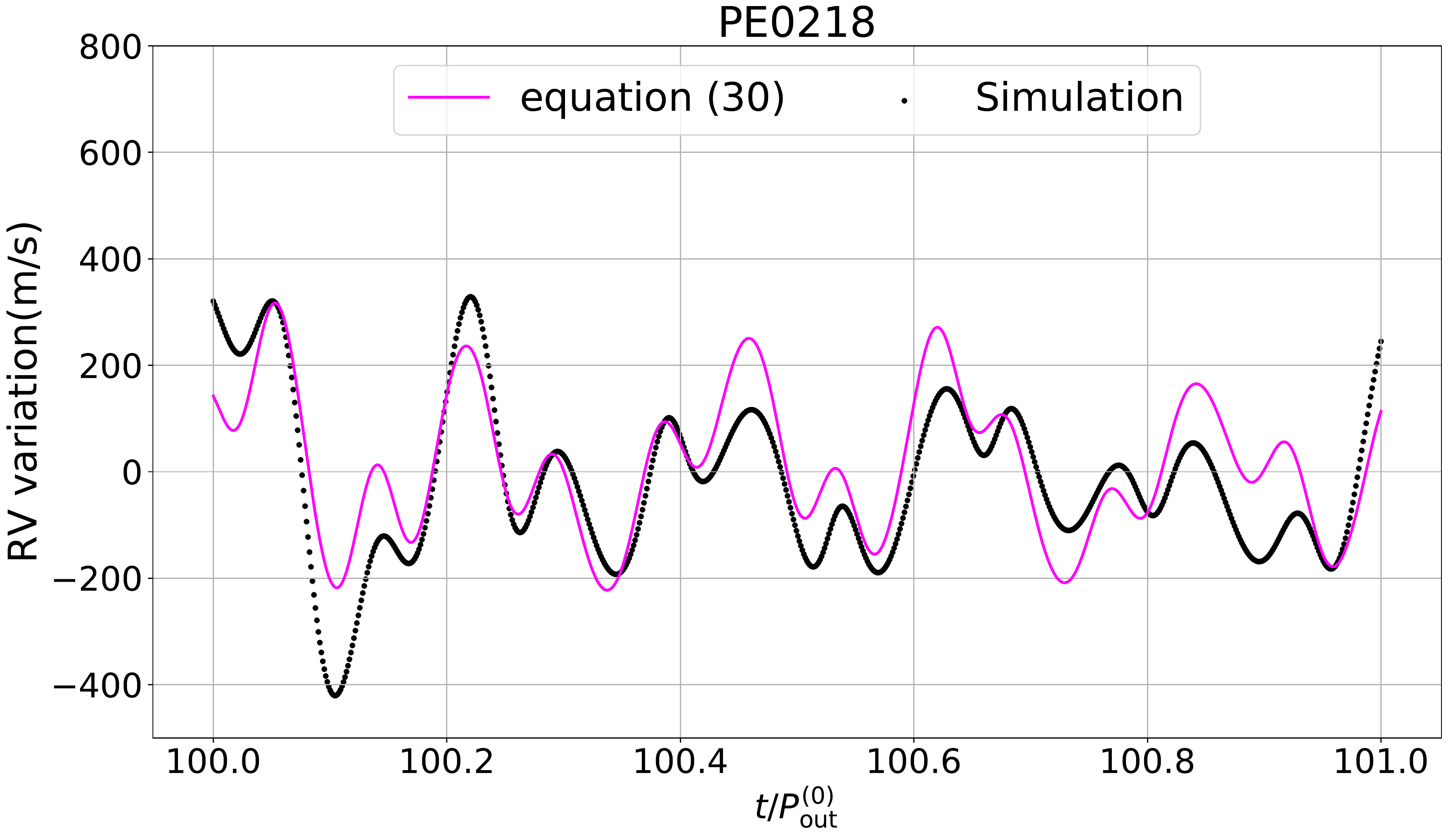} 
\includegraphics[clip,width=8.0cm]{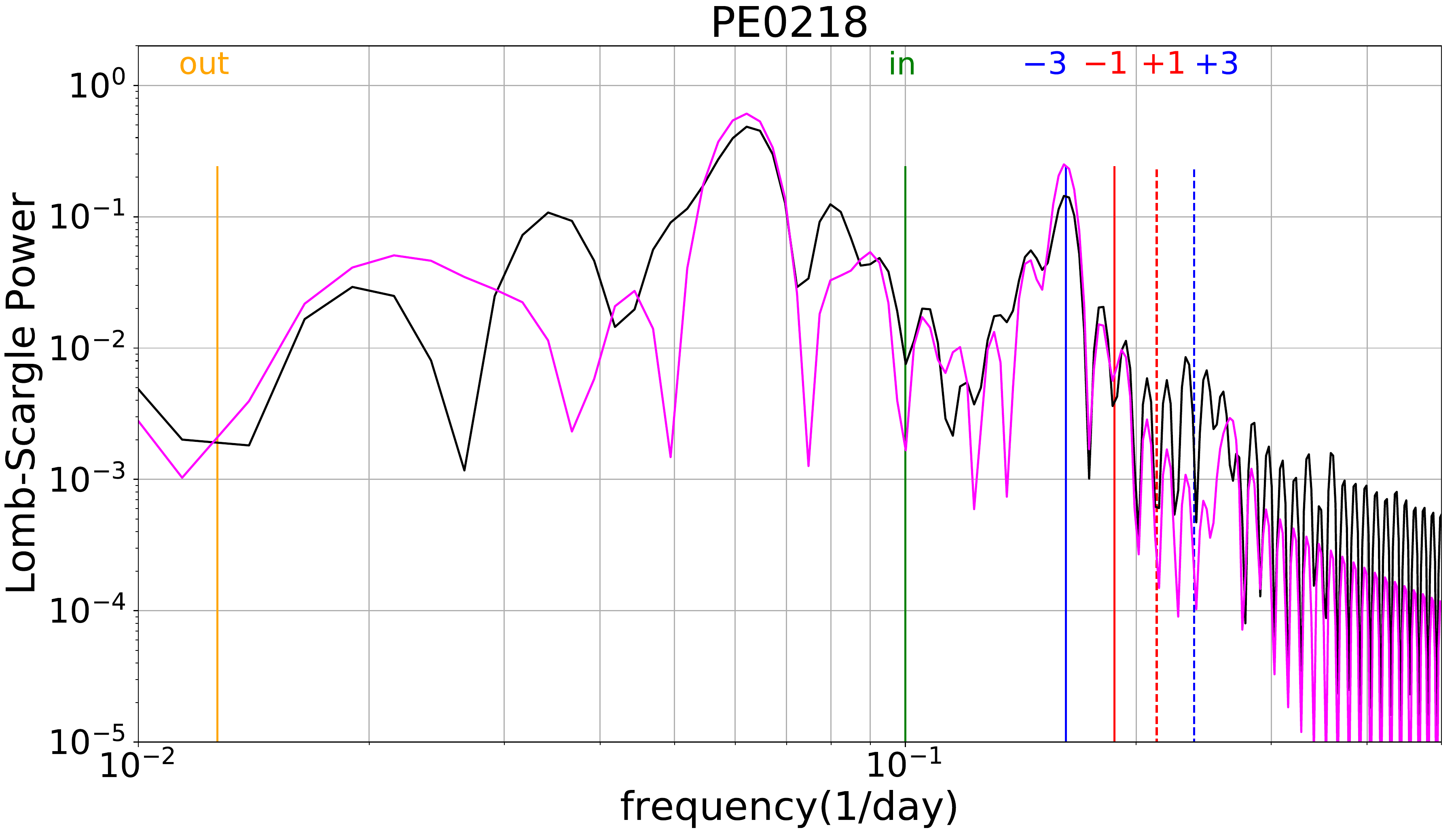}
\end{center}
\caption{Same as Figure \ref{fig:RV-variation} but for PE1010 and
  PE0218. The magenta lines show the approximate formula (equation
  (\ref{eq:bin_e})) evaluated at $t=100P^{(0)}_\oo$. The phase and
  constant offsets $(\delta t,C)$ are empirically determined to match the simulated residuals: $(\delta t,C)=$$(+0.115P^{(0)}_\oo,-480.0)$ and
  $(-0.27P^{(0)}_\oo,-430.0)$ for PE1010 and PE0218,
  respectively. \label{fig:RV-var-e}}
\end{figure*}

\citet{Morais2011} have derived an analytic approximation for the RV
variation in a coplanar eccentric triple,
to the lower order of $e_\ii$ and $e_\oo$:
\begin{eqnarray}
\label{eq:rv_e}
V_\mathrm{RV}(t)
= V^{(0)}_\kep(t) + \delta V_\kep(t) + V_\bin(t),
\end{eqnarray}
where $V^{(0)}_\kep(t)$ is the unperturbed Keplerian radial velocity, and $\delta V_\kep(t)$ in a coplanar eccentric case is now written as
\begin{eqnarray}
\label{eq:rv_e_kep}
\delta V_\kep(t) &=& 
K_1 \sin{I_\oo}
\cos \left( \nu_\oo\,t+\lambda_\ooz \right),
\end{eqnarray}
in terms of the initial mean longitude $\lambda_{j,0}$.  The true anomaly $f$ and
argument of pericenter $\omega$ can be expanded in terms of the small
eccentricity $e$ as
\begin{eqnarray}
\label{eq:f-omega-lambda}
f + \omega = \lambda + 2e\sin(\lambda-\omega) + \mathcal{O}(e^2) ,
\end{eqnarray}
\citep[e.g.][]{Murray2000}. Thus $f+\omega$ is identical to
$\lambda$ for a circular case, and equation (\ref{eq:delta-VKep})
reduces to equation (\ref{eq:rv_e_kep}). In an eccentric case, however,
equation (\ref{eq:f-omega-lambda}) is necessary to clarify the effect of
the eccentricities in a perturbative manner.

An analytic approximation for $V_\bin(t)$ in equation (\ref{eq:rv_e})
is derived by \citet{Morais2011}, which is
explicitly written as
\begin{eqnarray}
\label{eq:V-e-bin-MC}
V^{\mathrm{MC}}_\bin(t) &=&  \frac{3}{8}\,K_\bin \sin{I_\oo} \times
\left[ \frac{\nu_\ii}{2\,\nu_\ii-\nu_\oo}
  \cos[(2\,\nu_\ii-\nu_\oo)\,t +2\,\lambda_\iiz-\lambda_\ooz] \right. \cr 
  && -\frac{5\,\nu_\ii}{2\,\nu_\ii-3\,\nu_\oo}
  \cos[(2\,\nu_\ii-3\,\nu_\oo)\,t +2\,\lambda_\iiz-3\,\lambda_\ooz] \cr
  && +15\,e_\ii\,\frac{\nu_\ii}{\nu_\ii-3\,\nu_\oo}
  \cos[(\nu_\ii-3\,\nu_\oo)\,t+\lambda_\iiz-3\,\lambda_\ooz+\varpi_\ii] \cr 
  && +e_\ii\,\frac{\nu_\ii}{3\,\nu_\ii-\nu_\oo}
  \cos[(3\,\nu_\ii-\nu_\oo)\,t+3\,\lambda_\iiz-\lambda_\ooz -\varpi_\ii] \cr 
  && -5\,e_\ii\,\frac{\nu_\ii}{3\,\nu_\ii-3\,\nu_\oo}
  \cos[(3\,\nu_\ii-3\,\nu_\oo)\,t+3\,\lambda_\iiz-3\,\lambda_\ooz-\varpi_\ii] \cr 
  && -2\,e_\ii\,\frac{\nu_\ii}{\nu_\ii+\nu_\oo}
  \cos[(\nu_\ii+\nu_\oo)\,t+\lambda_\iiz+\lambda_\ooz -\varpi_\ii] \cr
  && -3\,e_\ii\,\frac{\nu_\ii}{\nu_\ii-\nu_\oo}
  \cos[(\nu_\ii-\nu_\oo)\,t +\lambda_\iiz-\lambda_\ooz +\varpi_\ii] \cr
  && +2\,e_\ii\,\frac{\nu_\ii}{\nu_\ii-\nu_\oo}
  \cos[(\nu_\ii-\nu_\oo)\,t+\lambda_\iiz-\lambda_\ooz-\varpi_\ii] \cr
  && +6\,e_\oo \frac{\nu_\ii}{2\,\nu_\oo}
  \cos \left(2\,\nu_\oo\,t+2\,\lambda_\ooz -\varpi_\oo \right) \cr
  && +e_\oo \frac{\nu_\ii}{2\,\nu_\ii}
  \cos \left(2\,\nu_\ii\,t+2\,\lambda_\iiz-\varpi_\oo \right) \cr
  && -25\,e_\oo \frac{\nu_\ii}{2\,\nu_\ii-4\,\nu_\oo}
  \cos[(2\,\nu_\ii-4\,\nu_\oo)\,t+2\,\lambda_\iiz-4\,\lambda_\ooz+\varpi_\oo] \cr
  && +3\,e_\oo\frac{\nu_\ii}{2\,\nu_\ii-2\,\nu_\oo}
  \cos[(2\,\nu_\ii-2\,\nu_\oo)\,t+2\,\lambda_\iiz-2\,\lambda_\ooz+\varpi_\oo] \cr
  && \left. +5\,e_\oo\frac{\nu_\ii}{2\,\nu_\ii-2\,\nu_\oo}
  \cos[(2\,\nu_\ii-2\,\nu_\oo)\,t+2\,\lambda_\iiz-2\,\lambda_\ooz-\varpi_\oo]
  \right] \ ,
\end{eqnarray}
where $\varpi_j = \omega_j + \Omega_j$.

In reality, however, our simulation results have an uncertain offset
relative to equation (\ref{eq:V-e-bin-MC}), and thus we model
$V_\bin(t)$ as
\begin{eqnarray}
\label{eq:bin_e}
V_\bin(t) = V^{\mathrm{MC}}_\bin(t) - V^{\mathrm{MC}}_\bin(0) + V_0 +C. 
\end{eqnarray}
In equation (\ref{eq:bin_e}), we define $V_0$ as the initial velocity
of the RV variation, and $C$ is an additional constant discussed below.

We first fit the simulation data using {\tt RadVel} to obtain
$V^{(0)}_\kep(t)$$+\delta V_\kep(t)$ at $t=\pzo=100P^{(0)}_\oo$. Thus
the residual RV variation from the simulation should correspond to
$V_\bin(t)$.  We also evaluate all of the orbital elements and $V_0$ at
$\pzo$, whose values are substituted into equation
(\ref{eq:V-e-bin-MC}). Since the quasi-Keplerian component
  estimated with our fitting procedure involves a time average over an
  outer orbital period, the residual RV variation $V_\bin(t)$ from the
  simulation should inevitably have a time shift relative to equation
  (\ref{eq:V-e-bin-MC}). Thus we introduce an empirical time shift
  $\delta t$ to match the analytical expression (\ref{eq:V-e-bin-MC})
  and the simulation result. This matching simultaneously requires the
  additional velocity offset term $C$, which is introduced in equation
  (\ref{eq:bin_e}).

Figure \ref{fig:RV-var-e} shows the resulting plot of RV variations
for PE1010 and PE0218.  We find that the simulated RV variations in
the left panel of Figure \ref{fig:RV-var-e} agree well with the
analytic approximation. Thus the presence of an inner binary can be
detected even in a moderately eccentric coplanar system, as long as
the observational data are sufficiently accurate to the level indicated
in Figure \ref{fig:RV-var-e}.

\subsection{noncoplanar orbits  \label{subsec:noncoplanar}}

Finally we examine how the noncoplanarity between the inner and outer
orbits affects the RV variation.  Since the general analysis of the
noncoplanar case is not realistic, we focus on two specific initial
configurations that we call inclined ($i_{\rm mut}=45^\circ$; denoted
by I) and orthogonal ($i_{\rm mut}=90^\circ$; denoted by O).

\begin{figure*}
\begin{center}
\includegraphics[clip,width=7.5cm]{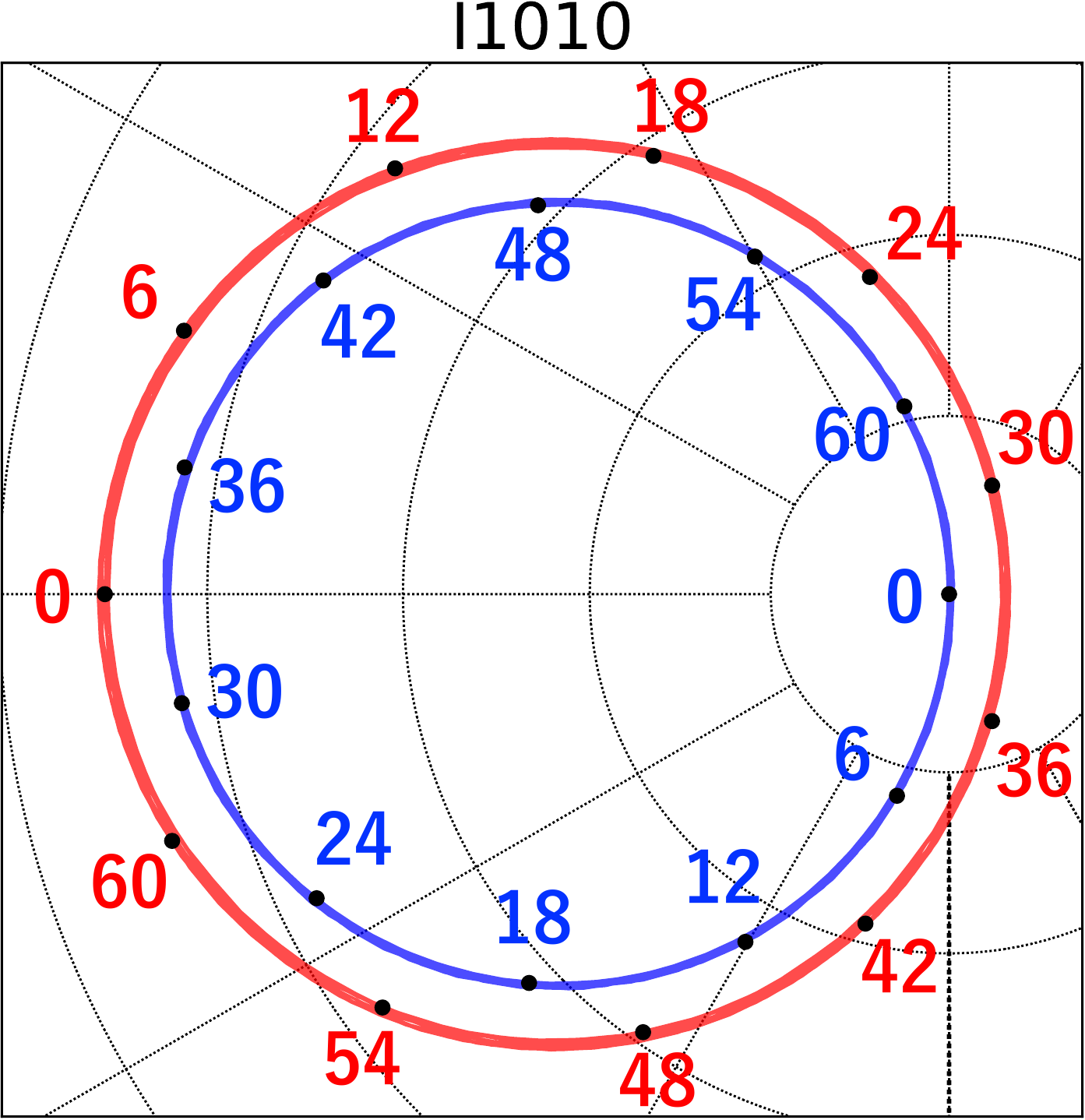} \hspace{14pt}
\includegraphics[clip,width=8.0cm]{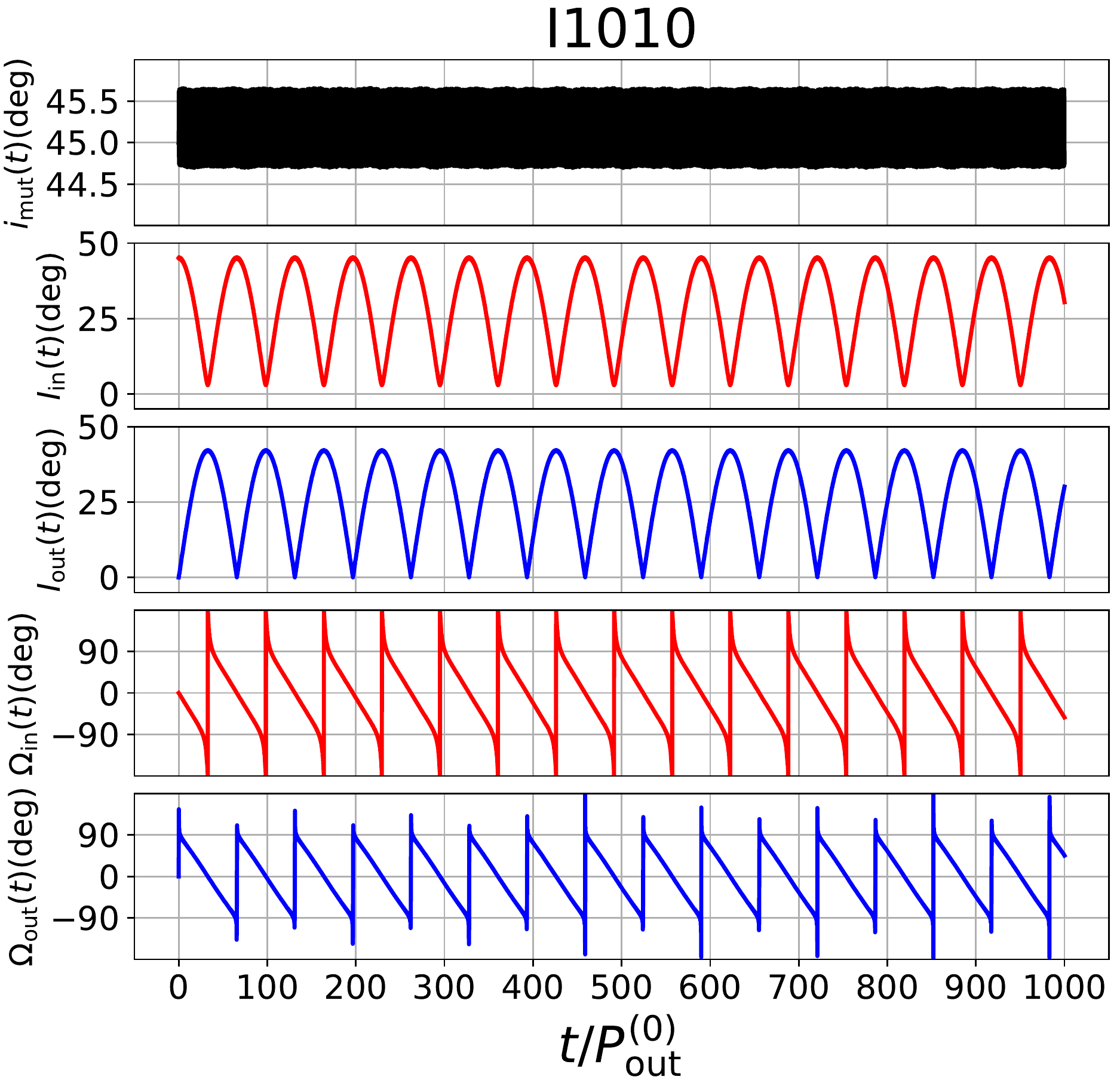}
\includegraphics[clip,width=7.0cm]{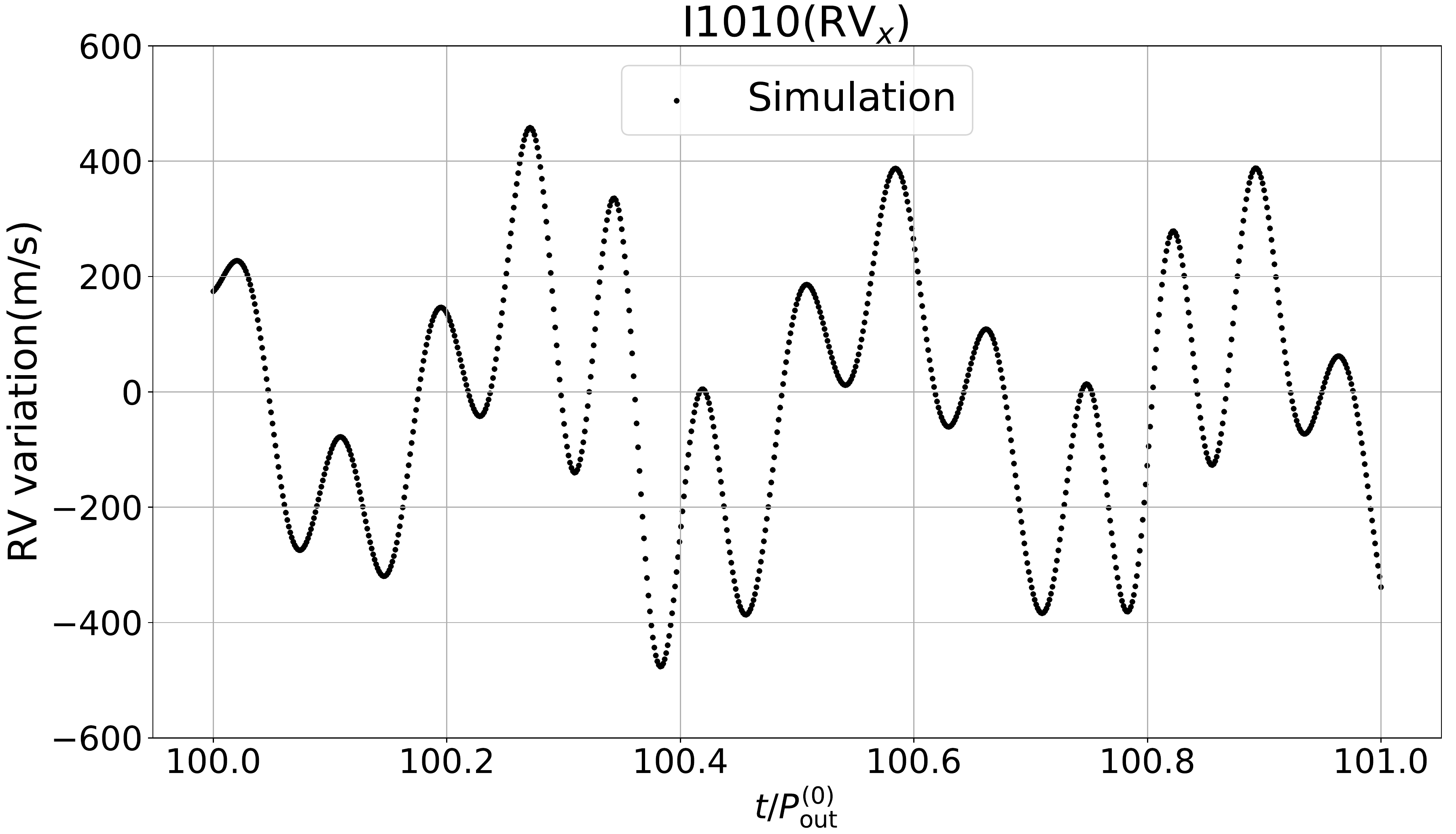}
\includegraphics[clip,width=7.0cm]{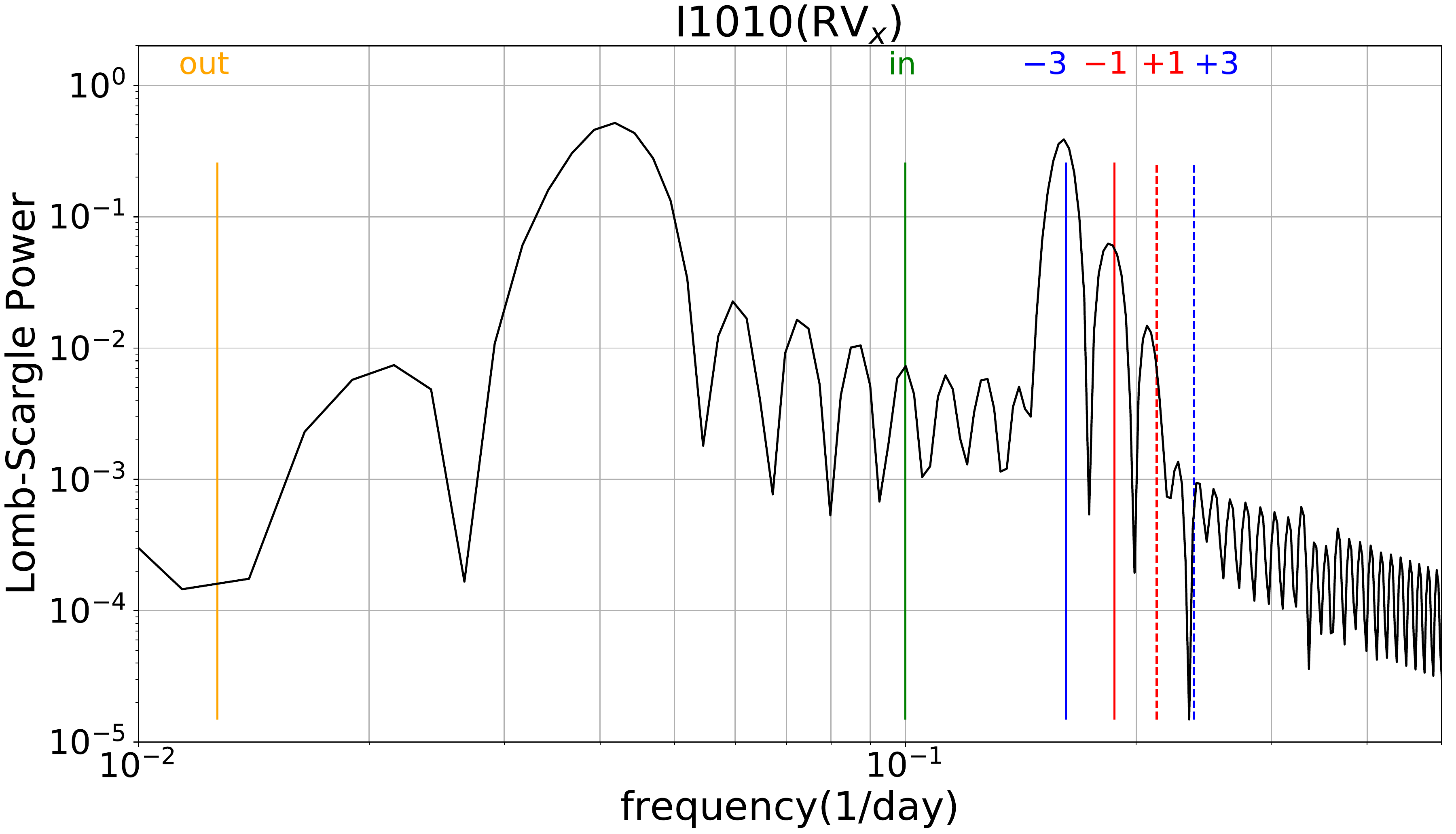}
\includegraphics[clip,width=7.0cm]{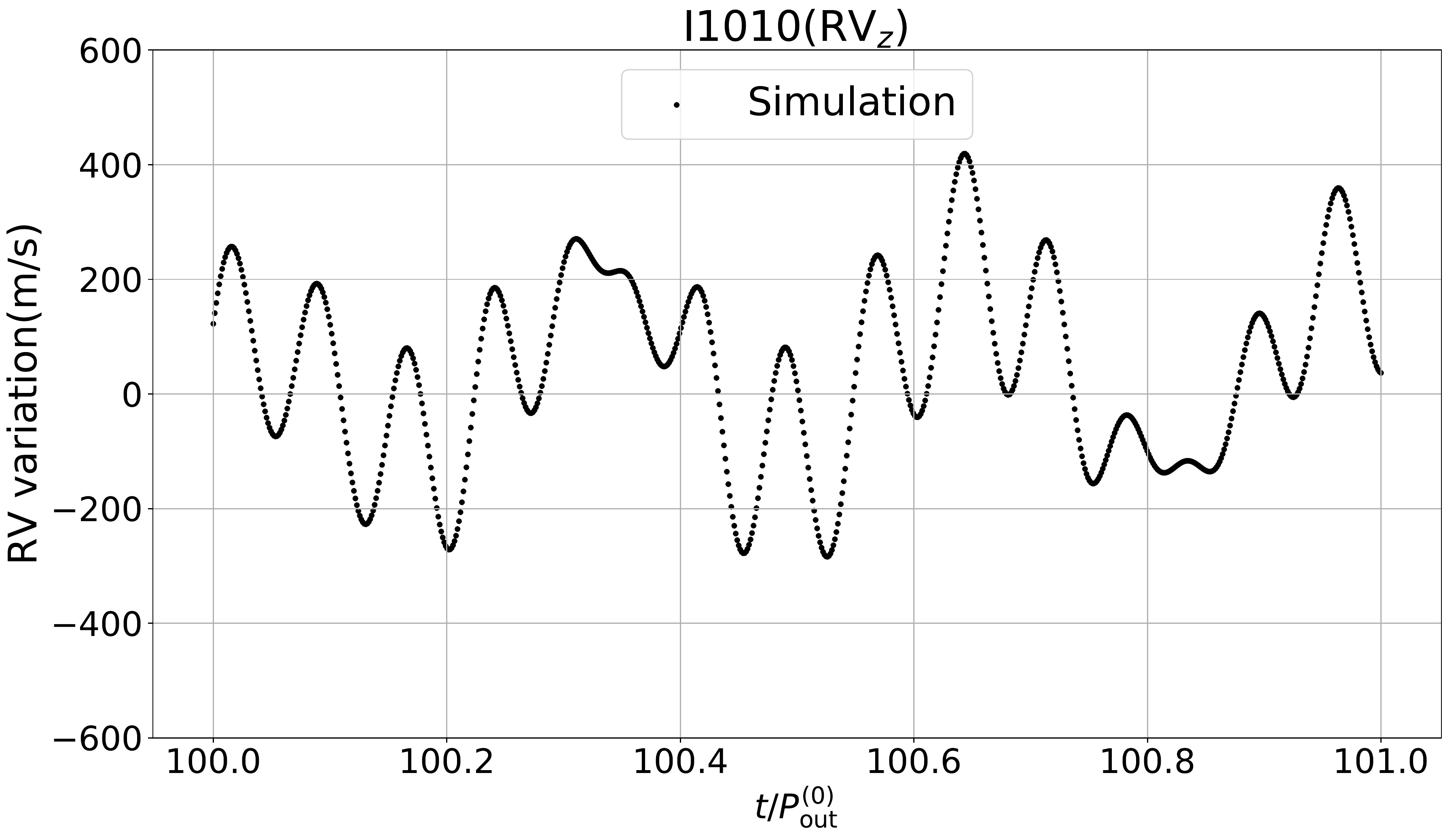}
\includegraphics[clip,width=7.0cm]{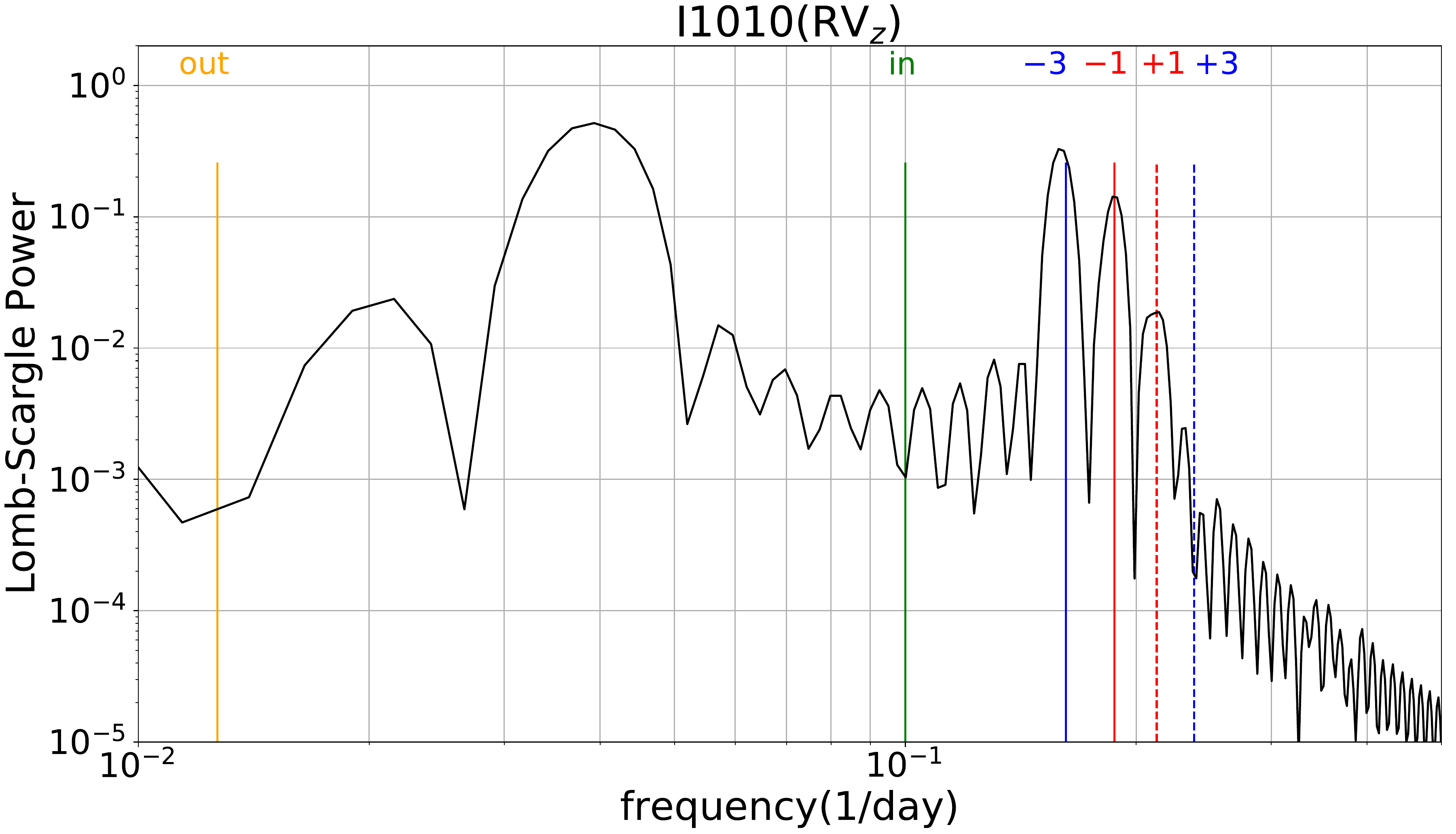}
\includegraphics[clip,width=7.0cm]{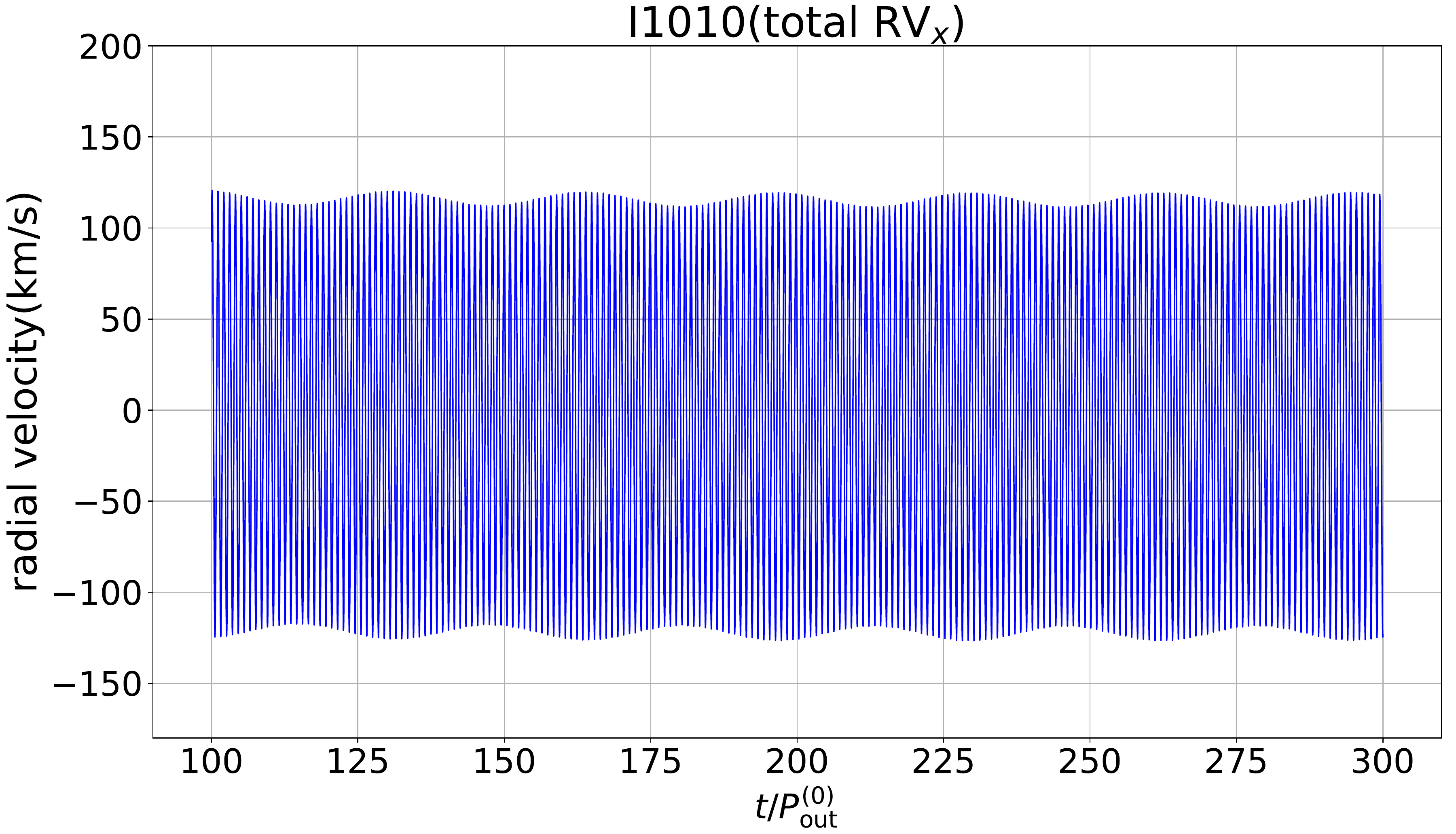}
\includegraphics[clip,width=7.0cm]{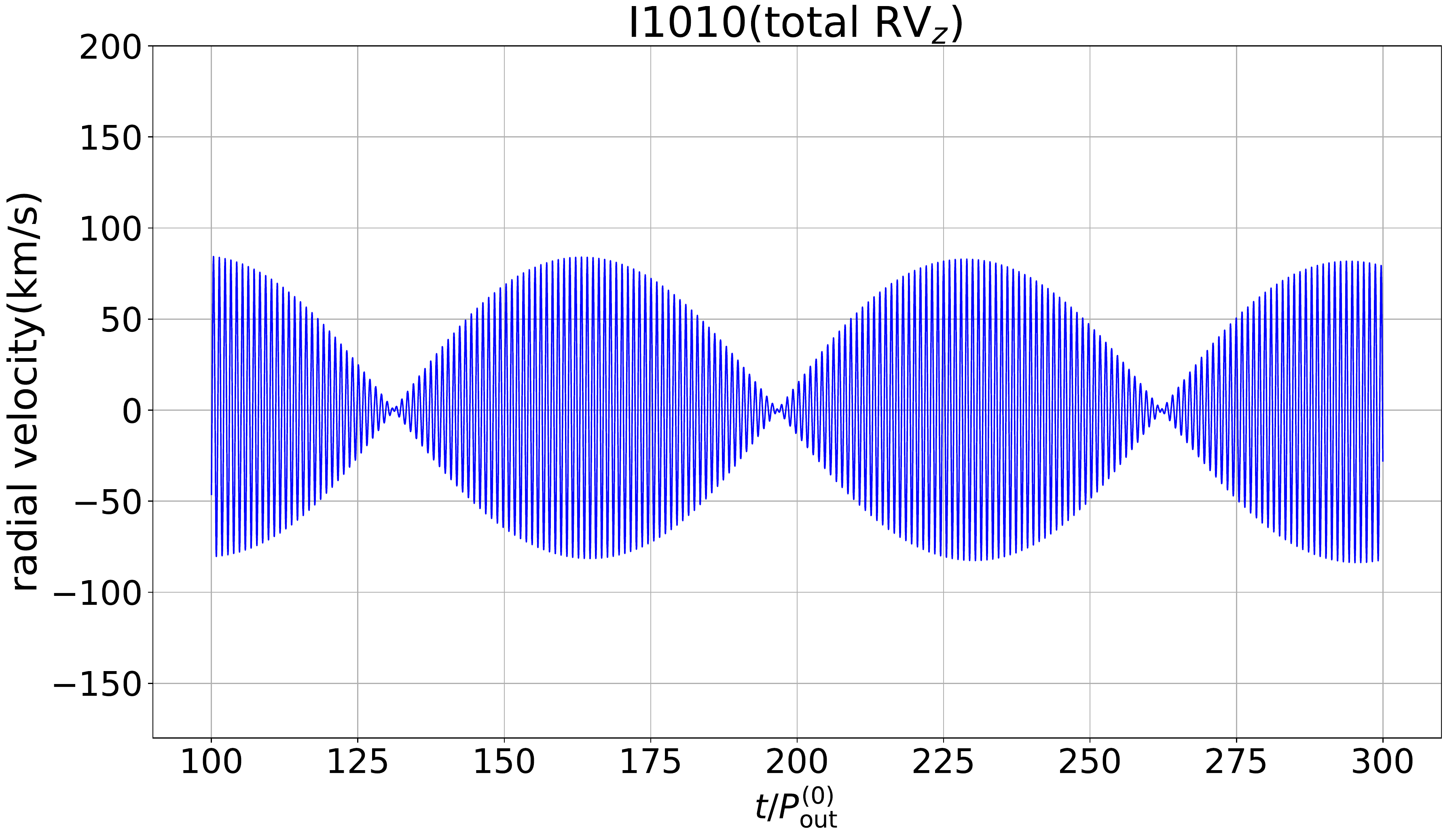}
\end{center}
\caption{Behavior of the noncoplanar triple I1010: {\it Top:} evolution of orientation of the inner and outer orbits (in red and blue, respectively). The longitudinal and latitudinal lines in the left panel are drawn every 30 and
  10 degrees, respectively. 
  {\it Middle:} time series of RV variations
  along $x$, $z$ axes, and the corresponding LS periodograms. 
  {\it Bottom:} total RV curves along the $x$ (left) and $z$ (right) axes.
\label{fig:I1010}}
\end{figure*}

\begin{figure*}
\begin{center}
\includegraphics[clip,width=7.5cm]{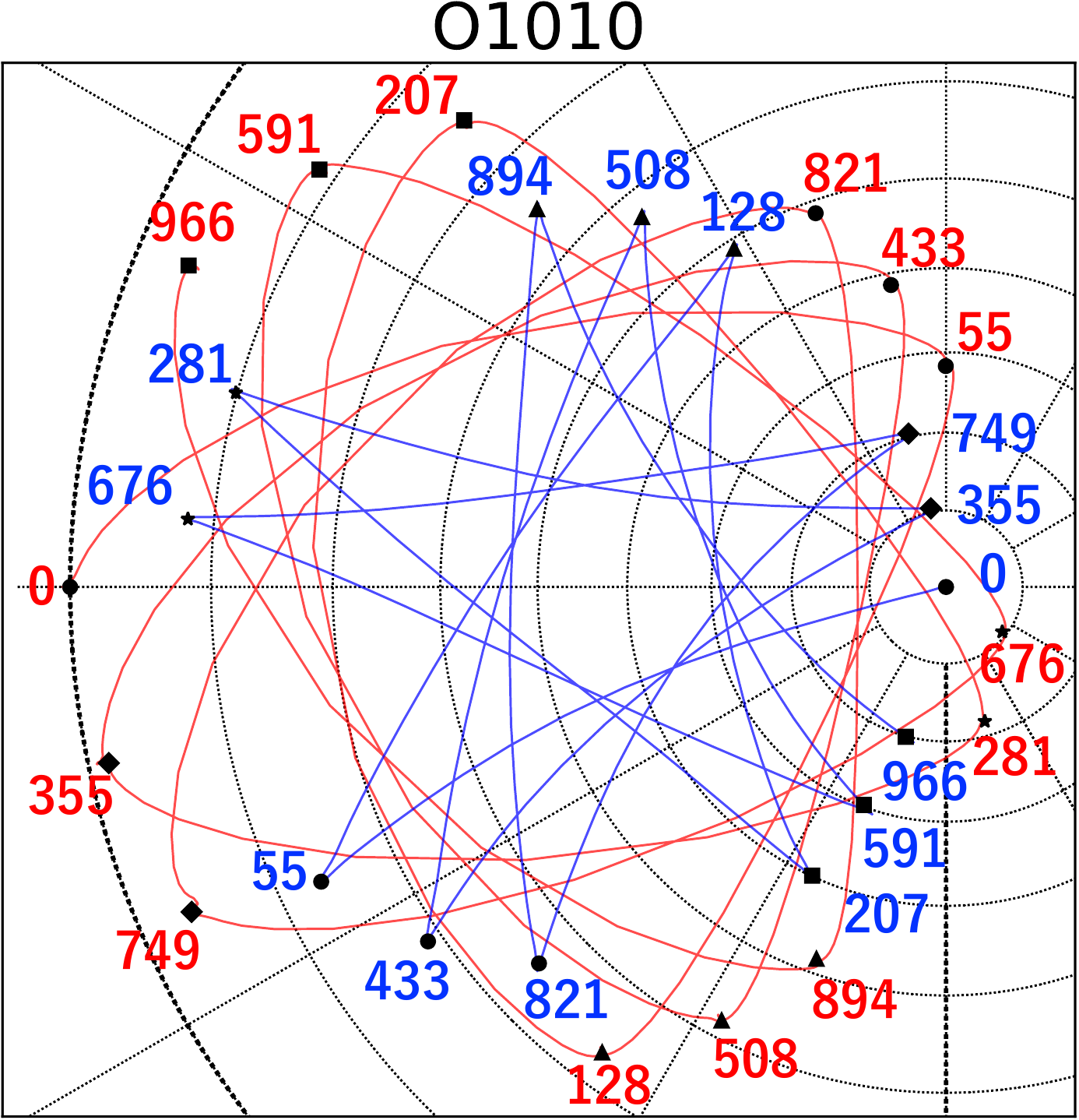} \hspace{14pt}
\includegraphics[clip,width=8.0cm]{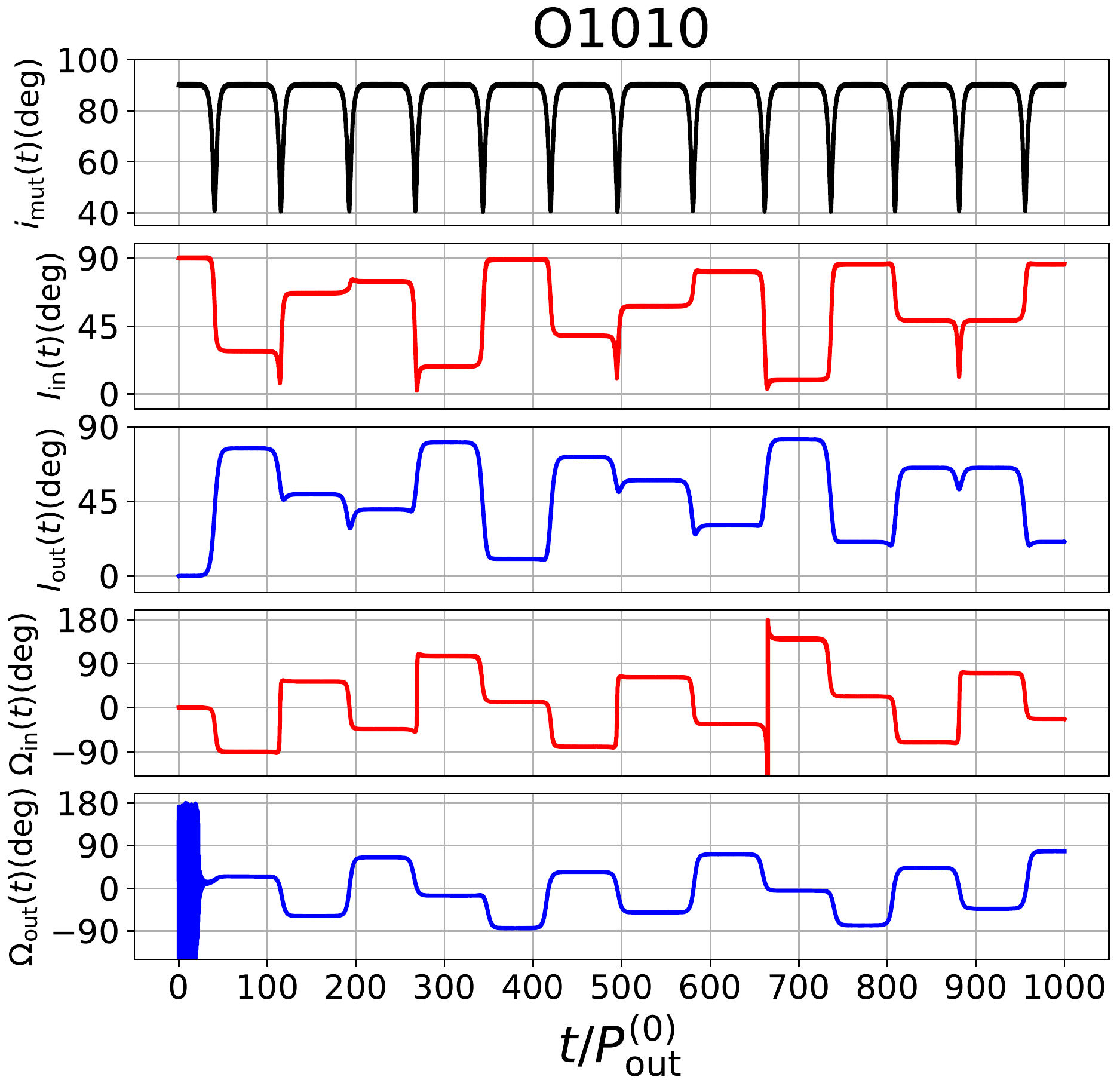}
\includegraphics[clip,width=7.0cm]{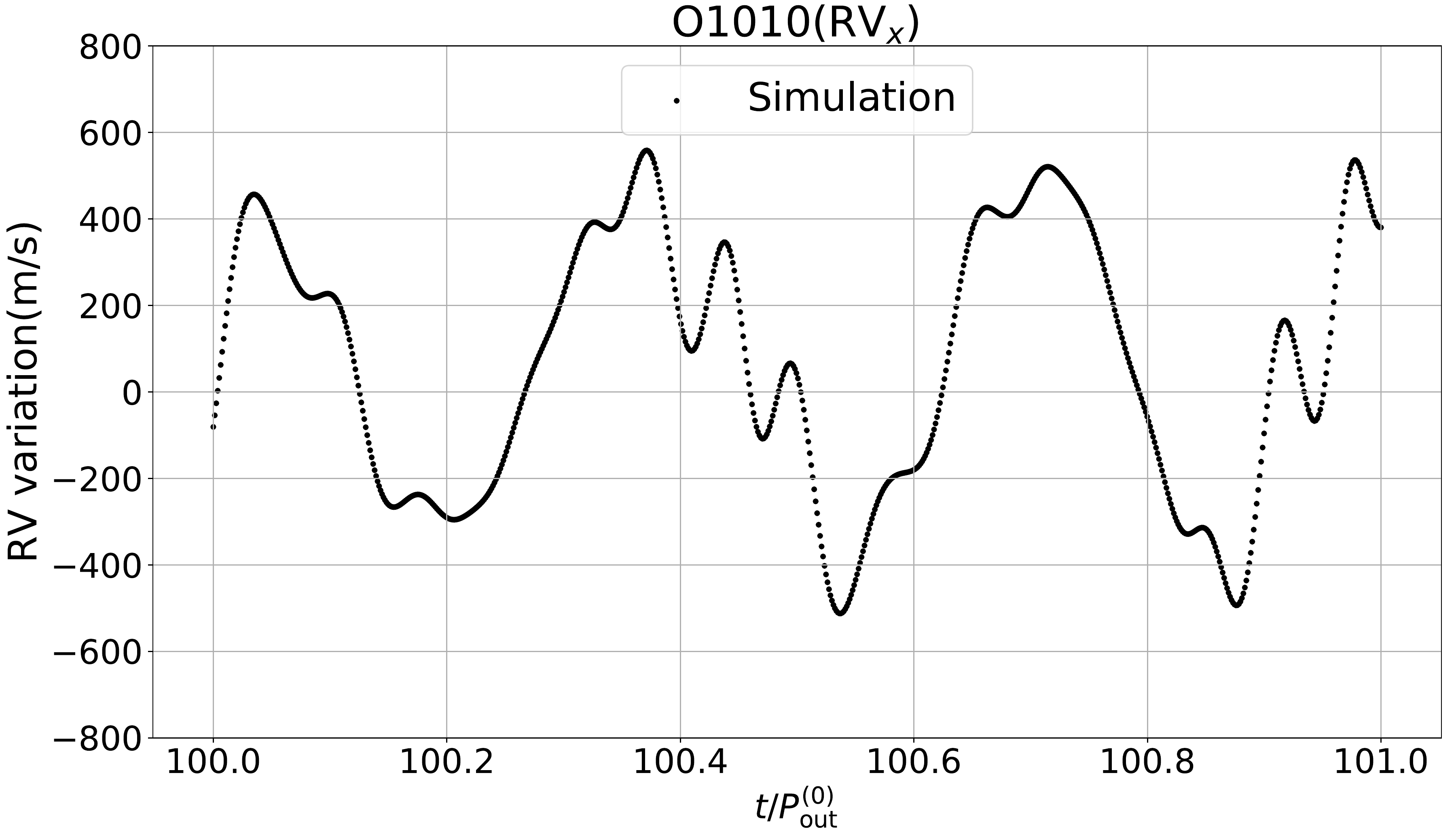}
\includegraphics[clip,width=7.0cm]{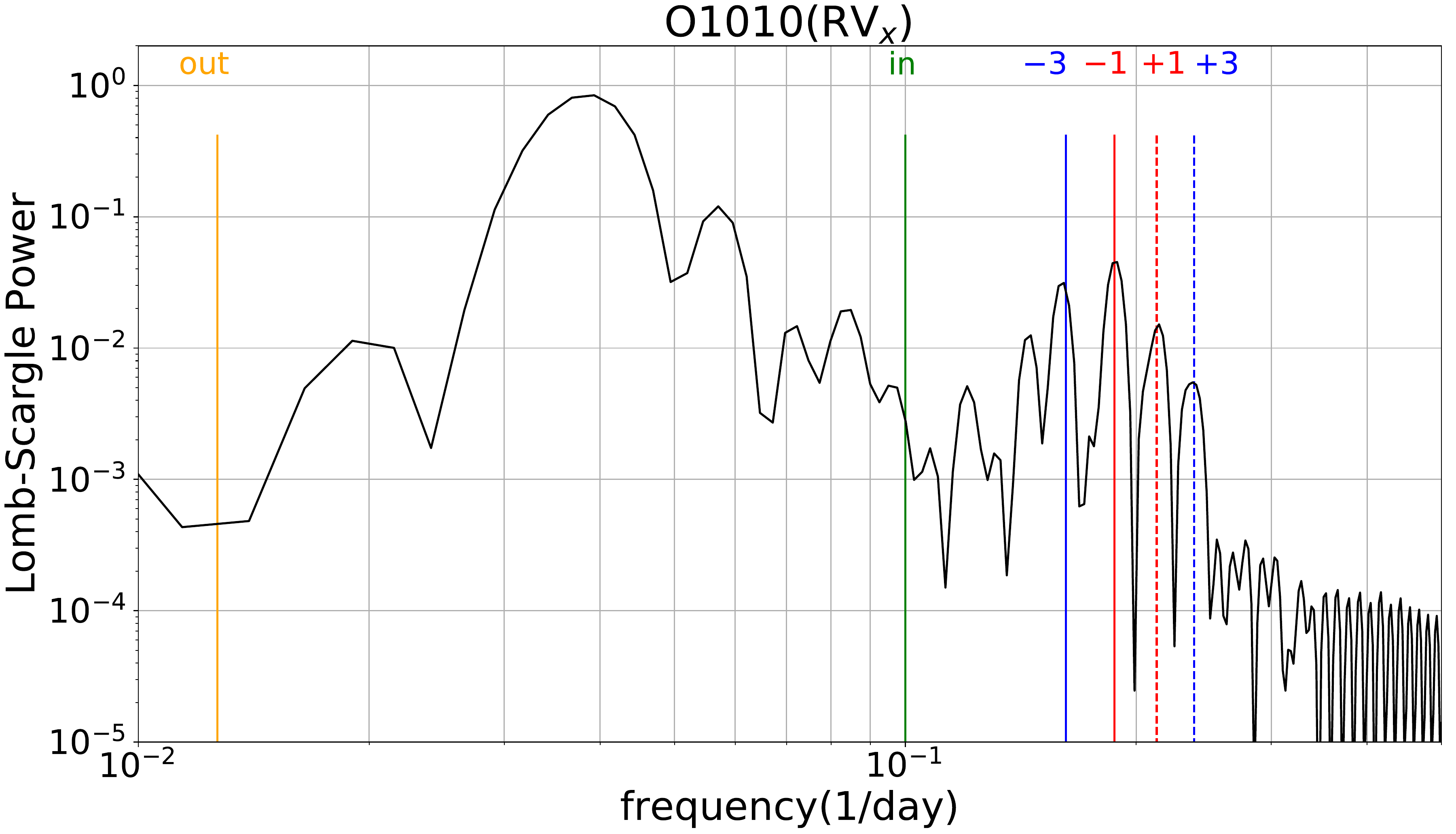}
\includegraphics[clip,width=7.0cm]{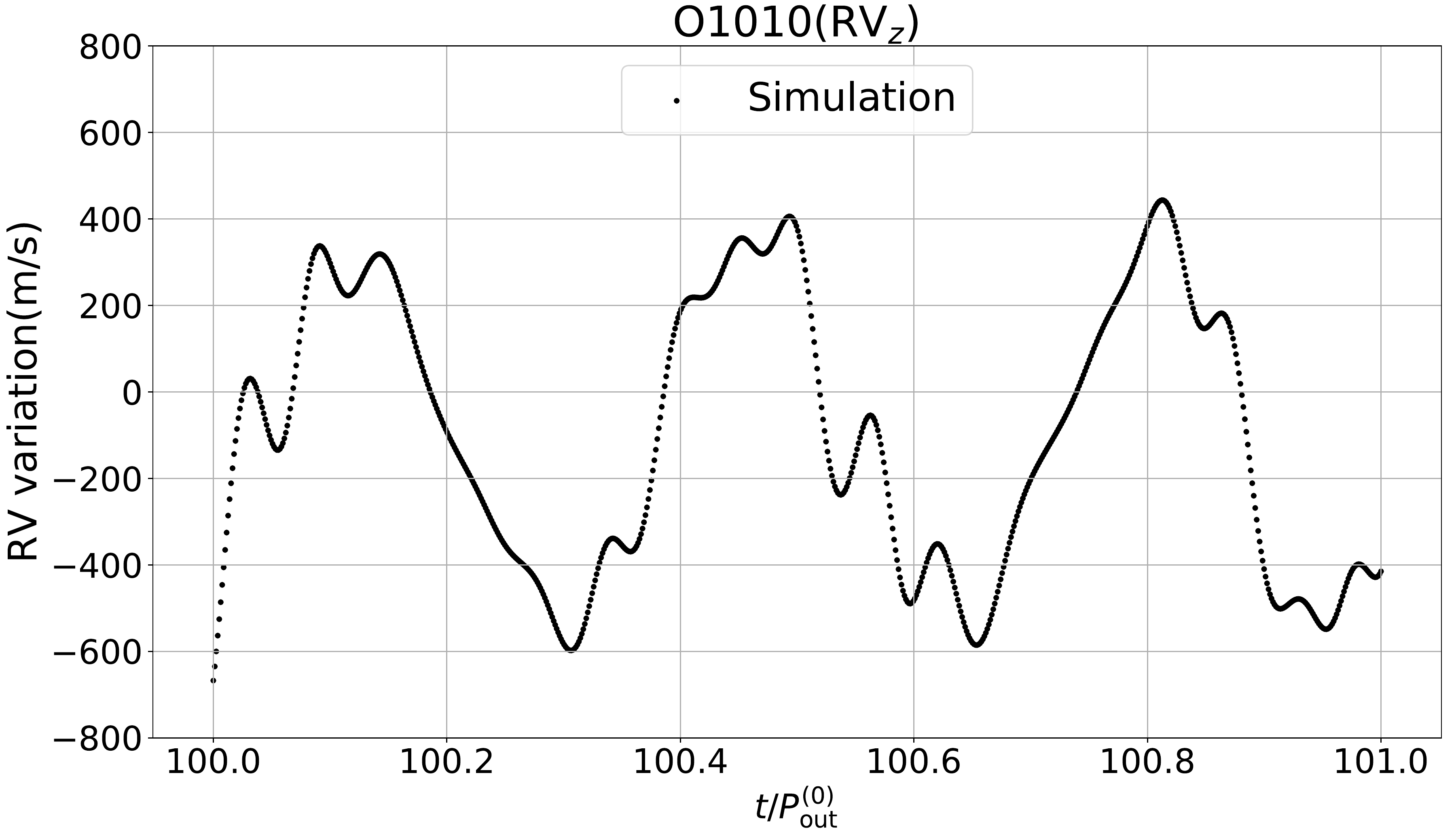}
\includegraphics[clip,width=7.0cm]{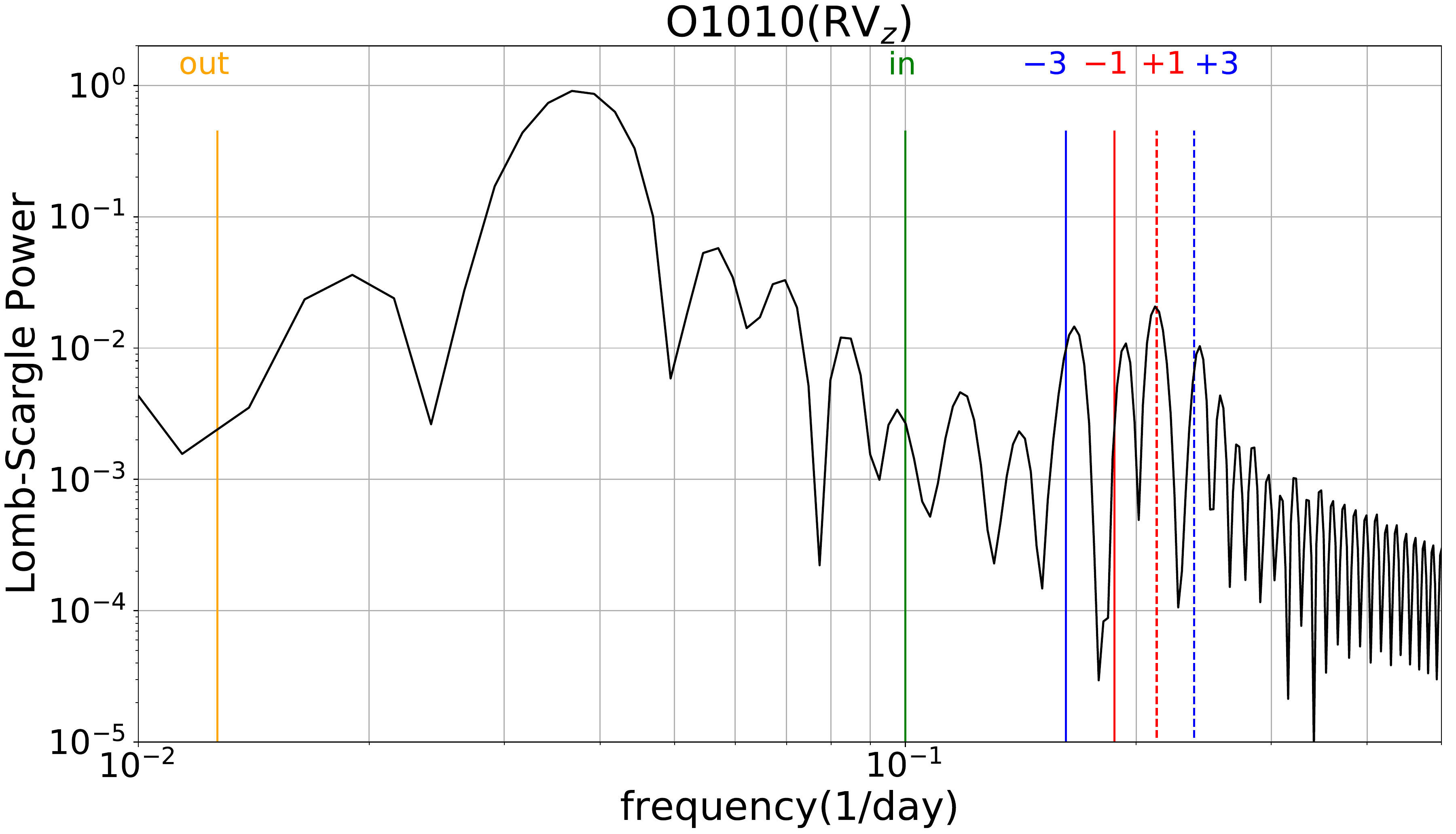}
\includegraphics[clip,width=7.0cm]{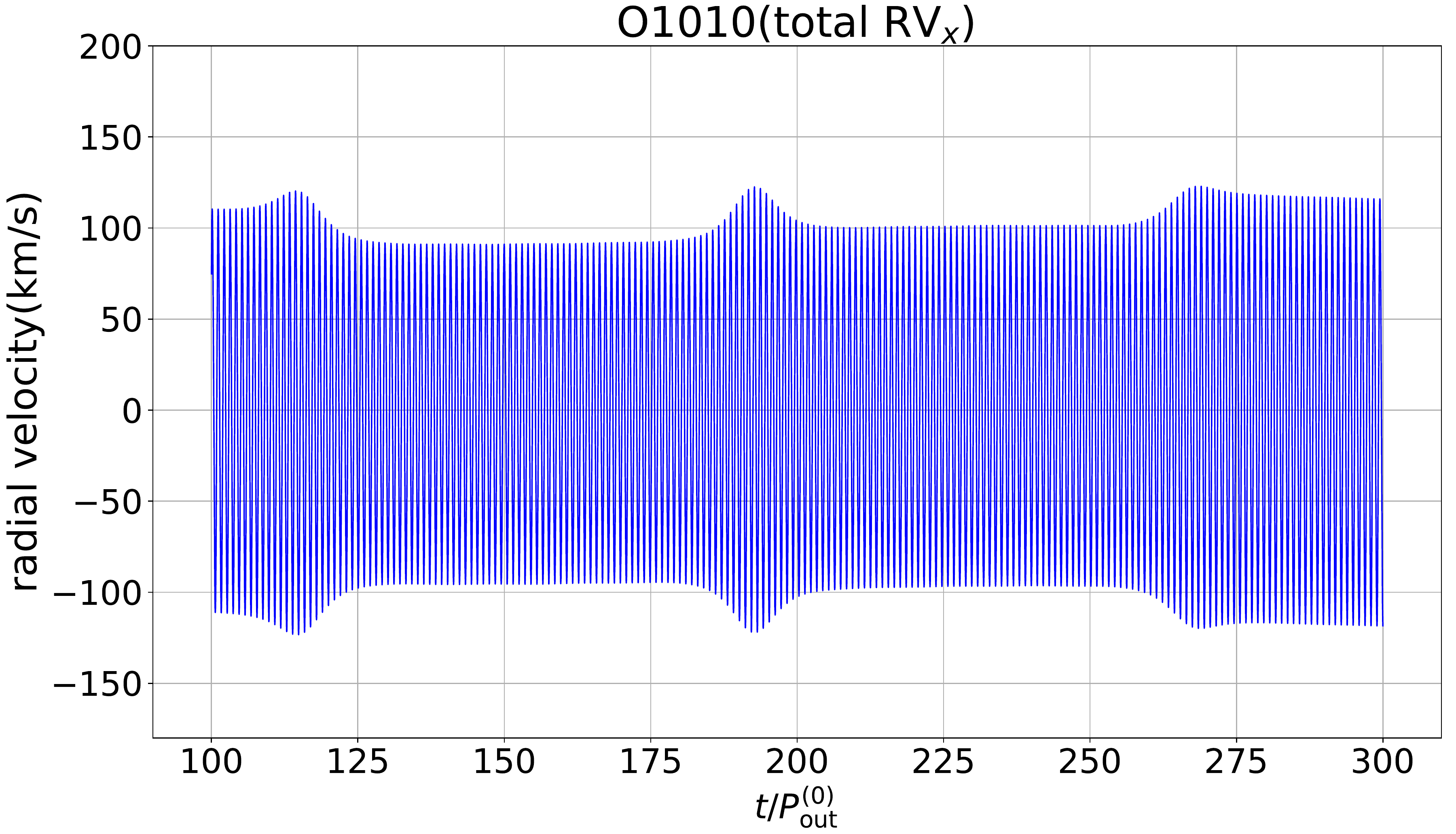}
\includegraphics[clip,width=7.0cm]{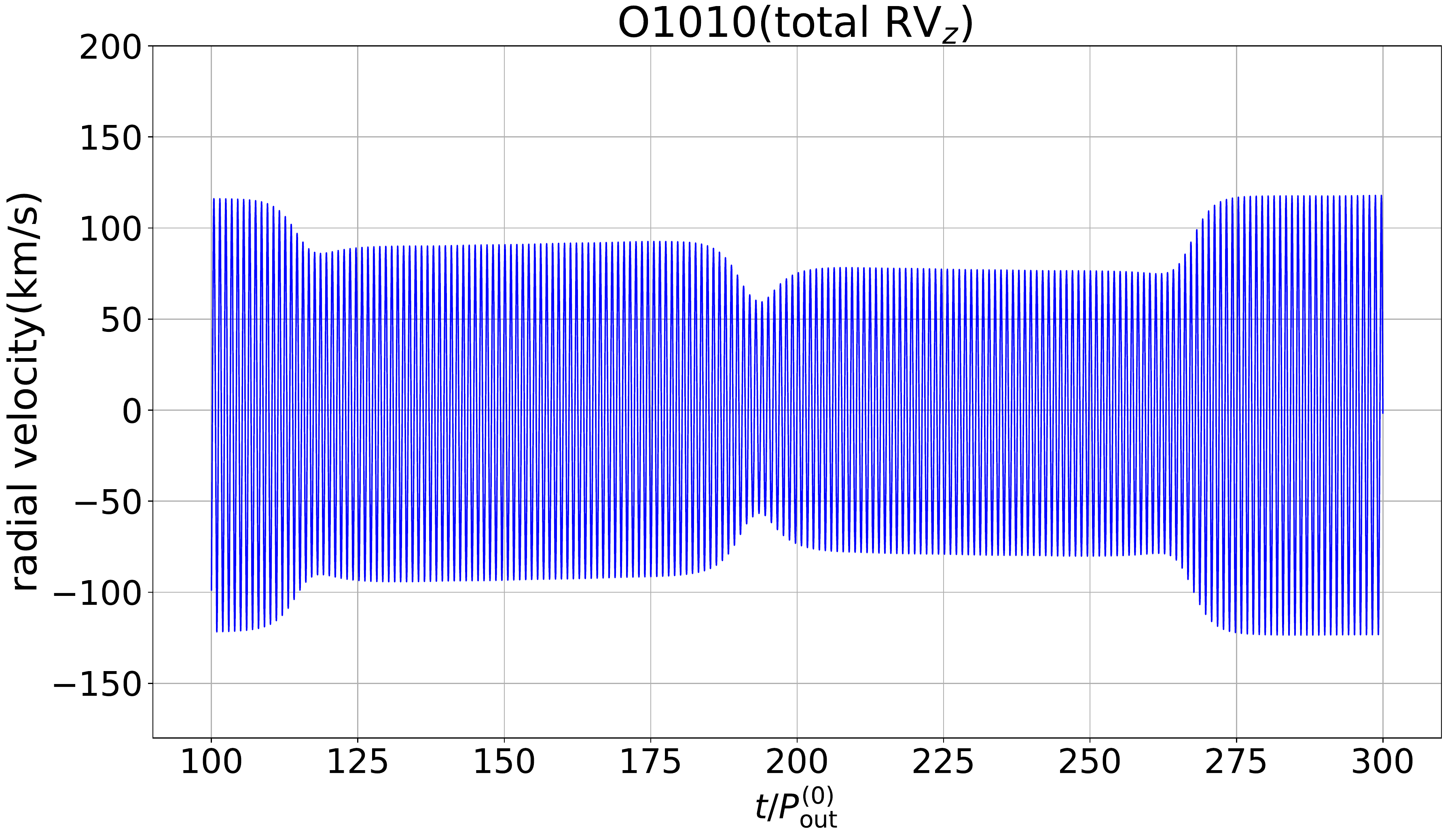}
\end{center}
\caption{Same as Figure \ref{fig:I1010} but for O1010.\label{fig:O1010}}
\end{figure*}

\begin{figure*}
\begin{center}
\includegraphics[clip,width=7.5cm]{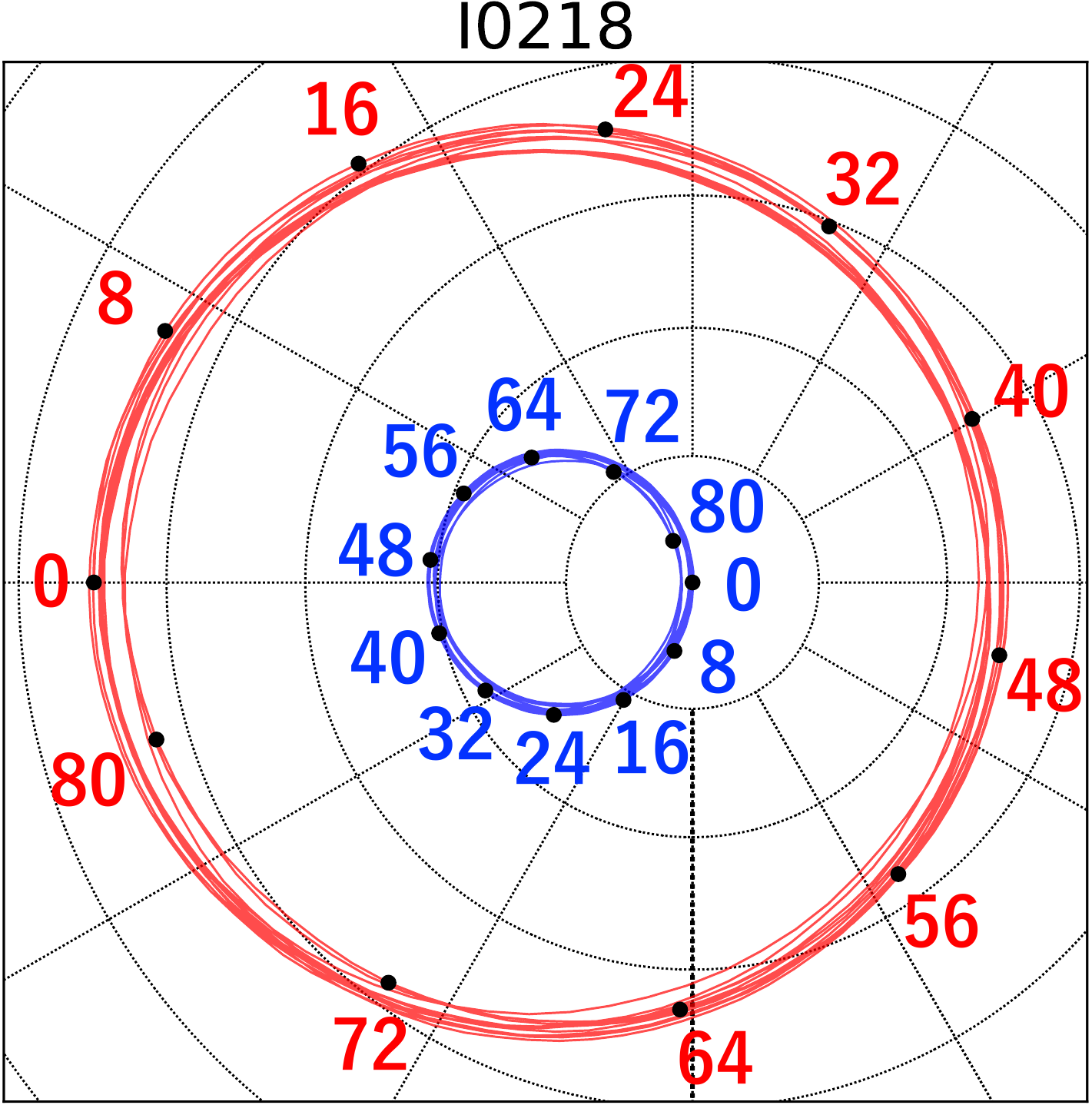} \hspace{14pt}
\includegraphics[clip,width=8.0cm]{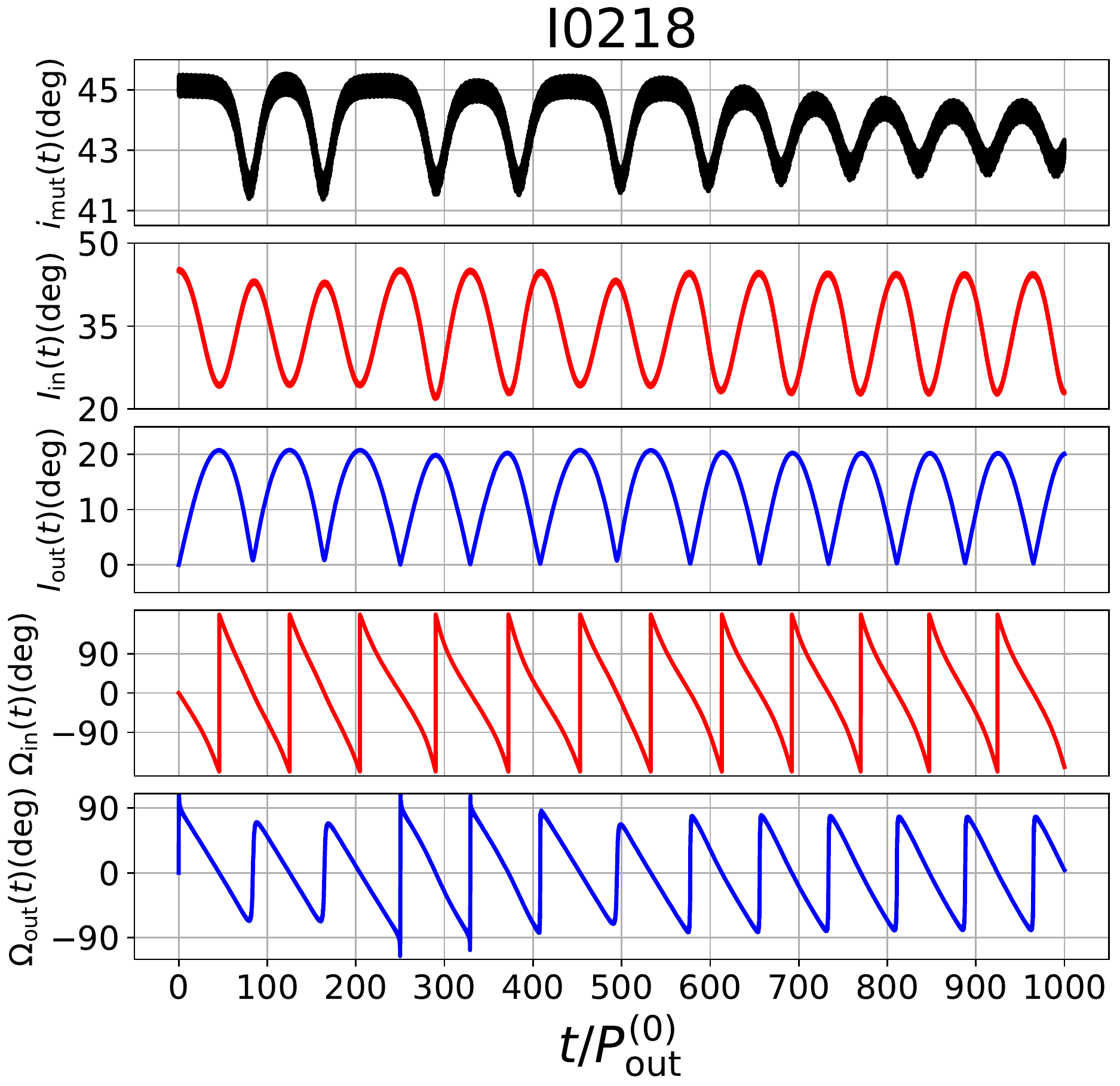}
\includegraphics[clip,width=7.0cm]{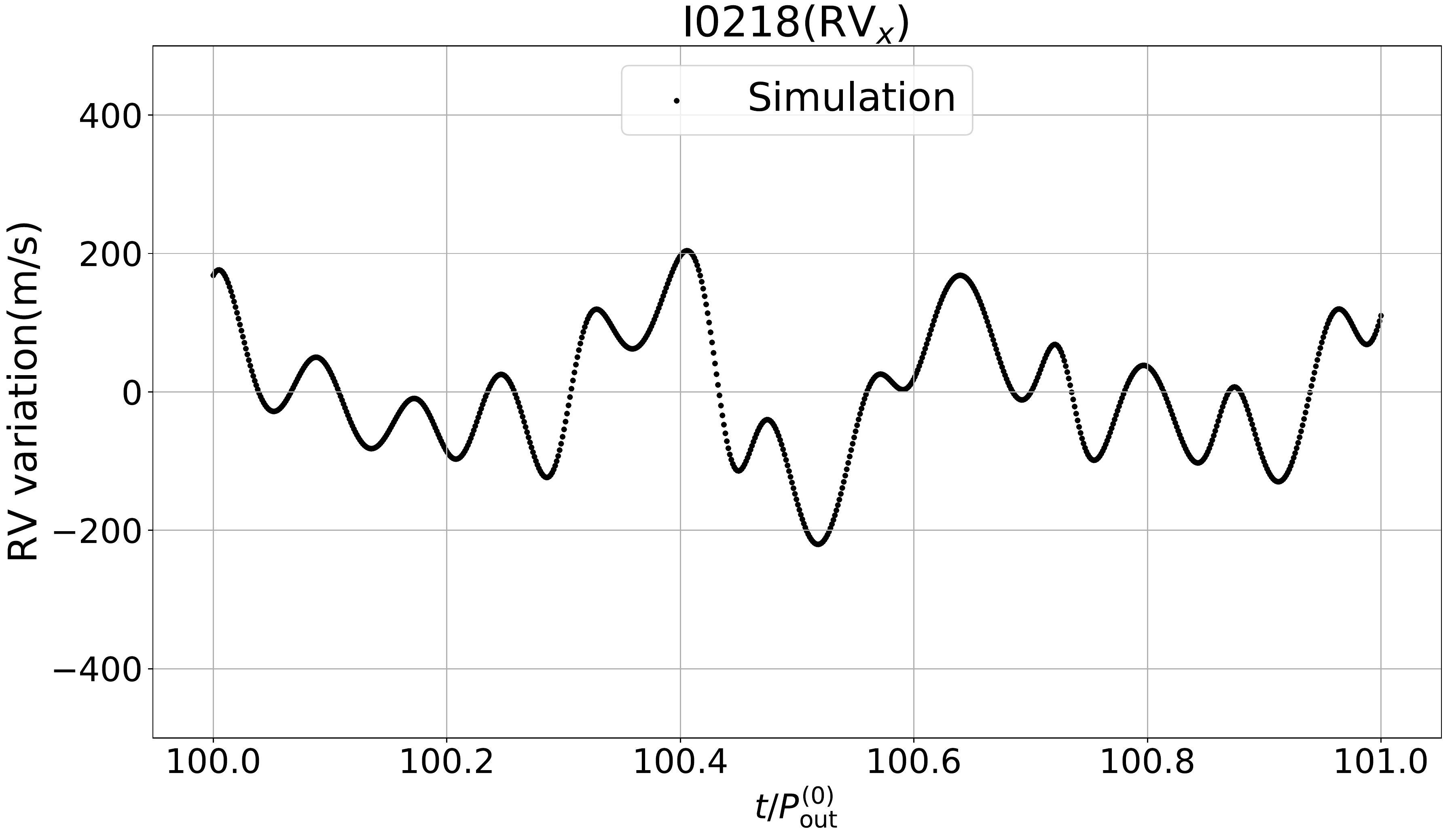}
\includegraphics[clip,width=7.0cm]{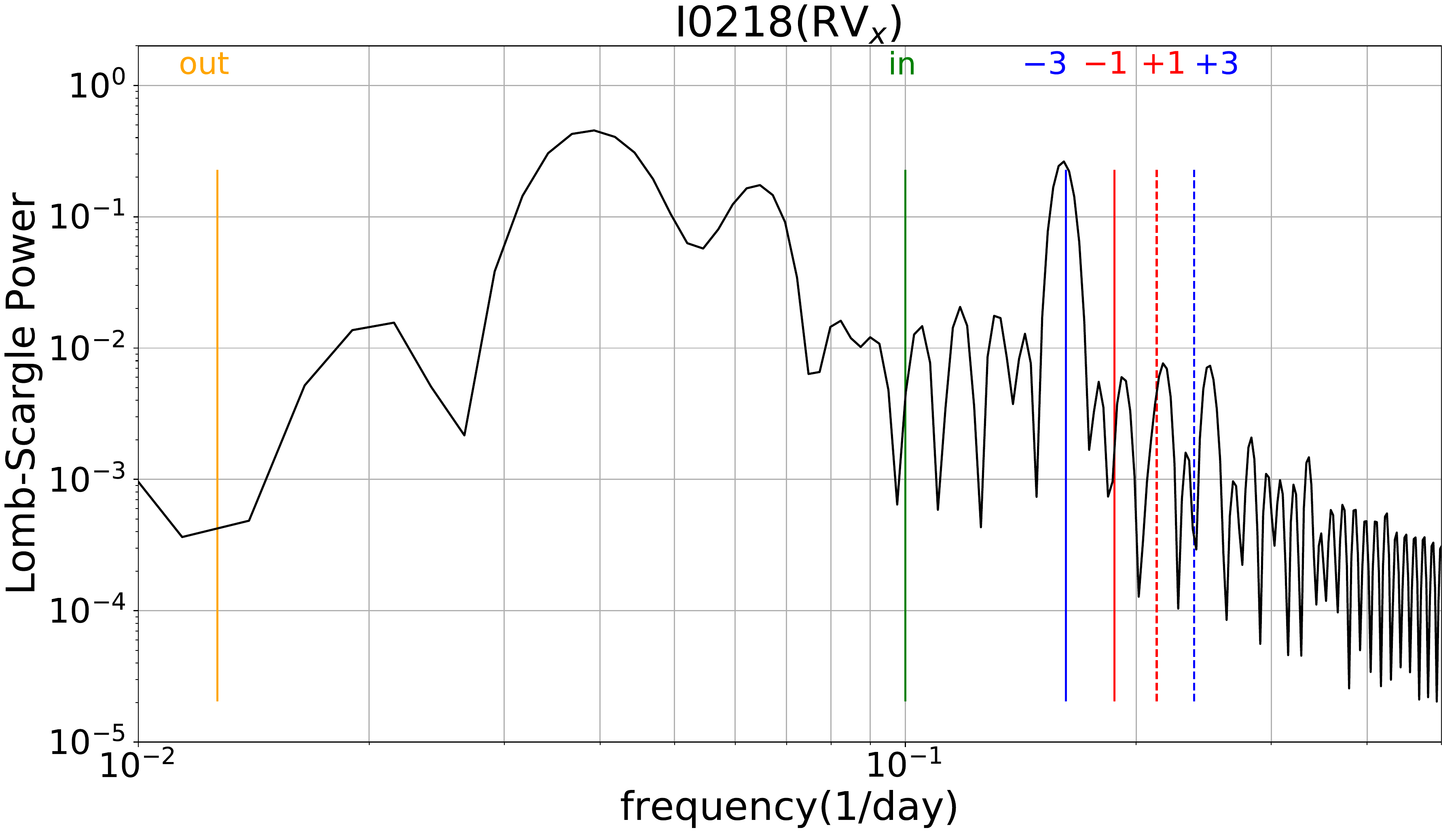}
\includegraphics[clip,width=7.0cm]{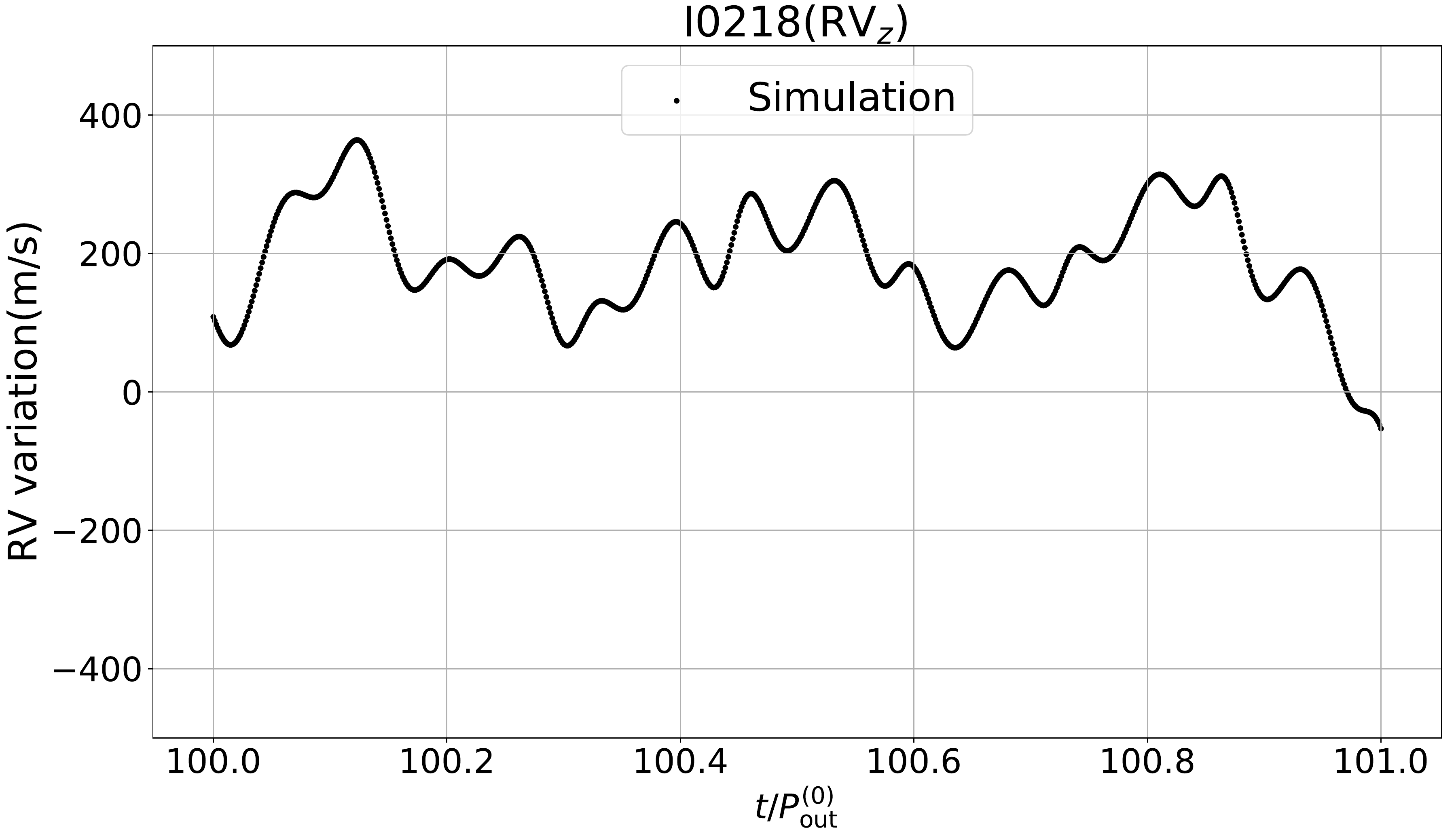}
\includegraphics[clip,width=7.0cm]{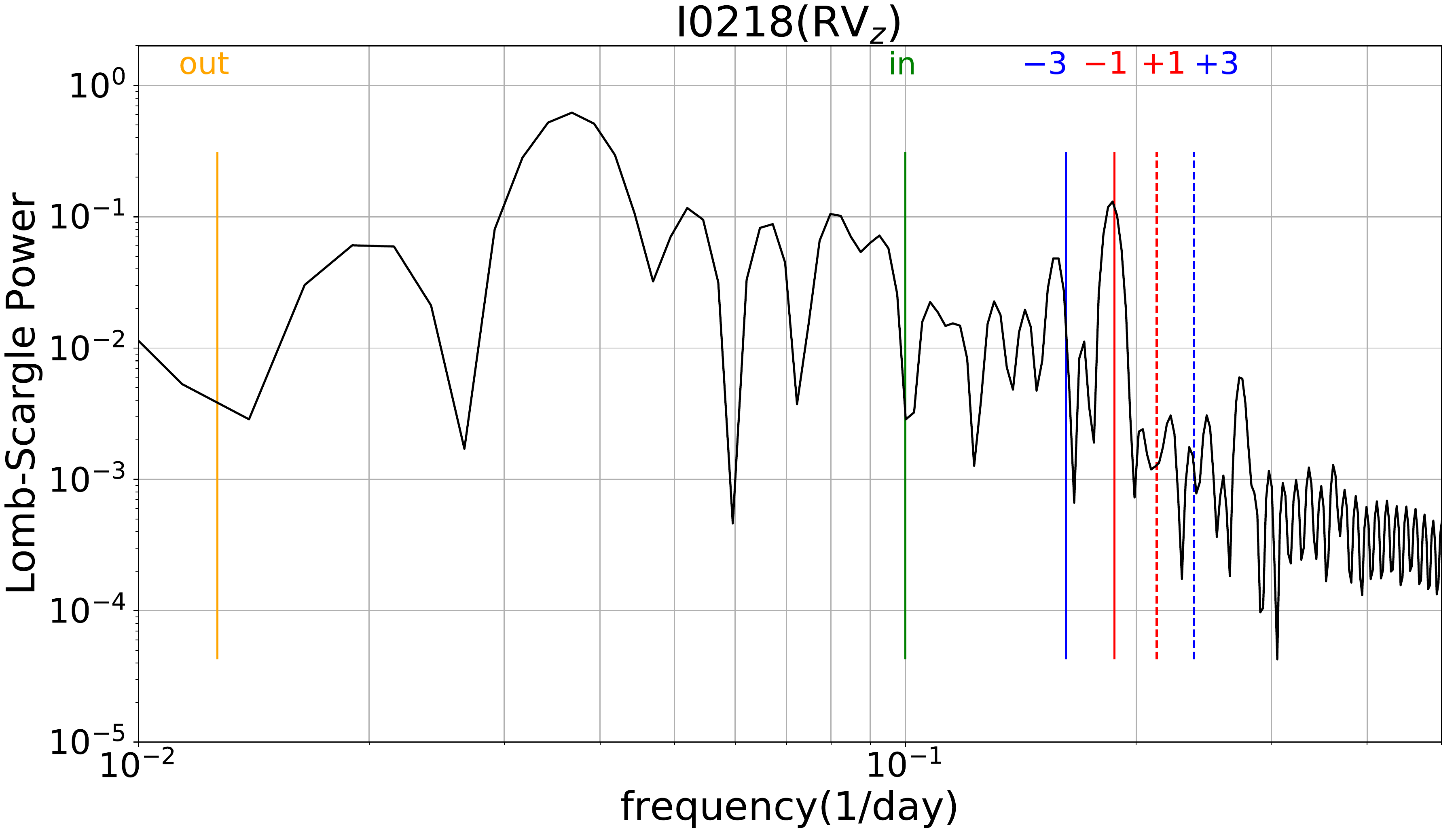}
\includegraphics[clip,width=7.0cm]{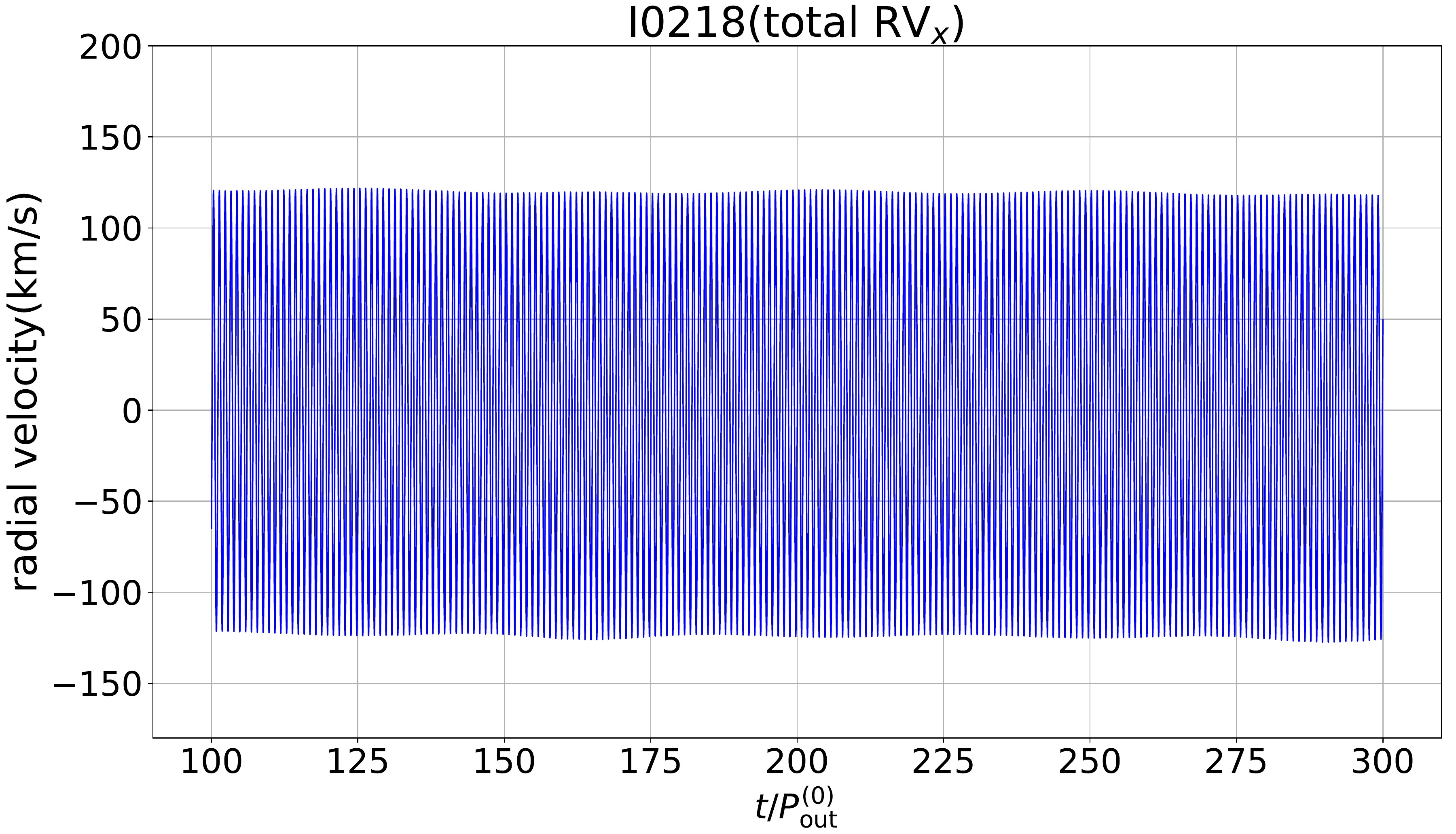}
\includegraphics[clip,width=7.0cm]{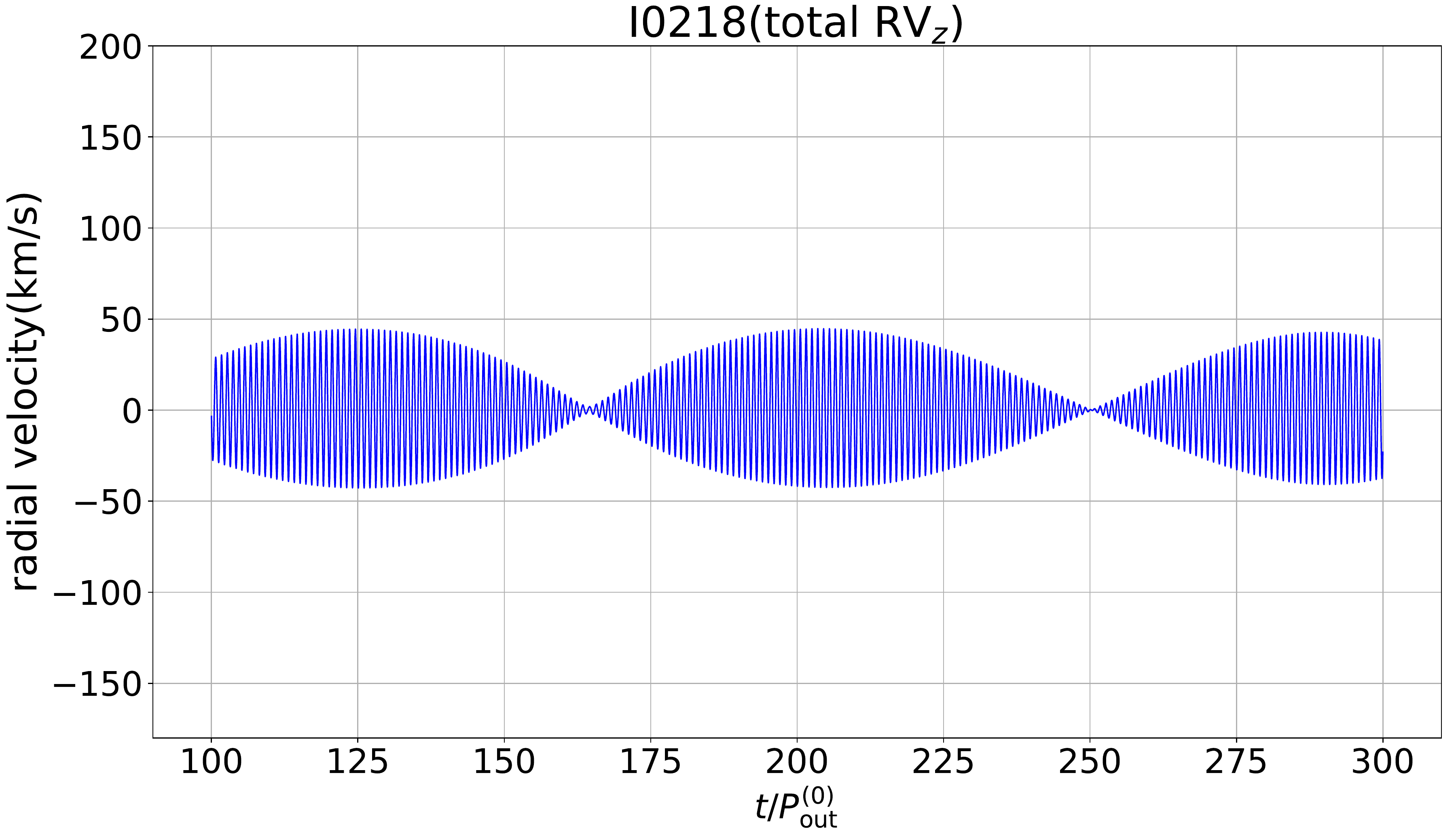}
\end{center}
\caption{Same as Figure \ref{fig:I1010} but for
  I0218. \label{fig:I0218}}
\end{figure*}

\begin{figure*}
\begin{center}
\includegraphics[clip,width=7.5cm]{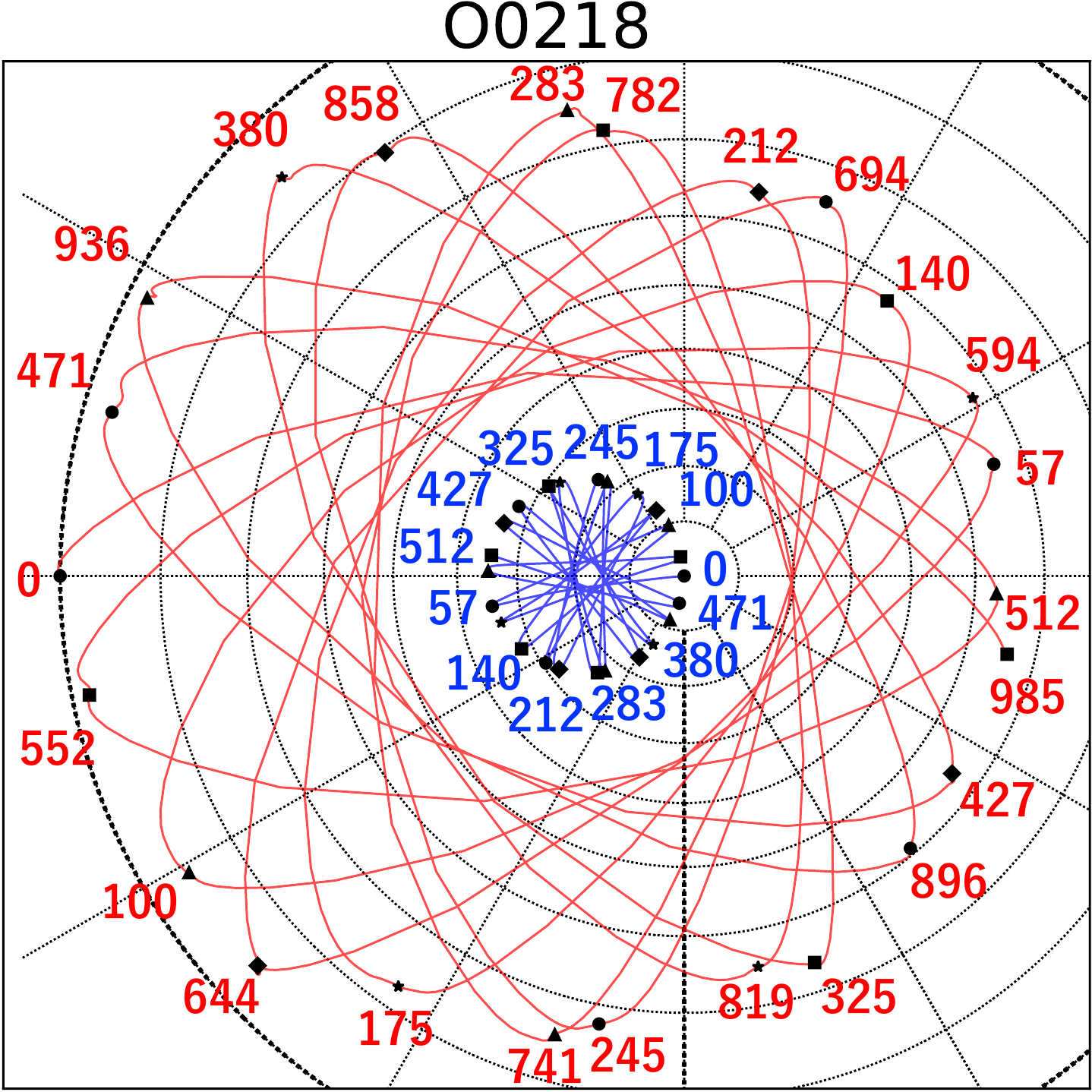} \hspace{14pt}
\includegraphics[clip,width=8.0cm]{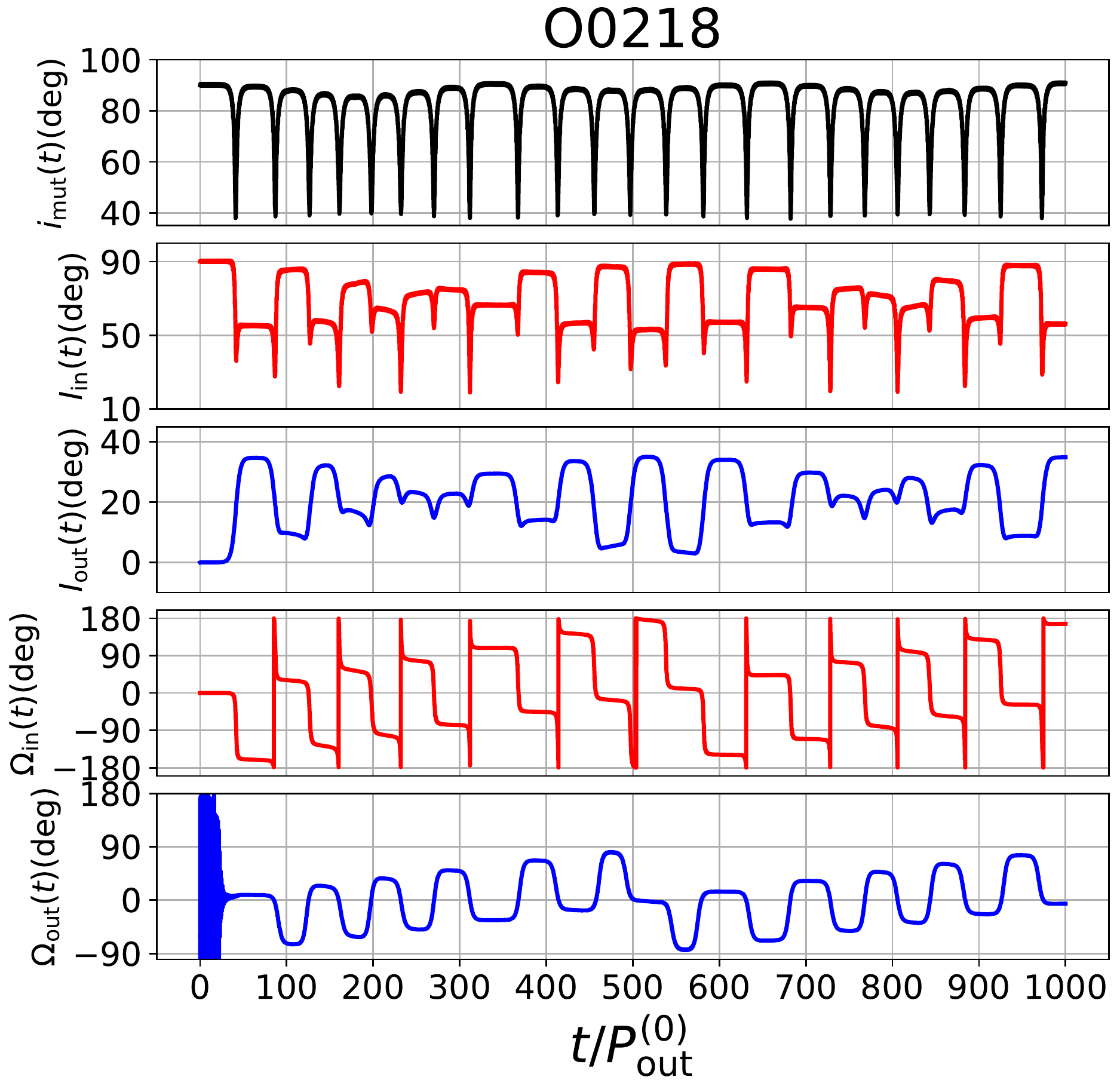}
\includegraphics[clip,width=7.0cm]{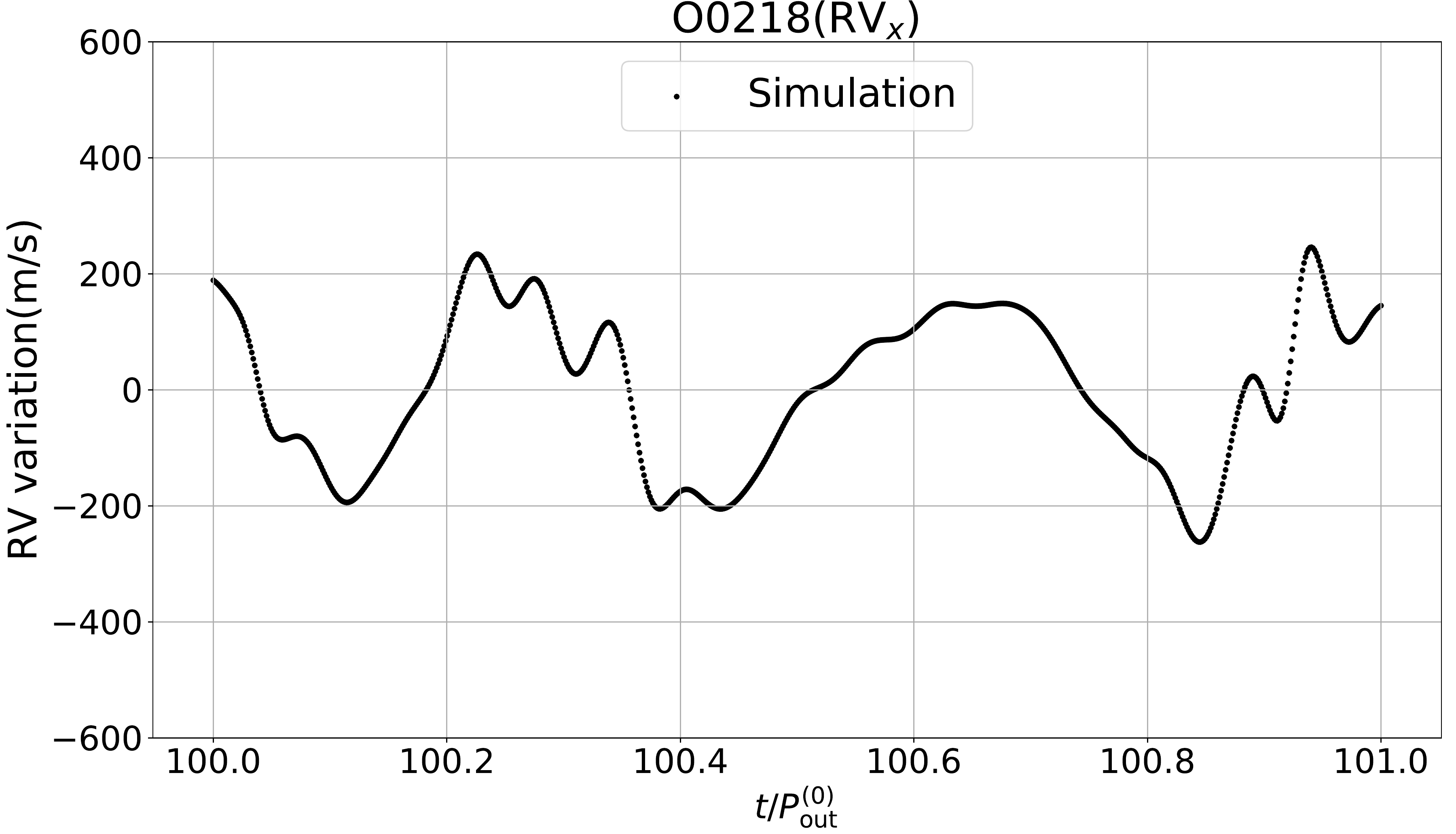}
\includegraphics[clip,width=7.0cm]{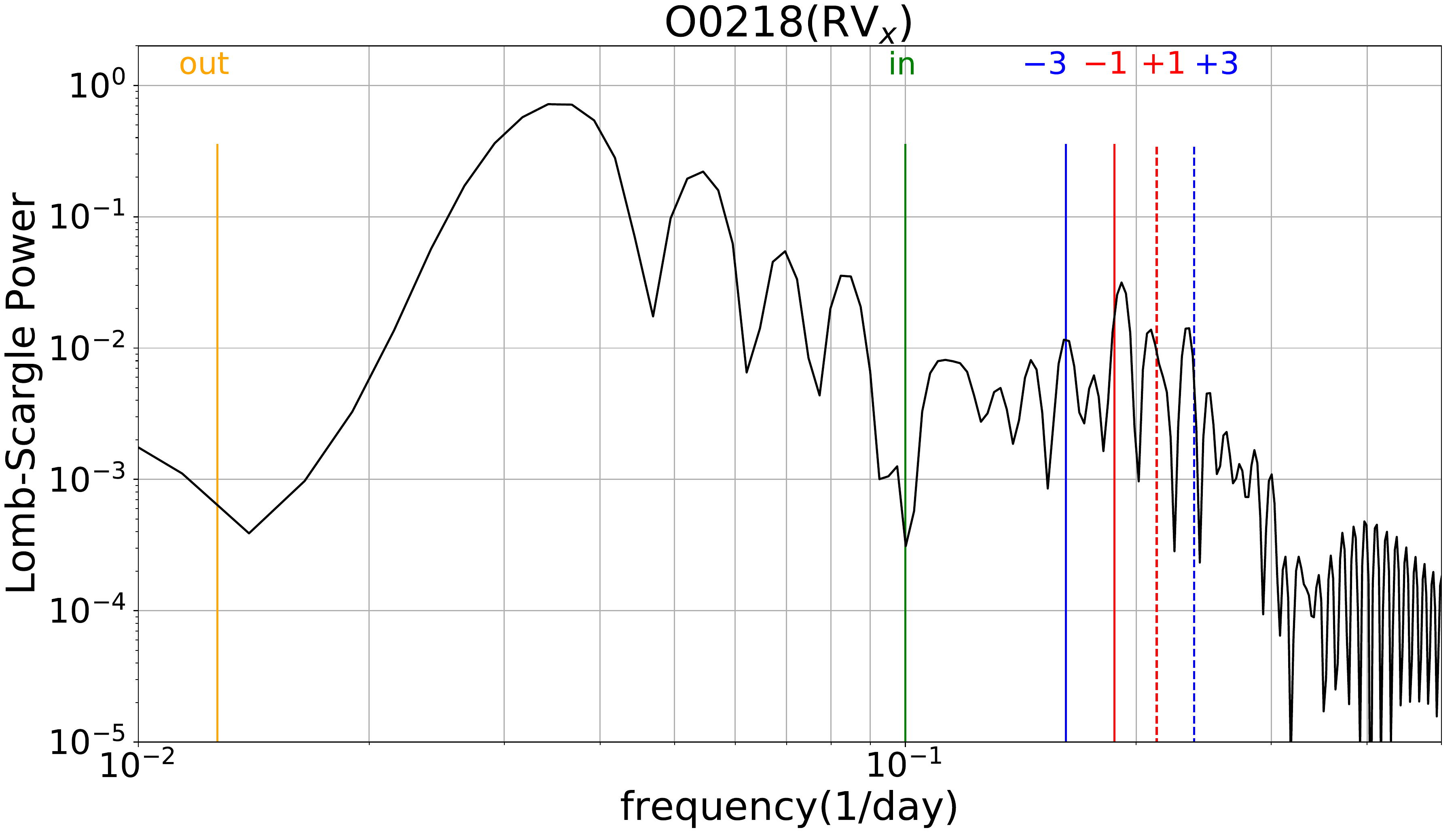}
\includegraphics[clip,width=7.0cm]{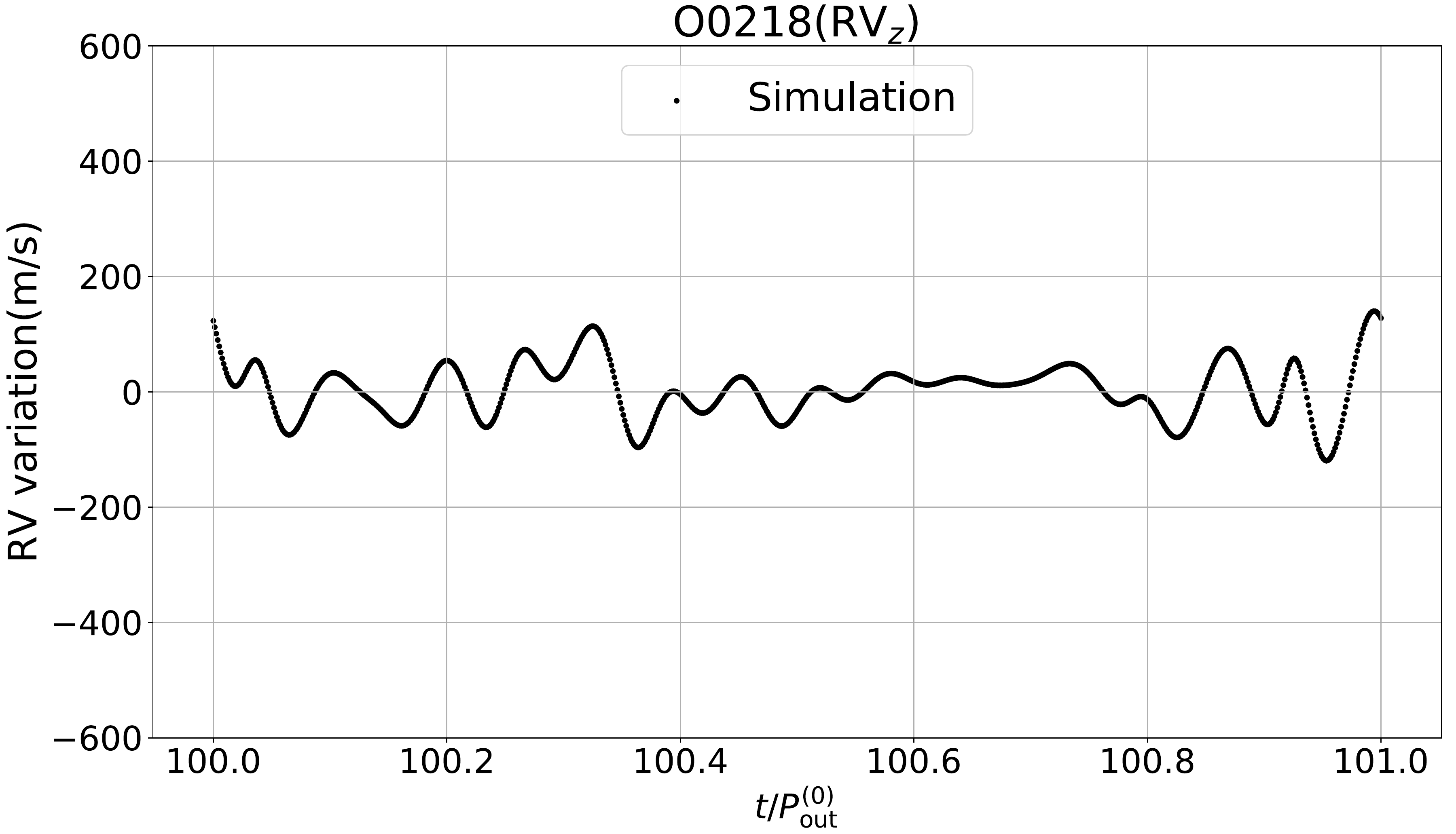}
\includegraphics[clip,width=7.0cm]{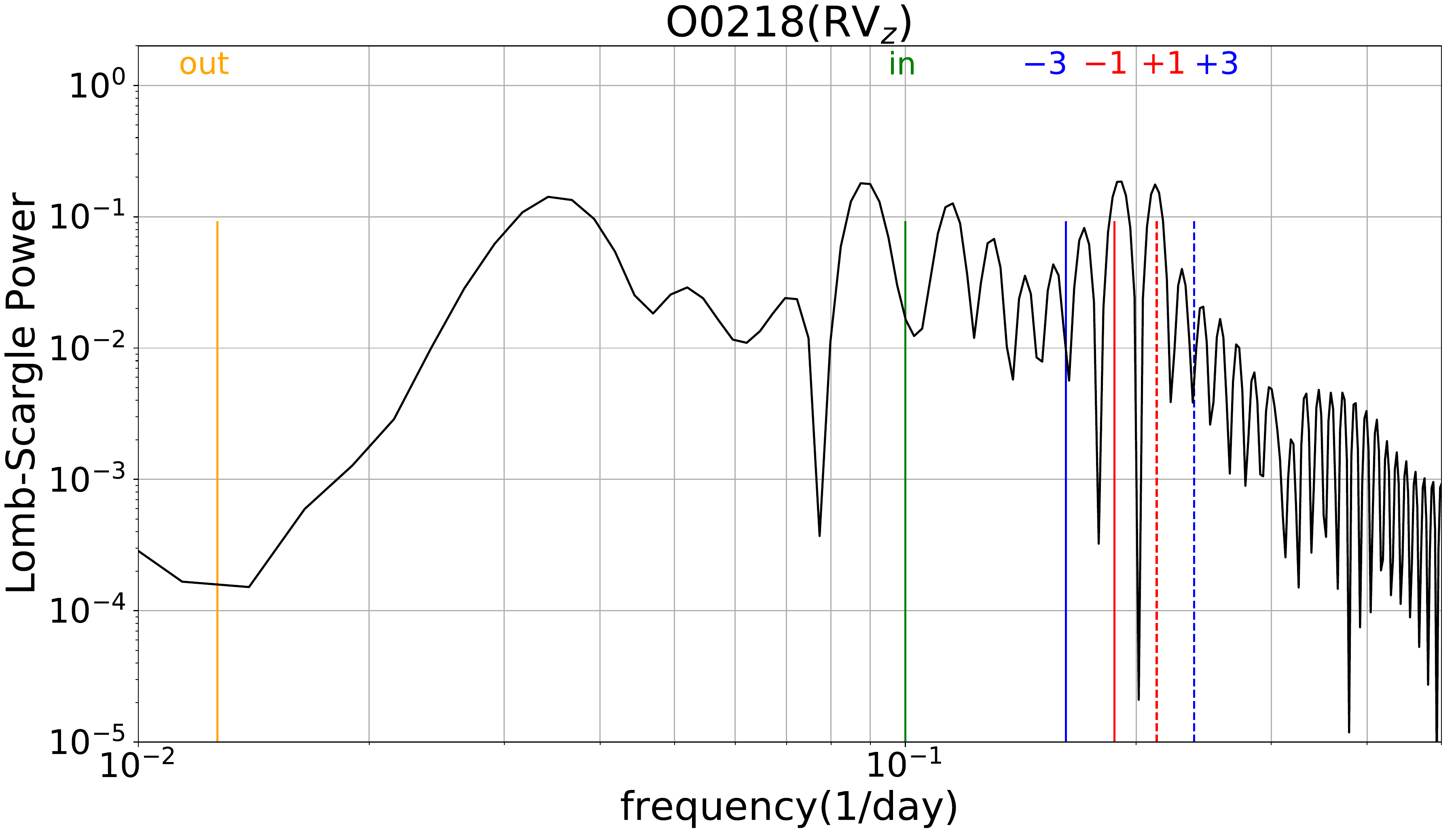}
\includegraphics[clip,width=7.0cm]{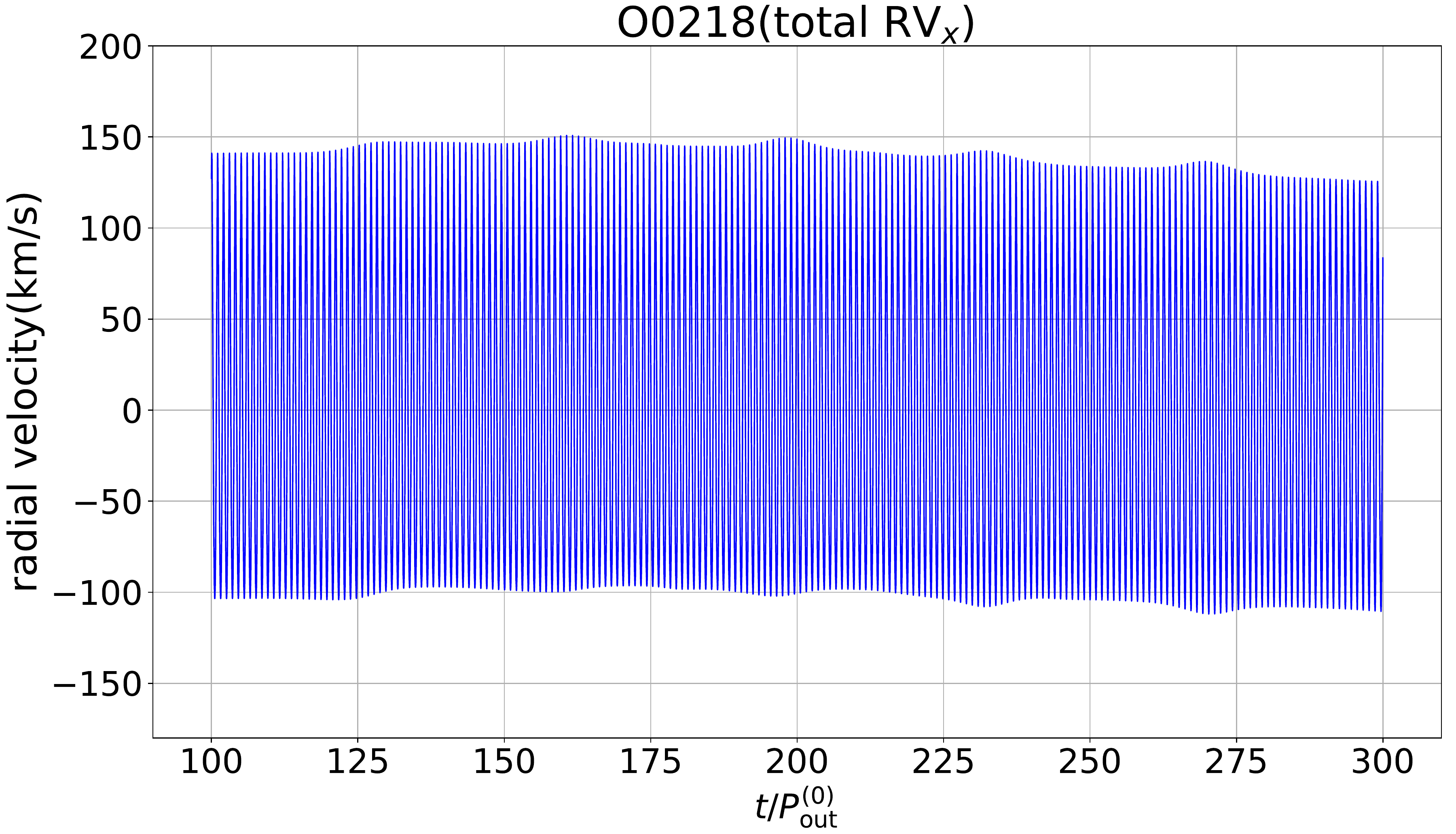}
\includegraphics[clip,width=7.0cm]{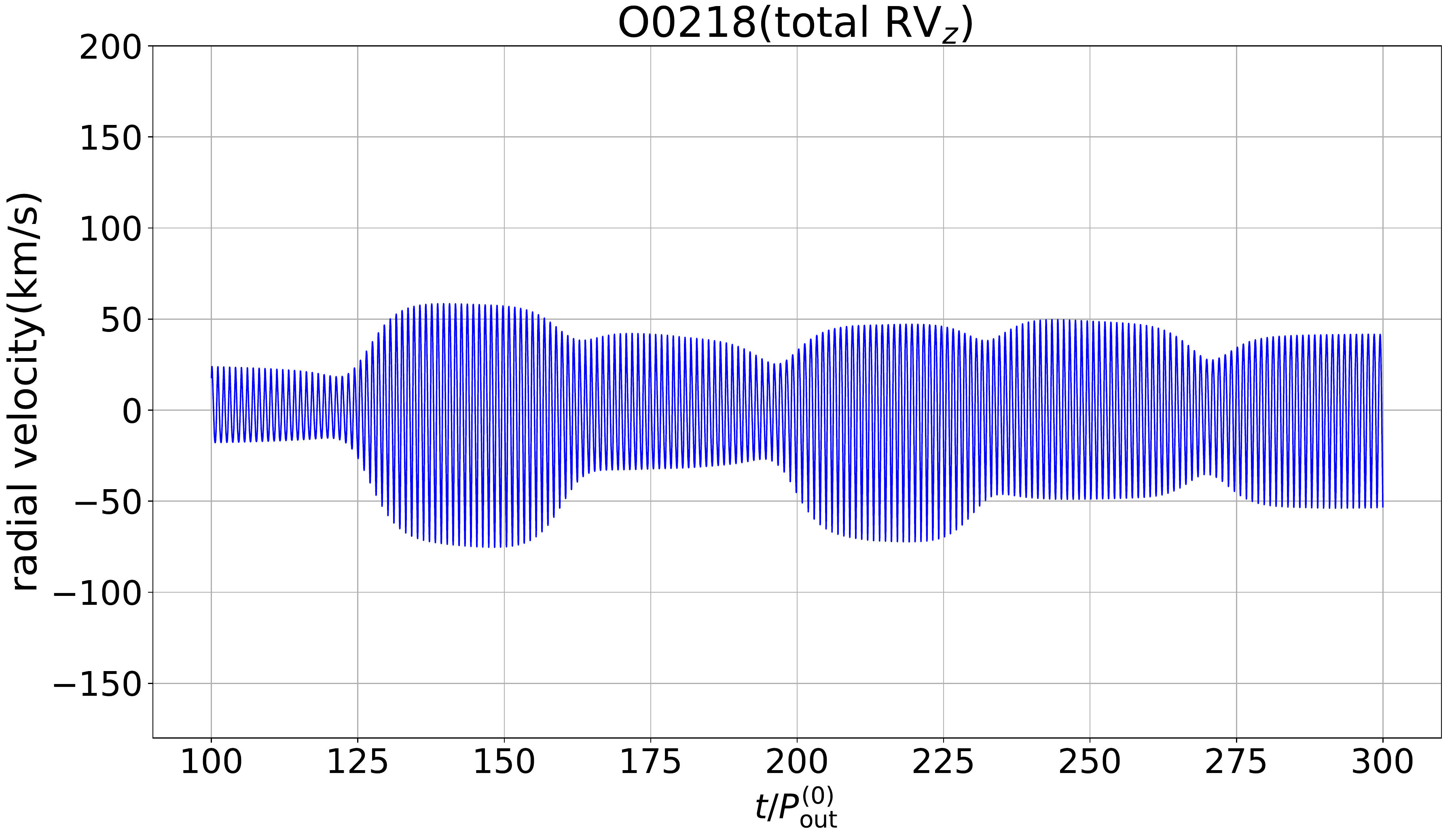}
\end{center}
\caption{Same as Figure \ref{fig:I1010} but for O0218.\label{fig:O0218}}
\end{figure*}

The results are plotted in Figure \ref{fig:I1010}, \ref{fig:O1010},
\ref{fig:I0218}, and \ref{fig:O0218} for I1010, O1010, I0218, and
O0218, respectively. Each figure has eight panels; the top left panels
display the trajectory of the direction of the angular momentum of the
inner (red) and outer (blue) orbits. The numbers indicate
$t/P_\oo^{(0)}$. The top right panels show the corresponding evolution of
the mutual inclination ($i_{\rm mut}$), orbital inclinations ($I_\ii$
and $I_\oo$), and longitudes of the ascending nodes ($\Omega_\ii$ and
$\Omega_\oo$). We plot the orbital parameters every one day output interval.
The middle panels plot the RV variations and the
corresponding LS periodograms viewed from the $x$ and $z$-axes of the
reference frame (Figure \ref{fig:orbits}). The bottom panels plot the
total RV curves, instead of the residual RV variations, viewed from
the $x$ and $z$-axes.

Consider first I1010, which has the mutual inclination of $i_{\rm
  mut}=45^\circ$ initially. As shown in the top panels of Figure
\ref{fig:I1010}, the inner and outer orbits precess around the total
angular momentum axis of the entire system in a periodic fashion. As described in Appendix A, this corresponds to the precession of the
inner and outer orbits around the total angular momentum axis of the
triple system. The period of $\approx 65 P^{(0)}_\oo$ is indeed well
explained by the approximate formula in equation (\ref{eq:POmega}).
This roughly corresponds to the Kozai-Lidov oscillation timescale
$T_{\rm KL}$\citep{Kozai1962,Lidov1962} (see Appendix A for details).
\clearpage

\clearpage

In this case, the evolution of the mutual inclination is fairly moderate,
except for the precession, and the inner and outer orbits remain to be
near-prograde with each other.  Therefore, although the additional periodic terms are present in the RV variations, both the time series
and the LS periodograms show clear modulations due to the inner binary
at frequencies of $\nu_{-3}$ and $\nu_{-1}$. Those trends should be
generic for moderately inclined systems with $i_{\rm mut}<45^\circ$.

An interesting observable feature is the significant modulation of the
Keplerian RV component over a timescale of $T_{\rm KL}$, induced by
the periodic variation of $I_\oo$. The bottom panels of Figure
\ref{fig:I1010} show the RV amplitude modulation from nearly edge-on
($x$-axis) and face-on ($z$-axis) lines of sight.
Since the time dependence of the total RV semi-amplitude
is approximately given as
\begin{eqnarray}
  K_\oo(t) \approx K_0\sin(I_\oo(t))
  \approx K_0\sin(I_\oo(t_0)) + K_0 \dot{I}_\oo(t_0)
  \cos(I_\oo(t_0)) (t-t_0) +\mathcal{O}((t-t_0)^2), 
\end{eqnarray}
large variations are expected especially for a nearly face-on
(i.e. $I_\oo(t_0)\approx0^\circ$) case.

If the mutual inclination of the two orbits is much larger, the
orbital orientations exhibit substantial dynamical evolution.  This is
illustrated in Figure \ref{fig:O1010} for the initially orthogonal
orbits.  In this case, the Kozai-Lidov oscillation
\citep[][]{Kozai1962,Lidov1962} plays an important role in the
evolution of inclinations themselves. Since the precession period
given by equation (\ref{eq:POmega}) is inversely proportional to $\cos
i_{\mathrm{mut}}$, the system stays for a long time at highly inclined
states of $i_{\mathrm{mut}}\approx \pi/2$.  Gradually then, the inner
eccentricity is extremely enhanced by the KL oscillation, and 
$i_{\mathrm{mut}}$ decreases rapidly. This behavior explains the
drastic modulation on the semi-amplitude of RV as shown in the bottom
panels of Figure \ref{fig:O1010}.

Figure \ref{fig:I1010} implies that the directions of the
  angular momentum vectors in the initially inclined orbits (I1010)
  evolve in a fairly periodic and regular fashion.  This is in marked
  contrast to the case of the initially orthogonal orbits; see the
  top left panel of Figure \ref{fig:O1010}.  The trajectories of the
  orientations of the inner and outer orbits for O1010 in the top left
  panel of Figure \ref{fig:O1010} seem to evolve in an irregular
  fashion.  They first stay at the initial location represented by the
  filled circles labeled with $0$ (blue and red for outer and inner
  orbits, respectively) until $t\approx 30P^{(0)}_\oo$. Then, they
  move along the trajectories rapidly and reach the next temporary
  stationary location at $t=55P^{(0)}_\oo$ as the top right panel
  indicates.  Then the orientations of the angular momenta stay in the
  same location until $t\approx105P^{(0)}_\oo$, and reach the next
  location at $t=128P^{(0)}_\oo$. This evolution pattern continues,
  while their mutual inclination $i_{\rm mut}$ oscillates between
  $40^\circ$ and $90^\circ$ in a regular and periodic fashion.

Independent of such complicated behavior of orbital angles, the RV
variations of frequencies $\nu_{\pm 3}$ and $\nu_{\pm 1}$ can be used
as a signature of inner binaries as indicated by the middle panels of
Figure \ref{fig:O1010}. This implies that we can use the same strategy
to detect an inner binary as well, even for a noncoplanar system.

Just for completeness, Figures \ref{fig:I0218} and \ref{fig:O0218}
show the results for noncoplanar and very unequal mass cases: I0218 and
O0218. The resulting figures support that the overall behavior is very
similar to equal-mass cases, except for higher-order effects, which may
come from the octupole disturbing function.  Since the angular
momentum of the inner binary is smaller than that in equal-mass cases,
the total angular momentum is dominated by that of the outer
orbit. Therefore, the outer orbital inclination is more stable. The RV
variations and LS periodograms confirm again that the basic strategy
for detecting an inner binary is valid also for unequal-mass
and noncoplanar triple systems.

Even a nondetection of such long-term RV variations induced
  by the precession or the KL oscillations can put constraints on the
  presence of the inner binary.  \citet{Liu2019}, for example,
  have observed the LB-1 system for 7 months ($\sim 3P^{(0)}_\oo$)
  over 1.5 years ($\sim 7P^{(0)}_\oo$), and found no systematic
  variation in the semi-amplitude of the total RV curve more than $\sim 1$ km/s. If a similar level of upper limits on the RV modulation is
  placed on a true star-BH binary, we can exclude the presence of an
  inner binary with moderate inclinations such as I1010 and I0218.
Therefore, an inner binary, if exists, should have either
near-coplanar (no appreciable precession) or very inclined (long
precession timescale) orbits.  For the latter case, the drastic change
in semi-amplitude of the RV might be observed after a characteristic
timescale of the KL oscillation $T_{\mathrm{KL}}$ (see Figures
\ref{fig:O1010} and \ref{fig:O0218}). This methodology is indeed
  successful at putting a constraint on the lower limit of mutual
  inclination for a stellar triple HD109648 from the detection of
  long-term RV variations: $5.4^\circ \leq i_\mathrm{mut}$ (prograde
  case) and $i_\mathrm{mut} \leq 174.6^\circ$ (retrograde case)
  \citep[][]{Jha2000}. Although the LB-1 system is most
  likely a stellar binary, there may be yet undetected similar star-BH
  systems for which the present methodology is applicable.  If an
  outer star of such systems has a relatively short orbital period, the
  longer-term monitoring of the total RV amplitude may reveal a
  possible noncoplanar inner binary.

\citet{Blaes2002,Liu2017,Liu2018,Thompson2011}, among others, have
suggested that the Kozai-Lidov oscillation acting on an inner BBH may
significantly accelerate the BBH merging timescale.  The detection of
noncoplanar triples containing a BBH, thus, would provide very
interesting opportunities to understand the formation pathway for the
population of BBHs that have been continuously detected with
gravitational wave signals.

\section{Discussion \label{sec:discussion}}


It is known that the GR precession of an inner binary
  suppresses the Kozai-Lidov (KL) oscillation effectively when its
  precession rate $\dot{\omega}_\mathrm{GR}$ exceeds the KL precession
  rate $\dot{\omega}_\mathrm{K}$. Their ratio is given by
\begin{eqnarray}
\label{eq:precession-ratio}
  \frac{\dot{\omega}_\mathrm{GR}}{\dot{\omega}_\mathrm{K}}=
\frac{3(1-e_\oo^2)^{3/2}}{\sqrt{1-e_\ii^2}}
\left(\frac{v_\ii}{c}\right)^2
\left(\frac{P_\oo}{P_\ii}\right)^2\frac{m_1+m_2+m_*}{m_*},
\end{eqnarray}
where $v_\ii \equiv \sqrt{\mathcal{G}(m_1+m_2)/a_\ii}$ corresponds to
the orbital velocity of the inner binary \citep[e.g.][]{Liu2015}.

\begin{figure*}
\begin{center}
\includegraphics[clip,width=8.0cm]{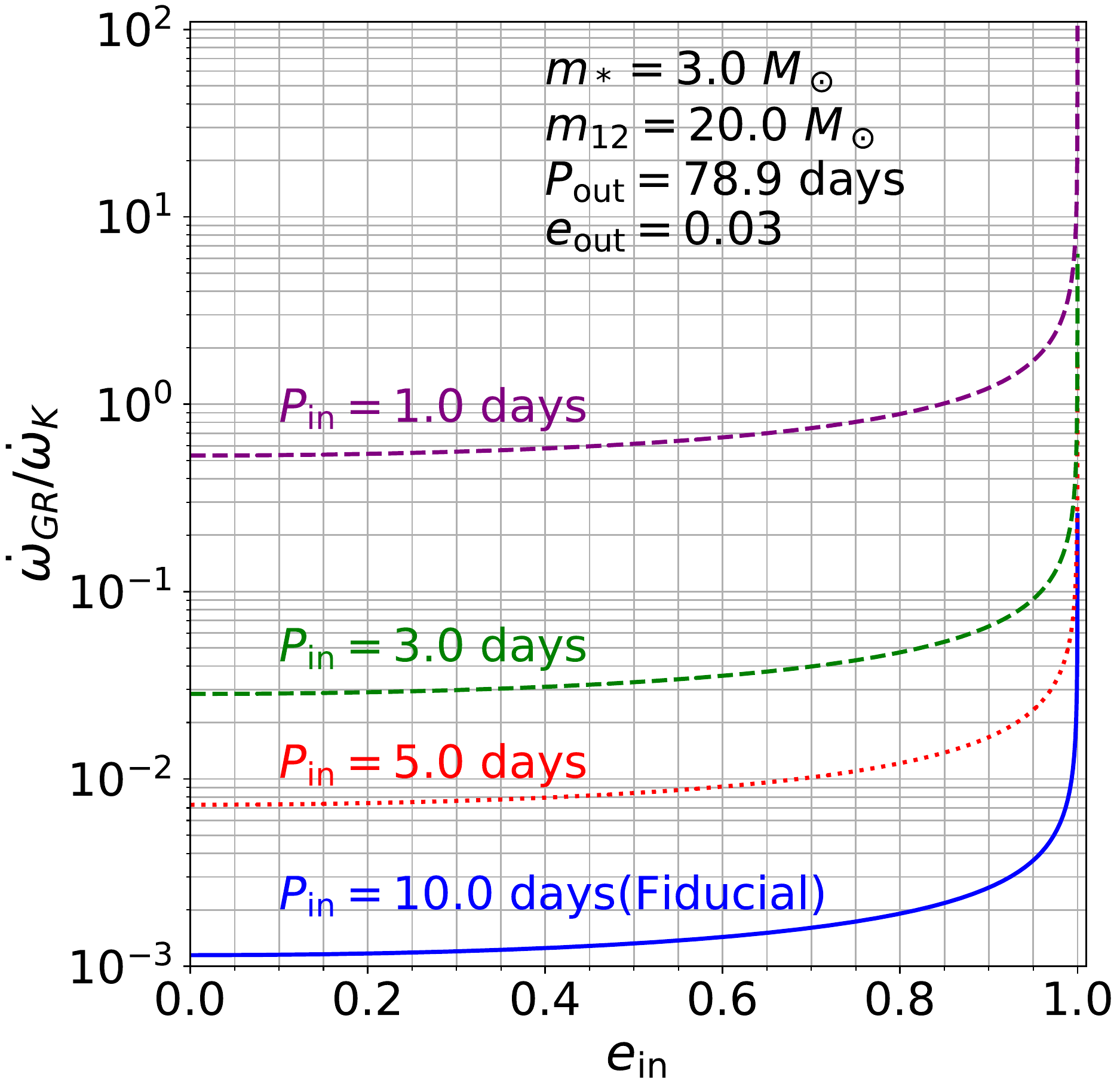}
\hspace{2.0pt}
\includegraphics[clip,width=8.0cm]{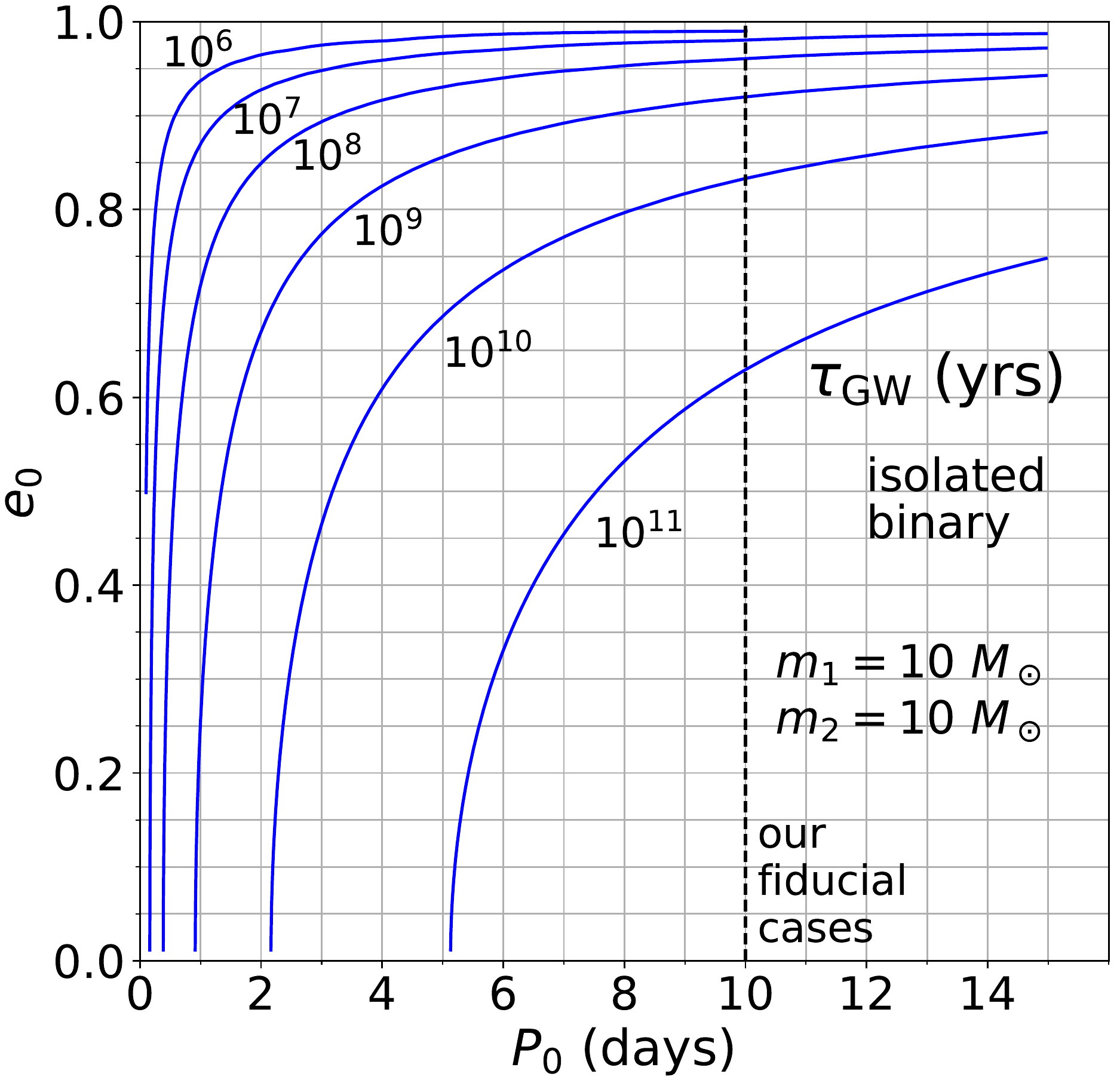}
\end{center}
\caption{Characteristic timescales of the GR corrections.
    {\it Left:} the ratio of GR and KL precession rates
    $\dot{\omega}_\mathrm{GR}/\dot{\omega}_\mathrm{K}$ of the inner
    pericenter arguments $\omega$ against the eccentricity of the
    inner binary $e_\ii$ for $P_\ii=1$, 3, 5 and 10 days. {\it Right:}
    contour plots of the merger time $\tau_\mathrm{GW}$ due to the
    gravitational wave emission on the $P_0-e_0$ plane
    for an isolated binary (neglecting the effect of the tertiary
    star). 
    \label{fig:timescale}}
\end{figure*}

The left panel of Figure \ref{fig:timescale} shows the
  precession ratio, equation (\ref{eq:precession-ratio}), against
  $e_\ii$ for $P_\ii=1$, 3, 5 and 10 days, where we adopt the fiducial
  values for the other parameters.  The plot indicates that the GR
  precession effect is safely neglected unless the inner binary is
  highly eccentric or has a very short orbital period.  Moreover,
  we performed simulations for noncoplanar models in Table
  \ref{tab:tab1} using {\tt REBOUNDx} with GR corrections,
  and made sure that the maximum inner eccentricity changes less than
  $3$ \% over $1000~P^{(0)}_\oo$ for both O1010 and O0218.
  Thus we conclude that our results based on purely Newtonian gravity
  are not affected by the GR precession.  The GR effect, however,
  might change the evolution of the triple over a much longer
  timescale, including the the secular stability of the system. This
  is an interesting problem on its own, but beyond the scope of this
  paper. We plan to study this problem in due course using the secular
  perturbation theory, instead of the direct N-body approach adopted
  here. 

 The gravitational wave (GW) emission may also affect the
  long-term stability of the system. The GW induced merger timescale
  for an eccentric isolated binary is analytically given by
  \citep[][]{Peters1964}
\begin{eqnarray}
\label{eq:tau-GW}
  \tau_\mathrm{GW} = \frac{12}{19}
    \frac{c_0^4}{\beta}\int^{e_0}_{0}{\frac{de~e^{\frac{29}{19}}
        \left(1+\frac{121}{304}e^2\right)^{\frac{1181}{2299}}}
{(1-e^2)^{3/2}}},
\end{eqnarray}
where
\begin{eqnarray}
  c_0 \equiv \frac{(1-e_0^2)}{e_0^{12/19}}\left(\frac{\mathcal{G}(m_1+m_2)P_0^2}{4\pi^2}\right)^{1/3}
  \left(1+\frac{121}{304}e_0^2\right)^{-\frac{870}{2299}}
  ~~~\mathrm{and}~~~
  \beta \equiv \frac{64}{5}\frac{\mathrm{G}^3m_1m_2(m_1+m_2)}{c^5}
\end{eqnarray}
with $P_0$ and $e_0$ being the initial orbital period and eccentricity,
respectively.

The right panel of Figure \ref{fig:timescale} plots the
  contour of $\tau_\mathrm{GW}$ as a function of the eccentricity and
  orbital period at the initial epoch, $e_0$ and $P_0$. Again, the GW
  emission is largely negligible unless the binary is highly eccentric or has a very short orbital period,
  which is not the case for our models considered here. This estimate,
  however, neglects the dynamical effect by the tertiary object on the
  inner binary, and may vary in a case where the KL oscillation excites
  significantly the eccentricity of the inner binary.

It is also possible that the orbit of the outer star is
  affected by the inverse-KL and other eccentricity-inclination
  resonances\citep[e.g.][]{deElia2019,Naoz2017,Naoz2020,Vinson2018},
  which may enhance the outer eccentricity effectively, depending on
  the initial conditions.  For example, \citet{Vinson2018} showed that
  the eccentricity of an outer test particle can be enhanced up to
  $\sim 0.2$ and $\sim 0.3$ by the inverse KL resonance and
  octupole resonance on $\varpi_\oo$$+\varpi_\ii$$-2\Omega_\oo$,
  respectively. They also pointed out that the outer eccentricity
  enhancement becomes stronger as the inner eccentricity increases,
  due to the octupole apse-aligned resonance.  The outer eccentricity
  enhancement may lead to the orbital crossing and the system may
  become unstable eventually.

For our fiducial cases, however, the amplitudes of the inner
  and outer orbital angular momenta are comparable. Thus the test
  particle approximation for the tertiary star is not valid. In this
  case, the total angular momentum conservation may prohibit the
  significant enhancement of the outer eccentricity.  Nevertheless,
  the inverse KL effect and other resonances may play an important
  role in the orbital evolution and secular stability of triples in
  certain sets of orbital parameters of the triples.

The long-term dynamical effects mentioned in the above (GW
  emission, normal and inverse KL oscillations, and other resonances)
  can also be examined by the secular perturbation analysis that we
  plan in future. Thus we do not discuss those further in the present
  paper, and we hope to report the detailed results elsewhere.

Finally, we briefly mention possible formation mechanisms of
  tight triples including compact binaries of our interest.  In
  reality, however, reliable formation scenarios are very difficult to
  construct, while several authors proposed possible evolution channels for triple systems \citep[e.g.][]{Toonen2016}.  Since the
  common-envelope evolution of binary stars significantly shrinks
  their orbital separations, tight binaries may be produced
  \citep[e.g.][]{Toonen2016,Zorotovic2010} if they survive the
  subsequent violent processes like supernovae. Many complex processes
  including, for instance, mass transfer in eccentric orbits
  \citep[e.g.][]{Dosopoulou2016}, mass-loss induced eccentric KL
  oscillation\citep[e.g.][]{Shappee2013,Michaely2014}, and supernova
  kicks \citep[e.g.][]{Pijloo2012,Toonen2016} have been discussed, and
  are recognized to play important roles in triple formation and
  evolution. While those formation scenarios of tight triples
  are still qualitative, they offer an interesting possibility that
  can be further examined with more quantitative long-term simulations.

In addition, the dynamical capture scenario of BBH formation
  \citep[e.g.][]{Zwart2000,OLeary2009,Rodriguez2016,Tanikawa2020,DiCarlo2020}
  predicts the formation of temporal triples \citep[see
    e.g.][]{Rodriguez2016}.  Thus, the dynamical capture would also
  provide a possible formation channel for triple systems including
  BBHs although it is currently not certain if such triples could
  survive after escaping from the star-dense region.

\section{Conclusion} \label{sec:summary}

It is expected that there are many star-BH binaries including
  unseen companions in our Galaxy
  \citep[e.g.][]{Breivik2017,Kawanaka2016,Mashian2017,Yamaguchi2018,Masuda2019}.
A fraction of them may be a triple system comprising an inner binary
black hole. Given the fact that LIGO has already detected many BBHs,
it is important to search for detached BBHs hidden in such systems
apart from the gravitational wave.

In the previous paper \citep[][]{Hayashi2020}, we proposed a novel
methodology for detecting an inner BBH in a triple system through the
periodic RV variations of the outer star, and presented the
observational feasibility using mock observations focused on coplanar
triple systems.  This paper has extended the study of
\citet{Hayashi2020} and considered more generic cases of coplanar
eccentric and noncoplanar inner binaries. In this paper, we
  adopt parameters of fiducial star-BBH triple systems largely
  inspired by LB-1 originally estimated by
  \citet[][]{Liu2019,El-Badry2020,Abdul-Masih2020}, and compute the
  expected RV variations using N-body simulations.  Although it is now
  unlikely that LB-1 is a star-BH binary \citep[][]{Shenar2020}, the
  results in this paper capture the basic behavior of such triples
  that remain to be detected in the near future.

Our main conclusions are summarized below:

(1) Coplanar inner binaries of $P_\ii > 2$ weeks in our fiducial
  triples with a few month outer orbital period are ruled out by the
dynamical stability condition. Conversely, an inner BBH of $P_\ii
\approx 1$ week should add an RV variation on the order of $100$ m/s
at roughly twice the orbital frequency of the inner binary.

(2) If a quasi-Kepler RV component on the order of $100$ km/s at the
outer orbital frequency is properly removed from the entire RV curve,
the residual RV variation can provide direct signatures of, or useful
constraints on, the presence of the inner binary.

(3) For coplanar triple systems, the shift of the outer pericenter
argument $\omega_\oo$ can be used to detect the inner binary. This is
basically the same idea as a hypothetical planet Vulcan inside
Mercury's orbit \citep[][]{LeVerrier1859}.

(4) For noncoplanar triple systems, the total RV semi-amplitude is
modulated periodically by the precession of the inner and outer
orbits over roughly the Kozai-Lidov oscillation timescale.  The
amplitude of the secular modulation depends on the observer's
line of sight, but can be on the order of $100$ km/s in principle.
The modulation timescale is usually long, but if the outer orbital
period is relatively short, to the order of months, it is quite
feasible to detect over 10 years for instance.  In addition, the RV
variation at roughly twice the orbital frequency of the inner binary
may be searched via short-cadence monitoring of the system, although
the behavior of the short-term RV variation becomes complicated for
noncoplanar triples. Incidentally, we note here that
  \citet{Jha2000} detected the RV semi-amplitude variation for a tight
  stellar triple HD109648, due to its nodal precession over their
  $8$ yr RV observation. Thus the similar detection for star-BBH
  triples should be quite feasible.

As we have stressed before, our proposed strategy to search for an
inner binary in a triple system is quite generic.  Even though we
adopt fiducial parameters of triple systems in this paper
according to the previous interpretation for the LB-1 system
  by \citet{Liu2019}, our methodology can be readily applied to
numerous star-black hole systems that are expected to be discovered in
the near future. Currently, there are many proposals to search for
star-black hole binaries with Gaia
\citep[e.g.][]{Breivik2017,Kawanaka2016,Mashian2017,Yamaguchi2018,Shikauchi2020_2}
and TESS \citep[][]{Masuda2019}. For instance, \citet{Yamaguchi2018}
predict that hundreds of such binaries will be discovered with Gaia in
its 5 year observation. \citet{Masuda2019} point out that dozens of
star- black hole systems would be detected through the detailed
analyses of TESS light curves.  Therefore, the detected number of
star-black hole systems could increase significantly in the near
future. Any other future observational missions should also contribute
much to such discoveries.

The dynamics of triple systems that we described here can be applied
to various other methods for detecting interesting astronomical systems.
For instance, probing the dynamics of binary pulsars in triple systems
using the pericenter shift \citep[e.g.][]{Suzuki2019}, the RV
variation of a star passing close to unseen companions, and the search for
binary planets \citep[e.g.][]{Lewis2015,Ochiai2014} in known
exoplanetary systems.

Finally, we would like to emphasize that the strategy proposed here is
no longer just a theoretical idea, but becomes an observationally feasible
methodology for searching for otherwise unseen astrophysical objects. In
the near future, this methodology is expected to help in detecting
not-yet-known populations of astronomical objects.

\acknowledgments

We thank an anonymous referee for several useful comments.
Simulations and analyses in this paper made use of {\tt REBOUND}, {\tt
  RadVel}, and {\tt Astropy}.  We gratefully acknowledge the support
from Grants-in Aid for Scientific Research by the Japan Society for
Promotion of Science (JSPS) No.18H01247 and No.19H01947, and from JSPS
Core-to-core Program ``International Network of Planetary Sciences''.

\vspace{5mm}

\software{{\tt Astropy} \citep[][]{astropy2013,astropy2018}, 
          {\tt RadVel} \citep{Fulton2018},
          {\tt REBOUND} \citep{Rein2012}
}


\appendix
\section{The secular Lagrange planetary equations for a triple system
  consisting of an inner binary and a tertiary
  star \label{sec:appendix}}

The noncoplanar results shown in subsection \ref{subsec:noncoplanar}
exhibit a precession-like behavior. We discuss the secular evolution of
orbital angles in noncoplanar triples from the Lagrange planetary
equations.

The orbit-averaged quadrupole Hamiltonian $\bar{F}$ is given by
\citep[e.g.,][]{Morais2012}:
\begin{eqnarray}
\label{eq:F}
  \bar{F} &=& C_{\mathrm{quad}}[2-12e_\ii^2-6(1-e_\ii^2)
    (\sin{I_\ii}\sin{I_\oo}\cos(\Delta\Omega)+\cos{I_\ii}\cos{I_\oo})^2+30e_\ii^2
    \cr
    && \times(-\sin{I_\oo}\cos{I_\ii}\sin{\omega_\ii}\cos(\Delta\Omega)
    -\sin{I_\oo}\cos{\omega_\ii}\sin(\Delta\Omega)
    +\sin{I_\ii}\sin{\omega_\ii}\cos{I_\oo})^2],
\end{eqnarray}
where 
\begin{eqnarray}
\label{eq:C}
  C_{\mathrm{quad}} &\equiv& \frac{\mathcal{G}}{16}
  \frac{m_1m_2}{m_1+m_2}\frac{m_*}{(1-e_\oo^2)^{3/2}}
  \left(\frac{a_\ii^2}{a_\oo^3}\right), \\
\label{eq:DeltaOmega}
\Delta\Omega &\equiv& \Omega_\ii-\Omega_\oo.
\end{eqnarray}
In equation (\ref{eq:C}) and throughout this appendix, we denote
Newton's gravitational constant by $\mathcal{G}$, since $G$ indicates
a canonical variable corresponding to an orbital angular momentum.

With the orbit-averaged Hamiltonian $\bar{F}$,
the secular evolution of orbital angles is explicitly written
as \citep[e.g.][]{Danby1988,Murray2000,Valtonen2006} 
\begin{eqnarray}
\label{eq:omega}
  \dot{\omega_j} &=&
  -\frac{\sqrt{1-e_j^2}}{\mu_j\nu_j a_j^2e_j}
  \frac{\partial \bar{F}}{\partial e_j}
  +\frac{\cos{I_j}}{\mu_j\nu_j a_j^2\sqrt{1-e_j^2}\sin{I_j}}
  \frac{\partial \bar{F}}{\partial I_j}, \\
\label{eq:Omega}
  \dot{\Omega_j} &=& -\frac{1}{\mu_j\nu_j a_j^2\sqrt{1-e_j^2}\sin{I_j}}
  \frac{\partial \bar{F}}{\partial I_j}, \\
\label{eq:I}
  \dot{I_j} &=& \frac{1}{\mu_j\nu_j a_j^2\sqrt{1-e_j^2}\sin{I_j}}
  \frac{\partial \bar{F}}{\partial \Omega_j}
  -\frac{\cos{I_j}}{\mu_j\nu_j a_j^2\sqrt{1-e_j^2}\sin{I_j}}
  \frac{\partial \bar{F}}{\partial \omega_j},
\end{eqnarray}
where $j (=\ii$ and $\oo)$. We define the corresponding reduced mass
as
\begin{eqnarray}
  \mu_\ii &\equiv& \frac{m_1m_2}{m_1+m_2} ,\\
  \mu_\oo &\equiv& \frac{m_*(m_1+m_2)}{m_1+m_2+m_*}.
\end{eqnarray}
We note that the Lagrange planetary equations are often written in
terms of the disturbing function $R \equiv -\bar{F}$
\citep[e.g.][]{Murray2000}.

Neglecting the $\mathcal{O}(e_\ii^2)$ terms in equation (\ref{eq:F}),
equations (\ref{eq:omega}), (\ref{eq:Omega}), and (\ref{eq:I})
for $j=\ii, \oo$ are explicitly written as follows:
\begin{eqnarray}
\label{eq:omegain_full}
\dot{\omega}_\ii &=& \frac{12C_{\mathrm{quad}}(1-e_\ii^2)}{G_\ii}
\left[2-\cos^2{i_\mathrm{mut}}
  -\frac{\cos{I_\ii}}{\sin{I_\ii}}\cos{i_\mathrm{mut}}
  (\cos{I_\ii}\sin{I_\oo}\cos{\Delta\Omega}-\sin{I_\ii}\cos{I_\oo}) \right. \cr
 && \left. -5(-\sin{I_\oo}\cos{I_\ii}\sin{\omega_\ii}
  \cos(\Delta\Omega)-\sin{I_\oo}\cos{\omega_\ii}
  \sin(\Delta\Omega)+\sin{I_\ii}\sin{\omega_\ii}\cos{I_\oo})^2 \right], \\
\label{eq:omegaout_full}
\dot{\omega}_\oo &=& \frac{6C_{\mathrm{quad}}}{G_\oo} \left[(3\cos^2{i_\mathrm{mut}}-1)
  -2\frac{\cos{I_\oo}}{\sin{I_\oo}}\cos{i_\mathrm{mut}}
  (\sin{I_\ii}\cos{I_\oo}\cos{\Delta\Omega}-\cos{I_\ii}\sin{I_\oo})\right], \\
\label{eq:Omegain_full}
\dot{\Omega}_\ii &=& 
\frac{12C_{\mathrm{quad}}}{G_\ii \sin{I_\ii}}\cos{i_\mathrm{mut}}
(\cos{I_\ii}\sin{I_\oo}\cos{\Delta\Omega}-\sin{I_\ii}\cos{I_\oo}), \\
\label{eq:Omegaout_full}
\dot{\Omega}_\oo &=& 
\frac{12C_{\mathrm{quad}}}{G_\oo \sin{I_\oo}}\cos{i_\mathrm{mut}}
(\sin{I_\ii}\cos{I_\oo}\cos{\Delta\Omega}-\cos{I_\ii}\sin{I_\oo}), \\
\label{eq:Iin_full}
\dot{I}_\ii &=& \frac{12C_{\mathrm{quad}}}{G_\ii}\cos{i_\mathrm{mut}}
\sin{I_\oo}\sin{\Delta\Omega}, \\
\label{eq:Iout_full}
\dot{I}_\oo &=& -\frac{12C_{\mathrm{quad}}}{G_\oo}\cos{i_\mathrm{mut}}
\sin{I_\ii}\sin{\Delta\Omega},
\end{eqnarray}
where $G_\ii$ and $G_\oo$ are the angular momenta of the inner and outer
orbits defined as
\begin{eqnarray}
  G_\ii &\equiv& \mu_\ii \nu_\ii a_\ii^2 \sqrt{1-e_\ii^2}, \\
  G_\oo &\equiv& \mu_\oo \nu_\oo a_\oo^2 \sqrt{1-e_\oo^2}.
\end{eqnarray}
Note that we use an arbitrary inertial frame to write down the
  equations, rather than the invariant plane. Equation
(\ref{eq:omegaout_full}) reduces to equation (\ref{eq:omega-dot}) for
coplanar prograde ($\Delta\Omega=0$ and $I_\ii=I_\oo$) and retrograde
($\Delta\Omega=0$ and $I_\ii=\pi+I_\oo$) systems.

Consider first the case of moderate mutual inclination
$i_\mathrm{mut}$ and small inner eccentricity $e_\ii$, in which the
Kozai-Lidov (KL) oscillation is not so effective and $e_\ii$ remains
negligibly small. In this case, the secular evolution is basically
described by the precession of the inner and outer angular momenta
around the total angular momentum axis with $G_\ii$, $G_\oo$, and
$G_{\rm tot}$ being constant, where
\begin{eqnarray}
G_\mathrm{tot} = \sqrt{G_\ii^2+G_\oo^2+2G_\ii G_\oo \cos{i_\mathrm{mut}}}.
\end{eqnarray}
Indeed such motion well explains those of I1010 and I0218, where the
normal directions of orbits move on the circles centered at the total
angular momentum direction.

Thus, its precession timescale can be computed by considering the
motion with respect to the invariant reference frame
($\Delta\Omega=\pi$, $i_\mathrm{mut}=I_\ii+I_\oo$).  Since
$G_\ii/\sin{I_\oo}=$$G_\oo/\sin{I_\ii}=$$G_\mathrm{tot}/\sin{I_\mathrm{mut}}$
holds in this case, equations (\ref{eq:Omegain_full}) and
(\ref{eq:Omegaout_full}) reduce to
\begin{eqnarray}
\dot{\Omega_j} = -\frac{12C_{\mathrm{quad}}G_\mathrm{tot}}{G_\ii G_\oo }\cos{i_\mathrm{mut}}.
\end{eqnarray}
The precession rate above is constant if we neglect the higher-order
variation of mutual inclination, and it is expressed analytically as
\begin{eqnarray}
\label{eq:POmega}
P_\Omega = \frac{2\pi}{\dot{\Omega}}&=&
\frac{\pi G_\ii G_\oo}{6C_{\mathrm{quad}}G_\mathrm{tot}\cos{i_\mathrm{mut}}} .
\end{eqnarray}

If we neglect the $\mathcal{O}(e_\ii^2)$ and $\mathcal{O}(e_\oo^2)$ terms,
equation (\ref{eq:POmega}) is further approximated as
\begin{eqnarray}
\label{eq:POmega2}
\frac{P_\Omega}{P_\oo} \approx
  \frac{80.7}{\cos{i_\mathrm{mut}}}
\left(\frac{m_1+m_2+m_*}{23~M_\odot}\right) 
\left(\frac{m_*}{3~M_\odot}\right)^{-1}
\left(\frac{P_\oo}{78.9~\mathrm{days}}\right)
\left(\frac{P_\ii}{10.0~\mathrm{days}}\right)^{-1}
\end{eqnarray}
for $G_\oo \gg G_\ii$, and
\begin{eqnarray}
\frac{P_\Omega}{P_\oo} \approx
\frac{92.0}{\cos{i_\mathrm{mut}}}
\frac{(m_1+m_2)^2}{4m_1m_2}
\left(\frac{m_1+m_2+m_*}{23~M_\odot}\right)^{\frac{2}{3}}
\left(\frac{m_1+m_2}{20~M_\odot}\right)^{-\frac{2}{3}}
\left(\frac{P_\oo}{78.9~\mathrm{days}}\right)^{\frac{4}{3}}
\left(\frac{P_\ii}{10.0~\mathrm{days}}\right)^{-\frac{4}{3}}
\end{eqnarray}
for $G_\oo \ll G_\ii$. We compute the periods for our four
noncoplanar models (in which $G_\oo \sim G_\ii$) from equation
(\ref{eq:POmega}). The values summarized in Table \ref{tab:tab3} are
in reasonable agreement with the results shown in Figures
\ref{fig:I1010} and \ref{fig:I0218}.
For comparison, we write down the conventional KL timescale for an inner test particle \citep[e.g.][]{Merritt2013}:
\begin{eqnarray}
\frac{T_\mathrm{KL}}{P_\oo} &=& \frac{m_1}{m_*}\left(\frac{P_\oo}{P_\ii}\right)(1-e_\oo^2)^{3/2} \nonumber \\
&& \approx 26 \left(\frac{m_1}{10~M_\odot}\right)
\left(\frac{m_*}{3~M_\odot}\right)^{-1}
\left(\frac{P_\oo}{78.9~\mathrm{days}}\right)
\left(\frac{P_\ii}{10~\mathrm{days}}\right)^{-1} ~ (e_\oo^2\ll1).
\end{eqnarray}
The timescale roughly agrees with equation (\ref{eq:POmega2}) within order estimation.

\begin{deluxetable}{lccccc}
\tablecolumns{6}
\tablewidth{1.0\columnwidth} 
\tablecaption{Parameters relevant to precession timescales for fiducial noncoplanar models} 
\tablehead{ case & $i_\mathrm{mut}$ (deg) & $G_\ii/C_{\mathrm{quad}}P^{(0)}_\oo$
  & $G_\oo/C_{\mathrm{quad}}P^{(0)}_\oo$ & $G_\mathrm{tot}/C_{\mathrm{quad}}P^{(0)}_\oo$ & $P_\Omega/P^{(0)}_\oo$}
\startdata
I1010 & $45$ & $153.8$ & $175.3$ & $304.2$ & $65.6$ \\
O1010 & $90$ & $153.8$ & $175.3$ & $233.2$ & $+\infty$ \\
I0218 & $45$ & $153.8$ & $486.9$ & $605.6$ & $91.6$ \\
O0218 & $90$ & $153.8$ & $486.9$ & $510.7$ & $+\infty$ 
\label{tab:tab3}
\enddata
\end{deluxetable}

While the orbital inclinations $I_\ii$ and $I_\oo$ are constant in the
invariant reference frame, i.e., defined with respect to the total
angular momentum axis, they also exhibit periodic variations due to
the $\Omega$ precessions for an an arbitrary line of sight.  Thus the
period of inclination variations is also given by equation
(\ref{eq:POmega}), which basically explains the behavior of I1010 and
I0218 shown in Figures \ref{fig:I1010} and \ref{fig:I0218}.

Consider next a larger mutual inclination like O1010 and O0218.  In
this case, the KL oscillation is efficient and increases the inner eccentricity significantly and periodically.  Since the
precession period, equation (\ref{eq:POmega}), is inversely
proportional to $\cos i_\mathrm{mut}$, the timescale of the
inclination change is very sensitive to the value of $i_\mathrm{mut}$.
As shown in Figures \ref{fig:O1010} and \ref{fig:O0218}, O1010 and
O0218 spend most of their time around $i_\mathrm{mut} \approx \pi/2$.
Then the KL oscillation gradually enhances the inner
eccentricity, and drastically changes the inclinations.  During such
transient time, $i_\mathrm{mut}$ becomes very small, but rapidly goes
back to $\approx \pi/2$ again.

A more quantitative estimate of the corresponding period is difficult
and generally requires numerical integration of a set of the Lagrange planetary equations including the eccentricity terms,
although several analytical and numerical results have been presented
in previous literature \citep[e.g.][]{Kinoshita1999,Merritt2013,
  Naoz2013,Antognini2015,Will2017,Vinson2018}.

\section{Orbital period of a modulated Keplerian motion \label{sec:appendix2}}
Equation (22) in the main text incorrectly ignored the time dependence of the initial true anomaly $f^{(0)}_\oo(t)$ in the approximation. The correct version of the equation should read
\begin{equation}\tag{22}
\label{eq:Pave}
\frac{P_\oo(t)}{P^{(0)}_\oo}
\approx \frac{2\pi}{\nu_\oo+\dot{\omega}_\oo+\dot{f}^{(0)}_\oo}
\approx 1-\frac{2\dot{\omega}_\oo P_\oo}{2\pi}\approx 0.978,
\end{equation}
where $\dot{\omega}_\oo$ is the precession rate of the outer pericenter argument $\omega_\oo$.
The detailed discussion of the approximation is described in appendix of 
\citet{Hayashi2021}.  As a result, $1-P_\oo(t)/P^{(0)}_\oo$ becomes
twice larger than that predicted in Figure 4 in the main text. Figure 4 should be 
replaced by Figure \ref{fig:b1} after this correction.

\renewcommand{\thefigure}{B1}
\begin{figure*}
\begin{center}
\includegraphics[clip,width=14.0cm]{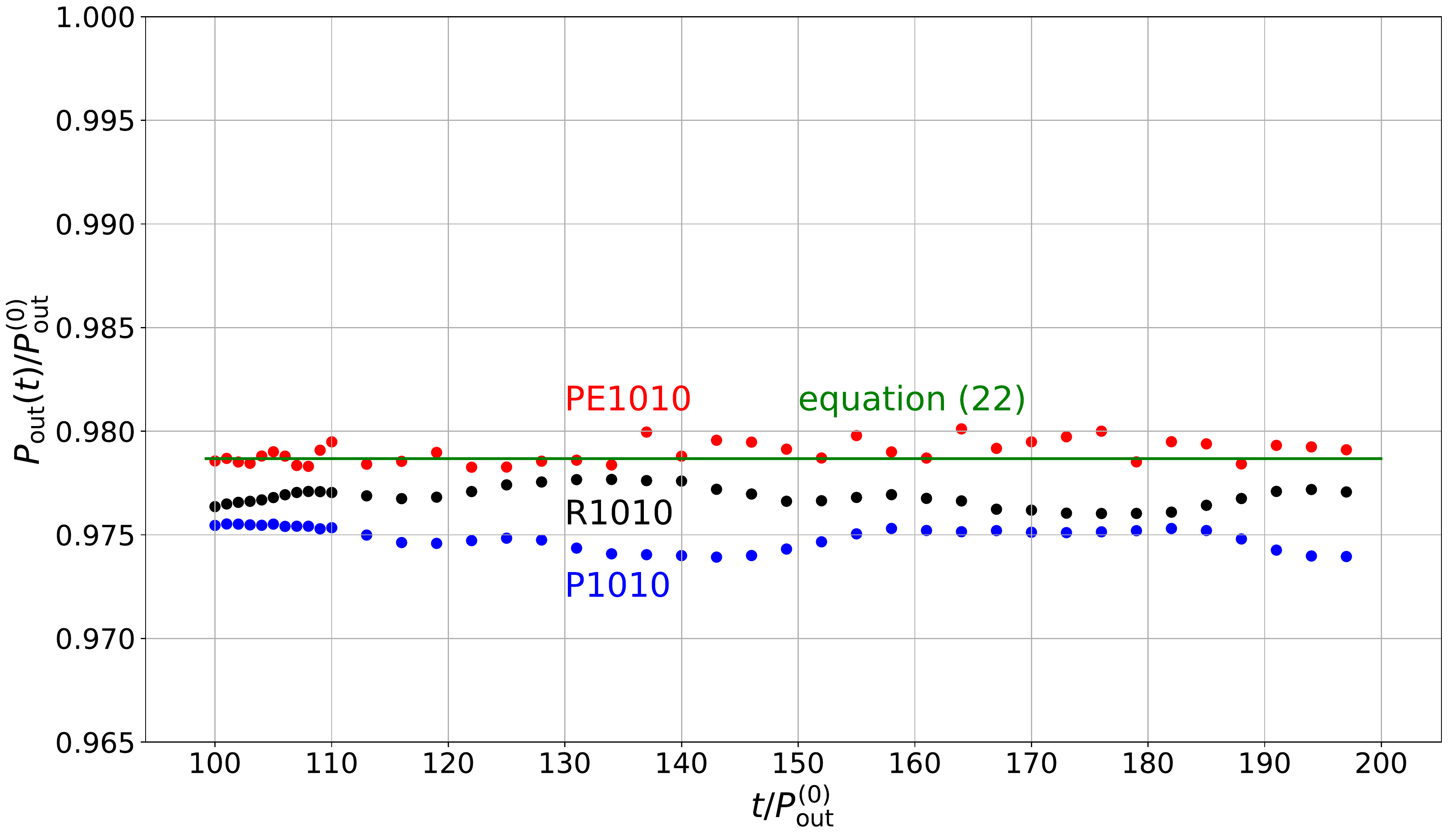}
\end{center}
\caption{Best-fit values of $P_\oo(t_n \equiv nP_\oo^{(0)})$ for
  coplanar systems. They are estimated with {\tt RadVel} using the
  0.1 day cadence simulated RV data over
  $nP_\oo^{(0)}<t<(n+1)P_\oo^{(0)}$ for $100 \leq n < 200$; P1010
  (blue), R1010 (black), and PE1010 (red).  The solid green line
  indicates the analytic prediction (see
  equation (22) in Appendix B). \label{fig:b1}}
\end{figure*}  

\section{The detailed evolution of mutual inclination in Figure 8}
In order to clarify the detailed evolution of mutual inclination in Figure 8, 
we here show the enlarged version of the top right panel of Figure 8. 
Figure \ref{fig:c1} shows enlarged panels of the evolution of mutual inclination for I1010. 
The first, second, third, and forth panels show the evolution over $0 - 1000P_\oo$, $0 - 50P_\oo$, $0 - 10P_\oo$, and $0 - 1.0P_\oo$, respectively. 
\clearpage

\renewcommand{\thefigure}{C1}
  \begin{figure*}
 	\begin{center}
 		\includegraphics[clip,width=12cm]{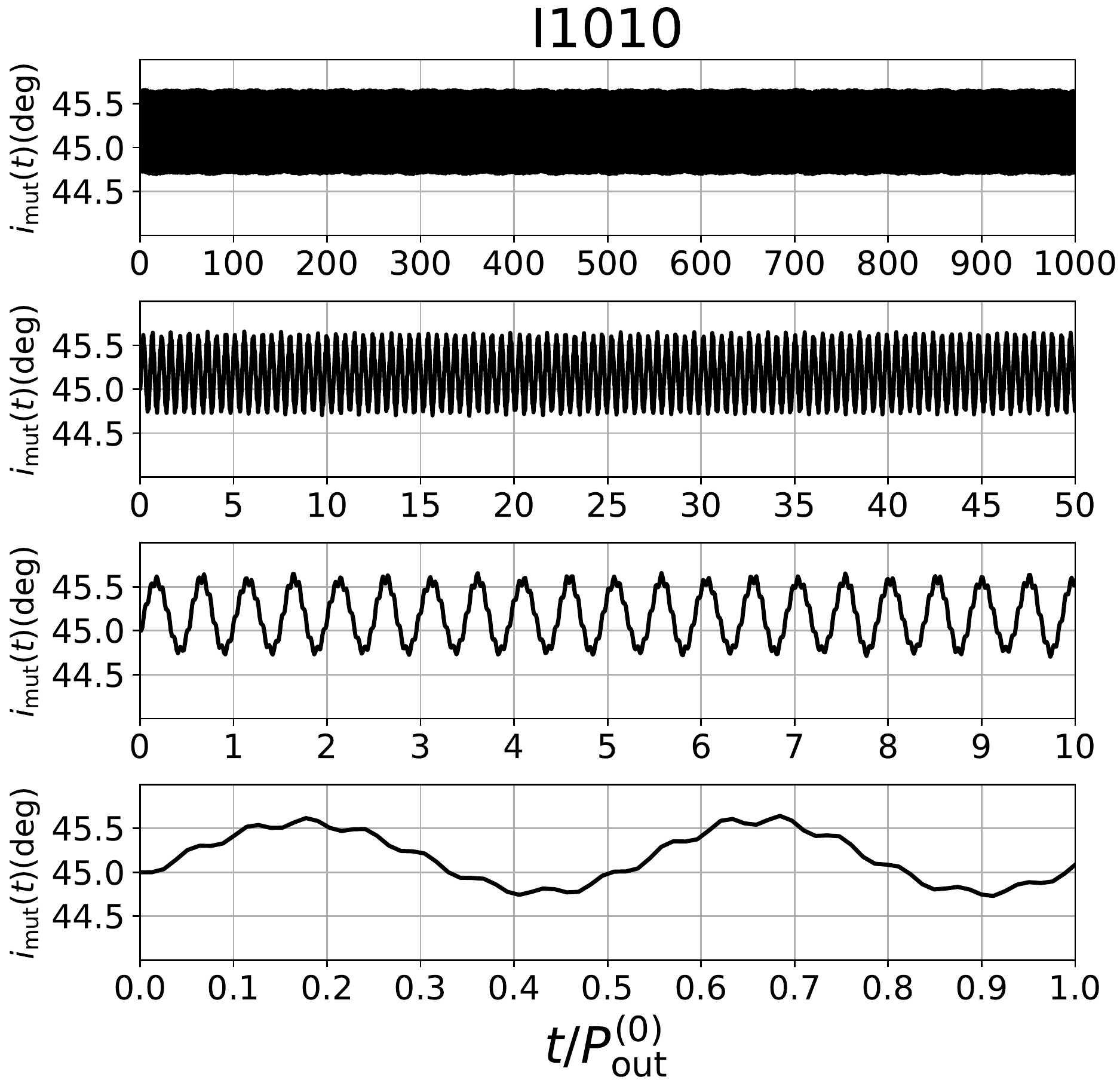}
 	\end{center}
 	\caption{The enlarged plot of the top right panel of Figure 8
for I1010 in order to clarify the detailed evolution of mutual
inclination.The first, second, third, and forth panels from top show 
the evolution over $0 - 1000P_\oo$, $0 - 50P_\oo$, $0 - 10P_\oo$, and $0 - 1.0P_\oo$, respectively. \label{fig:c1}}
 \end{figure*}


\end{document}